\documentstyle[preprint,tighten,floats,aps,epsf]{revtex}

\begin{document}
\title
{Structure of Light Unstable Nuclei Studied with Antisymmetrized Molecular
Dynamics}

\author{Yoshiko {\sc Kanada-En'yo} and Hisashi {\sc Horiuchi}$^{*}$
}
\address{Institute of Particle and Nuclear Studies, \\
High Energy Accelerator Research Organization,\\
1-1 Oho, Tsukuba, Ibaraki 305-0801, Japan\\
$^{*}$Department of Physics, Kyoto University, Kyoto 606-8502, Japan}

\maketitle

\abstract{Structures of light unstable nuclei, Li, Be, B, and C isotopes
are systematically studied with a microscopic method of antisymmetrized 
molecular dynamics.
The theoretical method is found to be very useful to study ground and 
excited states of various nuclei covering unstable nuclei.
The calculations succeed to reproduce many experimental 
data for nuclear structures; energies, radii, 
magnetic dipole moments, electric quadrupole moments, transition 
strength. In the theoretical results it is found that
various exotic phenomena in unstable nuclei 
such as molecular-like structures, neutron skin, and large deformations 
may appear in unstabel nuclei.
We investigate the structure change
with the increase of neutron number and with the increase of the excitation 
energies, and find the drastic changes between shell-model-like structures 
and clustering structures. 
The mechanism of clustering developments in unstable
nuclei are discussed.
}

\tableofcontents
\section{Introduction}
  Owing to the progress of the experimental technique, the information
of the ground and excited states of unstable nuclei is
increasing rapidly. There are various interesting subjects which are 
characteristics of unstable nuclei.
Our aim is to systematically study the nuclear structures of light nuclei
covering stable and unstable regions in order to understand the 
features of the nuclear many-body system. 
 In the light nuclear region, 
ground-state properties have become known experimentally
for the unstable nuclei up to the drip line.
The experimental data give various kinds of information such as
binding energies, radii, magnetic dipole moments, 
and electric quadrupole moments, and so on 
\cite{CONF,LOUV,KANAZAWA,NIIGATA,TANIHATAb,HALO,HALOb,ARNOLDa,ARNOLDb,ASAHIa,ASAHIc,MATSUTAa,C19}.
By the help of these experimental data, 
many interesting phenomena of the structures of the unstable nuclei 
have been suggested; neutron halo and skin structures, 
vanishing of the magic number, abnormal spin-parity of the ground state,
clustering structures, large deformations in unstable nuclei. 

 In many of the theoretical studies for light nuclei,
inert cores or clusters have been assumed. 
For example, three-body models
have been applied to $^{11}$Li by regarding it as a $^9$Li+$2n$ 
system and also applied to $^6$He as an $\alpha$+$2n$ system
\cite{THREE,ZZZZ,SUZUKIa}. In these studies the neutron-halo structure 
has been theoretically investigated.
In the studies of very light nuclei with $A\leq 10$ 
with extended cluster models in which $\alpha$ or $t$ cluster cores
and surrounding nucleons are assumed \cite{SUZUKIb,OGAWA,ITAGAKI}, 
it is suggested that the 
clustering structure may appear also in unstable nuclei.
However, they have not yet reached to the 
systematic investigations covering the wide region of nuclei,
since it is difficult to apply these models to the study of 
general heavier unstable nuclei. 
 For heavier unstable nuclei, systematic studies have been done
by using other theoretical frameworks such as shell model approaches 
\cite{UTSUNO} and the methods of mean field theory
\cite{TAJIMA,RING}.
Some of them suggested new features
such as large deformations and neutron skin structures.
However the applicability of mean-filed approaches is not
necessarily assured in light nuclei because of possible clustering 
structures.

 It is already well known that clustering structures 
\cite{PTPSC,ARIMAa,TANGa,NEUDC,HOIKTX}
appear in the ground states of ordinary light nuclei 
with $N\approx Z$ as seen in the 
$t+\alpha$ cluster structure of $^7$Li and in the $\alpha+\alpha$
cluster structure of $^8$Be, and in the $^{16}$O+$\alpha$ cluster structure
 of $^{20}$Ne.
Therefore, in light unstable nuclei one of the problems to be solved 
is the clustering structure.
Is the clustering seen also in unstable nuclei ?
If it is the case, what is the feature of the clusters. 
How does the developed clustering structure of stable nuclei 
change with the increase of the neutron number 
in a series of isotopes ?
It is naturally expected that the clustering structures of unstable 
nuclei are found not only in the ground state but also in the excited states. 
Recent experimental data suggest that 
the clustering structure may appear 
in excited states of unstable nuclei.  
For example, many excited states of $^{12}$Be have been found
with the experiments of the breakup reactions 
from $^{12}$Be to $^6$He+$^6$He and $^8$He+$^4$He channels
\cite{FREER}.
In theoretical studies, many groups 
have suggested clustering structures 
in neutron-rich nuclei and tried to discuss the features of the clustering
\cite{SUZUKIb,OGAWA,ITAGAKI,ENYOb,ENYOc}.
It is essential for the systematic studies of unstable
nuclei to describe both clustering aspects and 
shell-model-like aspects in one theoretical framework which is applicable to 
general nuclei. We are interested in how the structure changes with the
increase of the neutron number and the excitation energy.

 Our aim is to make systematic investigations of the structures of 
the ground and excited states of unstable nuclei with a microscopic model
which is free from such assumptions as the inert core or the existence 
of clusters.
Remarking the development of clustering structure,  
we try to understand many characteristic phenomena seen in  
unstable nuclei. We discuss the mechanism of clustering developments
of unstable nuclei. 

 In this paper, we adopt a theoretical method of 
antisymmetrized molecular dynamics(AMD).
Ono et al. have developed the method of AMD for the study of
nuclear reactions \cite{ONO,ONOa,ONOb,ONOc,ONOd}.
The framework of AMD has been extended by Kanada-En'yo
(one of the author of this paper) et al., and 
has been applied to the studies of nuclear 
structures \cite{ENYOb,ENYOc,ENYOa,DOTE,ENYOe,ENYOg}.
From these studies AMD has already proved to be a 
very useful theoretical approach for investigating the 
structure of the ground and excited states of 
light nuclei.
In the AMD framework, basis wave functions of the system are given by 
Slater determinants where the spatial part of each single-particle
wave function is a Gaussian wave packet.
One of the characteristics of AMD is the flexibility of the 
wave function which can represent various clustering structures 
and shell-model-like structures, which is
 because no inert cores and no clusters
are assumed. Another characteristic point of AMD is the 
frictional cooling method 
which is adopted in the energy variation
for obtaining the ground and excited states.
 In the case of the simplest version of the AMD \cite{ENYOb,ENYOc}
in this paper, the energy variation is made after the parity projection but 
before the total-spin projection. With this simplest method,
we can describe low-lying levels of 
the lowest bands with positive and negative parities.
  For the study of the excited states we adopt the AMD approach
which users the variation after the spin-parity projection (VAP).
VAP calculations in AMD framework have been already found to be advantageous
for the study of the excited states of light nuclei \cite{ENYOe,ENYOg}.
By the use of microscopic calculations with the obtained 
wave functions, we can easily acquire the theoretical values of 
energies, radii, magnetic 
dipole moments, electric quadrupole moments, and the 
transition strength of $E1$, $E2$, and $\beta$ which are useful 
quantities to deduce the informations of structures from the 
experimental data.

This paper is organized as follows. In the next section 
(Sec. \ref{sec:formura}) we explain the formulation of AMD for the 
study of nuclear structure. The effective interactions 
are described in Sec. \ref{sec:interatcion}.
In Sec. \ref{sec:amd} we show the results and give discussions of the
Li, Be, B, and C isotopes based on the calculations of the simplest 
version of AMD.
The study with VAP calculations in the framework of AMD 
is reported in Sec. \ref{sec:vap}, where we discuss the structure
of the excited states of neutron-rich Be isotopes. 
In Sec. \ref{sec:cluster}, we mention about the mechanism of clustering 
developments of Be isotopes. Finally we give the summary in 
Sec.\ref{sec:summary}. 

\section{Formulation } \label{sec:formura}
The AMD (antisymmetrized molecular dynamics) is a theory which is applicable 
to the studies of the  nuclear structure and the nuclear reaction.
Here we only explain the AMD framework for the study of nuclear 
structures. As for the AMD theory for the study of nuclear reaction,
the reader is referred to Ref.\cite{ONOa}.
\subsection{AMD wave function}
In AMD framework, the wave function of a system is written 
by a linear combination of AMD wave functions, 
\begin{equation}
\Phi=c \Phi_{AMD} +c' \Phi '_{AMD} + \cdots .
\end{equation}
An AMD wavefunction is a Slater determinant of Gaussian wave packets;
\begin{eqnarray}
&\Phi_{AMD}({\bf Z})=\frac{1}{\sqrt{A!}}
{\cal A}\{\varphi_1,\varphi_2,\cdots,\varphi_A\},\\
&\varphi_i=\phi_{{\bf X}_i}\chi_i\tau_i :\left\lbrace
\begin{array}{l}
\phi_{{\bf X}_i}({\bf r}_j) \propto
\exp\left 
[-\nu\biggl({\bf r}_j-\frac{{\bf X}_i}{\sqrt{\nu}}\biggr)^2\right],\\
\chi_{i}=
\left(\begin{array}{l}
{1\over 2}+\xi_{i}\\
{1\over 2}-\xi_{i}
\end{array}\right),
\end{array}\right. 
\end{eqnarray}
where $\chi_i$ is the intrinsic spin function parameterized by
$\xi_{i}$, and $\tau_i$ is the isospin
function which is up(proton) or down(neutron).
Thus an AMD wave function is parameterized by a set of complex parameters
${\bf Z}\equiv \{X_{ni},\xi_i\}\ (n=1\sim 3\ \hbox{and }  i=1 \sim A)$. 

If we consider a parity eigenstate projected from a Slater determinant 
the total wave function consists of two Slater determinants,
\begin{equation}
\Phi({\bf Z})=(1\pm P) \Phi_{AMD}({\bf Z}),
\end{equation}
where $P$ is a parity projection operator.
In case of total angular momentum eigenstates 
the wave function of a system
is represented by integral of rotated states,
\begin{equation}
\Phi({\bf Z})=P^J_{MK'}\Phi_{AMD}({\bf Z}) = 
\int d\Omega D^{J*}_{MK'}(\Omega)R(\Omega)\Phi_{AMD}({\bf Z}),
\end{equation}
The expectation values of operators by $\Phi({\bf Z})$
are numerically  calculated by 
a summation over mesh points of the Euler angles $\Omega$. 

In principle the total wave function can be a superposition of independent
AMD wave functions. 
We can consider a superposition 
of spin parity projected AMD wave functions $P^{J\pm}_{MK'}\Phi_{AMD}$'s, 
\begin{equation}
\Phi({\bf Z},{\bf Z}')=cP^{J\pm}_{MK'}\Phi_{AMD}({\bf Z})
+c'P^{J\pm}_{MK'}\Phi_{AMD}({\bf Z}')+\cdots.
\end{equation}

\subsection{Energy variation \label{subsec:variation}}

We make variational calculations
to find the state which minimizes the energy of the system;
\begin{equation}
{\cal E}=\frac{\langle\Phi|H|\Phi\rangle}{\langle\Phi|\Phi\rangle}
\end{equation}
by the method of frictional cooling.
Concerning with the frictional cooling method in AMD, the reader is referred to 
papers \cite{ENYOb,ENYOa}.
For the wave function $\Phi({\bf Z})$ 
parameterized by complex parameters ${\bf Z}$, 
the time development of the parameters is determined by the frictional cooling
equations,
\begin{eqnarray}
\frac{dX_{n k}}{dt}&=&
(\lambda+i\mu)\frac{1}{i \hbar} \frac{\partial 
{\cal E}({\bf Z},{\bf Z}')}{\partial X^*_{n k} }=
(\lambda+i\mu)\frac{1}{i \hbar} \frac{\partial}{\partial X^*_{n k} }
\frac{\langle \Phi({\bf Z})|H|\Phi({\bf Z})\rangle}{\langle \Phi({\bf Z})
|\Phi({\bf Z})\rangle},
\quad (n=1 \sim 3\quad k=1 \sim A)\\
\frac{d\xi_{k}}{dt}&=&
(\lambda+i\mu)\frac{1}{i\hbar}
\frac{\partial {\cal E}({\bf Z},{\bf Z}')}{\partial\xi^*_{k}}=
(\lambda+i\mu)\frac{1}{i\hbar}
\frac{\partial}{\partial\xi^*_{k}}
\frac{\langle \Phi({\bf Z})|H|\Phi({\bf Z})\rangle}{\langle \Phi({\bf Z})
|\Phi({\bf Z})\rangle},
\quad (k=1 \sim A)
\end{eqnarray}
with arbitrary real numbers $\lambda$ and $\mu < 0$. It is easily proved that 
 the energy of the system decreases with time as follows,
\begin{eqnarray}
\frac{d{\cal E}}{dt}&=&\sum^A_{i=1}\left(
\frac{\partial {\cal E}}{\partial 
{\bf X}_i}\cdot \frac{d {\bf X}_i}{dt}+
\frac{\partial {\cal E}}{\partial 
{\bf X}^*_i}\cdot \frac{d {\bf X}^*_i}{dt}+
\frac{\partial {\cal E}}{\partial \xi_i}\cdot \frac{d \xi_i}{dt}+
\frac{\partial {\cal E}}{\partial 
\xi^*_i}\cdot \frac{d \xi^*_i}{dt}\right)\\
&=&\frac{2\mu}{\hbar}\sum^A_{i=1}
\left(\frac{\partial {\cal E}}{\partial 
{\bf X}_i}\cdot
\frac{\partial {\cal E}}{\partial 
{\bf X}^*_i}+
\frac{\partial \cal E}{\partial 
\xi_i}\cdot
\frac{\partial \cal E}{\partial 
\xi^*_i}\right)<0.
\end{eqnarray}
After sufficient time
steps for cooling, the wave function of the minimum-energy state is obtained.

\subsection{Angular momentum projection}
Expectation values of a given tensor operator $T^k_q$(rank $k$) for the 
total-angular-momentum projected states are written as follows,
\begin{eqnarray}
%\begin{equation}
&&\langle P^{J_1}_{M_1K_1}\Phi_1|T^k_q|P^{J_2}_{M_2K_2}\Phi_2\rangle \\
&&=\frac{2J_2+1}{8\pi^2} (J_2M_2kq|J_1M_1)
\sum_{K\nu}(J_2Kk\nu |J_1K_1) \int d\Omega D^{J_2*}_{KK_2}(\Omega)
\langle\Phi_1|T^k_\nu R(\Omega)|\Phi_2\rangle, 
\end{eqnarray}  
%\end{equation}  
where $D^J_{MK}$ are the well-known Wigner's D functions and $R(\Omega)$
stands for the rotation operator with Euler angles $\Omega$.
In the practical calculations, the three-dimensional integral is evaluated 
numerically by taking a finite number of mesh points of the Euler angles
$\Omega$.

\subsection{Simplest version of AMD for the study of nuclear structure} 
In the simplest version of AMD for the study of nuclear structure,
the  ground state wave function of a system is constructed 
by the energy variation of the parity 
eigenstate projected from a Slater determinant.
Furthermore, the directions of intrinsic spins of single particle 
wave function are fixed to be up and down as
$\xi_i=\pm \frac{1}{2}$ for simplicity. 
Therefore the spin-isospin functions of single-particle 
wave function are chosen as $p\uparrow$, $p\downarrow$, $n\uparrow$,
and $n\downarrow$ in the initial state and are fixed  
in the energy variation.
In this case the total wave function of a system is parameterized only by 
${\bf X}\equiv \{{\bf X}_1,{\bf X}_2,\cdots,{\bf X}_A\}$ which are the 
centroids of Gaussian wave packets in the phase space,
\begin{equation}
\Phi({\bf X})=(1\pm P) \Phi_{AMD}({\bf X}).
\end{equation}
We regard the minimum-energy state 
obtained with the energy variation 
(described in \ref{subsec:variation}) for the parity projected state   
as the intrinsic state of the system. In order to compare with experimental
data, we project the intrinsic wave function to the total-angular-momentum
eigenstates and calculate the expectation values of operators.
In that sense, ``the simplest version of AMD'' stands for
 the variational calculations after the parity projection 
but variation before projection
(VBP) with respect to the total-angular momentum in this paper.
In the same way as the ground state, the lowest non-normal parity state
is calculated by energy variation for the non-normal parity projected 
state.

\subsection{Variation after projection}
The wave function of the system should 
be a total-angular-momentum eigenstates.
We can perform energy variation after the spin-parity 
projection(VAP) with the method of 
frictional cooling for the trial function $\Phi=P^{J\pm}_{MK'}
\Phi_{AMD({\bf Z})}$ \cite{ENYOe}. 

First we make VBP calculation to prepare
an initial state $\Phi_{AMD}({\bf Z}_{init})$ for the VAP calculation.
We choose an appropriate $K'$ quantum number for each spin parity 
that makes the energy expectation value for the spin parity eigenstate
$\langle P^{J\pm}_{MK'}\Phi_{AMD}({\bf Z}_{init})|H|P^{J\pm}_{MK'}
\Phi_{AMD}({\bf Z}_{init}) \rangle/ 
\langle P^{J\pm}_{MK'}({\bf Z}_{init})|P^{J\pm}_{MK'}({\bf Z}_{init})
 \rangle$ minimum.
$K'$ is a component of the total angular momentum along
the approximately principal axis on the intrinsic system.
In order to obtain the wave function for the lowest $J^\pm$ state, 
we perform VAP calculation
for $\langle P^{J\pm}_{MK'}\Phi_{AMD}({\bf Z})|H|P^{J\pm}_{MK'}
\Phi_{AMD}({\bf Z}) \rangle/ 
\langle P^{J\pm}_{MK'}\Phi_{AMD}({\bf Z})|P^{J\pm}_{MK'}\Phi_{AMD}({\bf Z})
 \rangle$ with the appropriate $K'$
quantum number chosen for the initial state.
In the VAP procedure, the principal $z$-axis 
of the intrinsic deformation is not assumed to equal with the
 $3$-axis of Euler angle in the total angular momentum
projection. In general the principal $z$-axis is automatically 
determined in the energy variation. That is to say, the spin 
parity eigenstate $P^{J\pm}_{MK'}\Phi_{AMD}$ obtained by VAP with 
a given $K'=\langle J_3\rangle$ can be the state with 
so-called $K=\langle J_z\rangle$ quantum number mixing 
in terms of the intrinsic deformation.

\subsection{Higher excited states \label{subsec:excited}}
As mentioned above,
with the VAP calculation for $\Phi({\bf Z})=P^{J\pm}_{MK'}\Phi_{AMD}({\bf Z})$
of the $J^\pm$ eigenstate with $K'$
we obtain the wave function for the lowest $J^\pm$ state, which is 
represented by the set of parameters ${\bf Z}={\bf Z}^{J\pm}_1$. 
To search the parameters for the higher excited $J^\pm_n$ states, 
the wave functions are superposed so as to be orthogonal 
to the lower states as follows. The parameters ${\bf Z}^{J\pm}_n$ 
for the $n$-th
$J^\pm$ state are reached by varying the energy of the orthogonal
component to the lower states;
\begin{equation}
\Phi({\bf Z})=P^{J\pm}_{MK'}\Phi_{AMD}({\bf Z})-\sum^{n-1}_{k=1}
{\langle P^{J\pm}_{MK'}\Phi_{AMD}({\bf Z}^{J\pm}_k)
|P^{J\pm}_{MK'}\Phi_{AMD}({\bf Z})\rangle 
\over
\langle P^{J\pm}_{MK'}\Phi_{AMD}({\bf Z}^{J\pm}_k)
|P^{J\pm}_{MK'}\Phi_{AMD}({\bf Z}^{J\pm}_k)\rangle} 
P^{J\pm}_{MK'}\Phi_{AMD}({\bf Z}^{J\pm}_k).\label{eqn:excite}
\end{equation}

In the present paper, we call the variational calculation 
after the spin parity projection (mentioned in previous subsection) 
and the calculation for the higher excited
states described in this subsection as VAP calculations.

\subsection{Diagonalization in VAP}

After VAP calculations for various $J^\pm_n$ states mentioned above, 
the intrinsic states
$\Phi^1_{AMD}, 
\Phi^2_{AMD},\cdots, 
\Phi^m_{AMD}$, 
which correspond to the $J^\pm_n$ states, 
are obtained as much as the number 
of the calculated levels.
Finally we construct the improved wave functions for the $J^\pm_n$ states
by diagonalizing the Hamiltonian matrix 
$\langle P^{J\pm}_{MK'} \Phi^i_{AMD}
|H|P^{J\pm}_{MK''} \Phi^j_{AMD}\rangle$
and the norm matrix
$\langle P^{J\pm}_{MK'} \Phi^i_{AMD}
|P^{J\pm}_{MK''} \Phi^j_{AMD}\rangle$
simultaneously with regard to ($i,j$) for 
all the intrinsic states and ($K', K''$).
In comparison with the experimental data such as energy levels and
 $E2$ transitions, the theoretical values are calculated with the 
final states after diagonalization.

\section{Interaction } \label{sec:interatcion}
For the effective two-nucleon interaction, we adopt the Volkov No.1 force
\cite{VOLKOV}
as the central force. The adopted parameters in this paper 
contain only Wigner and Majorana 
components. For some nuclei we have performed calculations by adding 
appropriate Bartlett and Heisenberg components to the Volkov force. However
the results have proved to be not so much affected by 
the additional components except for the binding energies.
Instead of the Volkov force, we also adopt case (1) and case (3) of 
MV1 force \cite{TOHSAKI}, which contain the zero-range three-body 
force $V^{(3)}$ as density dependent terms in addition to the two-body 
interaction $V^{(2)}$, 
\begin{eqnarray}
 V_{DD}&=&V^{(2)}+V^{(3)}\\
V^{(2)}&=&(1-m+b P_\sigma-h P_\tau -m P_\sigma P_\tau )
\left\lbrace 
V_A \exp\left[-\left(\frac{r}{r_A}\right)^2\right]+ 
V_R \exp\left[-\left(\frac{r}{r_R}\right)^2\right]\right\rbrace,\\
&& b=h=0,\nonumber\\
&& V_A = -83.34 \hbox{MeV}, r_A = 1.60 \hbox{fm},\nonumber\\
&& V_R = 104.86 \hbox{MeV (case 1)},\qquad 99.86 
\hbox{MeV (case 3)}, r_R = 0.82 \hbox{fm},\nonumber\\
 V^{(3)}& =& v^{(3)}\delta({\bf r}_1-{\bf r}_2)
\delta({\bf r}_1-{\bf r}_3),\nonumber\\
&&v^{(3)} = 4000 \hbox{MeV fm$^6$ for (case 1)},\qquad 5000 
\hbox{MeV fm$^6$ (case 3)}\nonumber,
\end{eqnarray}
where $P_\sigma$ and $P_\tau$ stand for spin and isospin exchange 
operators, respectively, and 
$r$ denotes $|{\bf r}|\equiv |{\bf r}_1-{\bf r}_2|$.
As for the two-body spin-orbit force $V_{LS}$, we adopt the G3RS force 
\cite{LS} ;
\begin{eqnarray}
&& V_{LS}= \left\{ u_I \exp\left(-\kappa_I r^2\right) +
u_{II} \exp\left(-\kappa_{II} r^2\right)\right\} P(^3O)
{\bf L}\cdot ({\bf S_1}+{\bf S_2}),\\
&& {\bf L}\equiv {\bf r}\times \left(-i\frac{\partial}{\partial {\bf r}}
\right),\qquad \kappa_I = 5.0 \hbox{fm}^{-2}, 
 \kappa_{II} = 2.778 \hbox{fm}^{-2}, 
\end{eqnarray}
with $P(^3O)$
denoting the projection operator onto the triplet odd$(^3O)$ two-nucleon state.
The Coulomb interaction is approximated by a sum of seven Gaussians.

\section{Study with simplest version of AMD} \label{sec:amd}

Since the wave function should be a total-angular-momentum eigenstate,
it is expected that the VAP calculation gives better results than the simplest
version of AMD. It is, however, not easy to perform VAP calculation
because of the three-dimensional integral for the total-spin projection
which is evaluated by 
taking a large number of mesh points of the Euler angles.
In order to study systematically the structures of ground states of light 
nuclei covering from the ordinary region to the exotic unstable region, we
perform the simplest version of AMD calculations (energy variation for the 
parity projected state) for even-odd, odd-even 
and even-even isotopes of Li, Be, B and C. 
The obtained states are projected to the 
total-angular-momentum eigenstates in calculating the expectation 
values to compare the results with the experimental data. 
Fortunately, 
in many nuclei with some exceptions, 
the obtained structures are not so much different 
from the ones obtained with VAP calculations.
The results with VAP will be
shown in the next section. 

\subsection{Results \label{subsec:amdres}}
In this section the theoretical results of the simplest version of AMD
are compared with the experimental
data.

We have used the Volkov force with 
Majorana parameter $m=0.56$ ((a)), and 
the case 1 of MV1 force with $m=0.576$ ((b)) for the Li and Be isotopes.
We have also adopted the MV1 force with mass
dependent $m$ parameters ((c)) for B isotopes, and 
the MV1 force with $m=0.63$ ((d)).
The adopted parameters are listed in Table \ref{tbl:intparam} and
\ref{tbl:intparamb}.
The details are explained in each place.
The optimum width parameters $\nu$ are shown in Table \ref{tbl:libenu},
\ref{tbl:bnu},\ref{tbl:cnu}.
The expectation values for the operators of observable quantities
are calculated by projecting the intrinsic states
obtained with the simplest version of AMD into the eigenstates of 
parity and total-angular-momentum.
Most theoretical values except for the energy levels 
are obtained without the mixing of $K$-quantum which 
is the component of the spin $J$ along the approximate principal
axis of the intrinsic system. Instead, we choose an appropriate 
$K$ which gives the minimum energy for each spin of a system.   
We have diagonalized the Hamiltonian matrix with respect to the
$K$ quantum number within the spin $J$ projected 
states to calculate energy levels, and found only small mixing of the $K$
quantum number which implies that 
$K$ is approximately a good quantum number in the lowest $J^\pm$ states
projected from the intrinsic states.

\begin{table}
\caption{\label{tbl:intparam} The adopted interaction parameters.}

\begin{center}
\begin{tabular}{cccc}
(a)& Volkov No.1 & $m=0.56$, $b=h=0$ & $u_I=-u_{II}=900$ MeV \\ 
(b)& MV1 case(1)& $m=0.576$, $b=h=0$ & $u_I=-u_{II}=900$ MeV \\ 
(d)& MV1 case(1)& $m=0.63$, $b=h=0$ & $u_I=-u_{II}=900$ MeV \\ 
(e)& MV1 case(1)& $m=0.336$, $b=-0.2$, $h=0.4$ & $u_I=-u_{II}=1500$ MeV \\ 
\end{tabular}
\end{center}

\caption{\label{tbl:intparamb} The mass dependent interaction parameters 
of the force (c) adopted for B isotopes .}
\begin{center}
\begin{tabular}{cccc}
$^{11}$B, $^{13}$B 
& MV1 case(1)& $m=0.576$, $b=h=0$ & $u_I=-u_{II}=900$ MeV \\ 
$^{15}$B
 & MV1 case(1)& $m=0.63$, $b=h=0$ & $u_I=-u_{II}=900$ MeV \\ 
$^{17}$B, $^{19}$B 
& MV1 case(1)& $m=0.65$, $b=h=0$ & $u_I=-u_{II}=900$ MeV \\ 
\end{tabular}
\end{center}

\end{table}

\begin{table}
\caption {\label{tbl:libenu}The width parameters $\nu$ of Gaussian 
wave packets adopted in the calculations of the simplest version of AMD
for Li and Be isotopes.
The optimum width parameter is chosen for each nucleus and each set of 
interactions.
}
\begin{center}
\begin{tabular}{cc|ccc}
& &\multicolumn{2}{c}{width parameter $\nu$ (fm$^{-2}$)}&\\
 & interaction  & (a) 
& (b)& \\
\hline
& & & &\\
& $^6$Be$(+)$   & 0.215        & 0.195         &\\
& $^7$Be$(-)$   & 0.230        & 0.200         &\\
& $^8$Be$(+)$   & 0.250        & 0.205         &\\
& $^9$Be$(-)$   & 0.245        & 0.195         &\\
& $^9$Be$(+)$   & 0.235        & 0.200         &\\
& $^{10}$Be$(+)$   & 0.230        & 0.190         &\\
& $^{10}$Be$(-)$   & 0.225        & 0.190         &\\
& $^{11}$Be$(-)$   & 0.220        & 0.180         &\\
& $^{11}$Be$(+)$   & 0.220        & 0.180         &\\
& $^{12}$Be$(+)$   & 0.215        & 0.175         &\\
& $^{12}$Be$(-)$   & 0.210        & 0.180         &\\
& $^{13}$Be$(+)$   & 0.205        & 0.170         &\\
& $^{14}$Be$(+)$   & 0.210        & 0.170         &\\
&                 & ¡¡            &               &\\
\hline
&  $^{7}$Li$(-)$   & 0.230        & 0.200         &\\
&  $^{9}$Li$(+)$   & 0.210          & 0.180         &\\
& $^{{1}{1}}$Li$(+)$   &  0.195         & 0.170         &\\
\end{tabular}
\end{center}
\end{table}

\begin{table}
\caption {\label{tbl:bnu}The optimum width parameters $\nu$ of Gaussian 
wave packets adopted for B isotopes.
}
\begin{center}
\begin{tabular}{cccccc}
 & \multicolumn{5}{c}{Width parameter $\nu$ (fm$^{-2}$)} \\
\hline
Interaction & $^{11}$B & $^{13}$B & $^{15}$B & $^{17}$B & $^{19}$B \\
\hline
$m$=0.576 & 0.185 & 0.175 & 0.175 & 0.165 & 0.160 \\
$m$=0.63 &0.170 &0.160 & 0.155 &0.150 &0.145 \\
$m$=0.65 & $-$ &0.155 &0.150 &0.150 &0.135\\
\end{tabular}

\end{center}
\end{table}

\begin{table}
\caption {\label{tbl:cnu}The optimum 
width parameters $\nu$ of Gaussian wave packets adopted for C isotopes.
}
\begin{center}
\begin{tabular}{cc|ccc}
& &\multicolumn{2}{c}{width parameter $\nu$ (fm$^{-2}$)}&\\
 & interaction  & (b) 
& (d)& \\
\hline
& & & &\\
& $^9$C$(-)$   &  0.18  & 0.170               &\\
& $^{10}$C$(+)$ & 0.19  & 0.180            &\\
& $^{11}$C$(-)$ & 0.19  & 0.175             &\\
& $^{12}$C$(+)$ & 0.19  & 0.175             &\\
& $^{13}$C$(-)$ & 0.18  & 0.170             &\\
& $^{14}$C$(+)$ & 0.18  & 0.165            &\\
& $^{15}$C$(+)$ & 0.17  & 0.160             &\\
& $^{16}$C$(+)$ & 0.17  & 0.160            &\\
& $^{17}$C$(+)$ & 0.17  & 0.155             &\\
& $^{18}$C$(+)$ & 0.17  & 0.150            &\\
& $^{19}$C$(+)$ & 0.16  & 0.150           &\\
& $^{20}$C$(+)$ & 0.16  & 0.145           &\\
& $^{22}$C$(+)$ & 0.16  & 0.140           &\\
&                 & ¡¡            &               &\\
\end{tabular}
\end{center}
\end{table}

\subsubsection{Energies}\label{subsub:energy}
\begin{figure}
\caption{\label{fig:libeene}
The binding energies of Li and Be isotopes.
The circles(triangles) are the theoretical values with 
the interaction (a) ((b)).
The square points indicate the experimental data.
}
\centerline{
\epsfxsize=1.0\textwidth
\epsffile{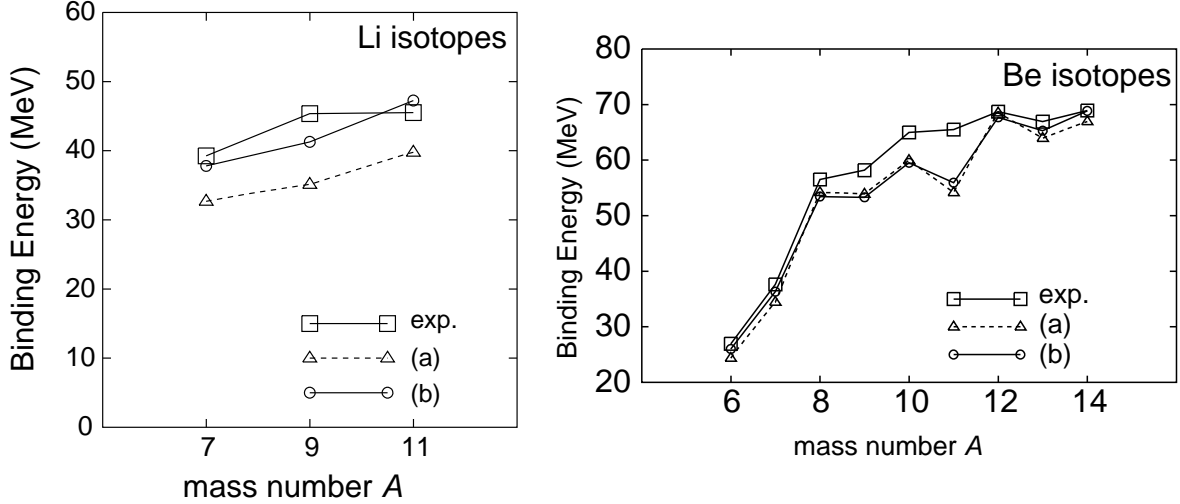}
}
\end{figure}

Figure \ref{fig:libeene} shows the binding energies of the ground states
of Li and Be isotopes. 
With both of the interactions (a) and (b), 
the binding energies of Li and Be isotopes are qualitatively 
reproduced.
The calculated result of $^{11}$Be is the binding energy of the lowest 
$\frac{1}{2}^+$ state, though the normal-parity $\frac{1}{2}^-$ state is 
lower than $\frac{1}{2}^+$ state in these calculations.
Detailed discussion of the energy levels and the parity of the ground state
of $^{11}$Be will be given later.

%%%%%%%%%%%% FIGURE %%%%%%%%%%%%%%%%%%%%%%%%%
%%%%%%%%%%%% FIGURE %%%%%%%%%%%%%%%%%%%%%%%%%

%%%%%%%%%%%% FIGURE %%%%%%%%%%%%%%%%%%%%%%%%%
\begin{figure}
\caption{\label{fig:bene}
The binding energies of B isotopes.
The circles are the theoretical values calculated with 
the MV1 force with $m=0.576$, $0.63$, and $0.65$.
The square points indicate the experimental data.
}
\centerline{
\epsfxsize=0.6\textwidth
\epsffile{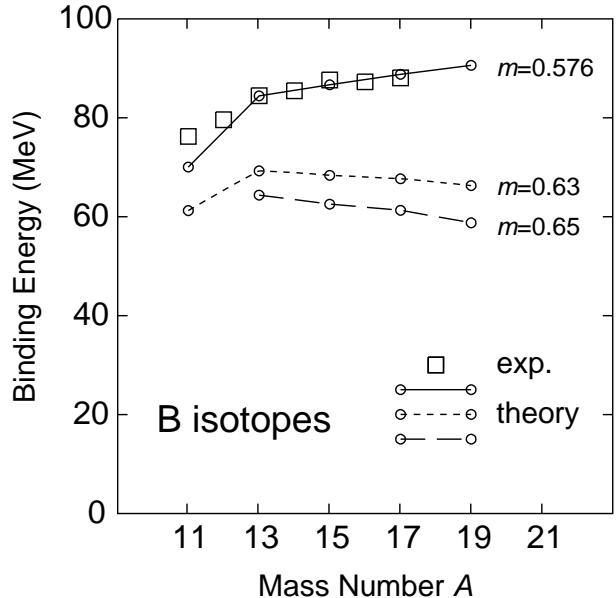}
}
\end{figure}
%%%%%%%%%%%% FIGURE %%%%%%%%%%%%%%%%%%%%%%%%%

%%%%%%%%%%%% FIGURE %%%%%%%%%%%%%%%%%%%%%%%%%
\begin{figure}
\caption{\label{fig:cene}
The binding energies of C isotopes.
The lines are the theoretical values calculated with 
the MV1 force with $m=0.576$ and $0.63$.
The figure also shows the calculations with the Volkov No.1 force with
$m=0.6$.
The square points indicate the experimental data.}
\centerline{
\epsfxsize=0.6\textwidth
\epsffile{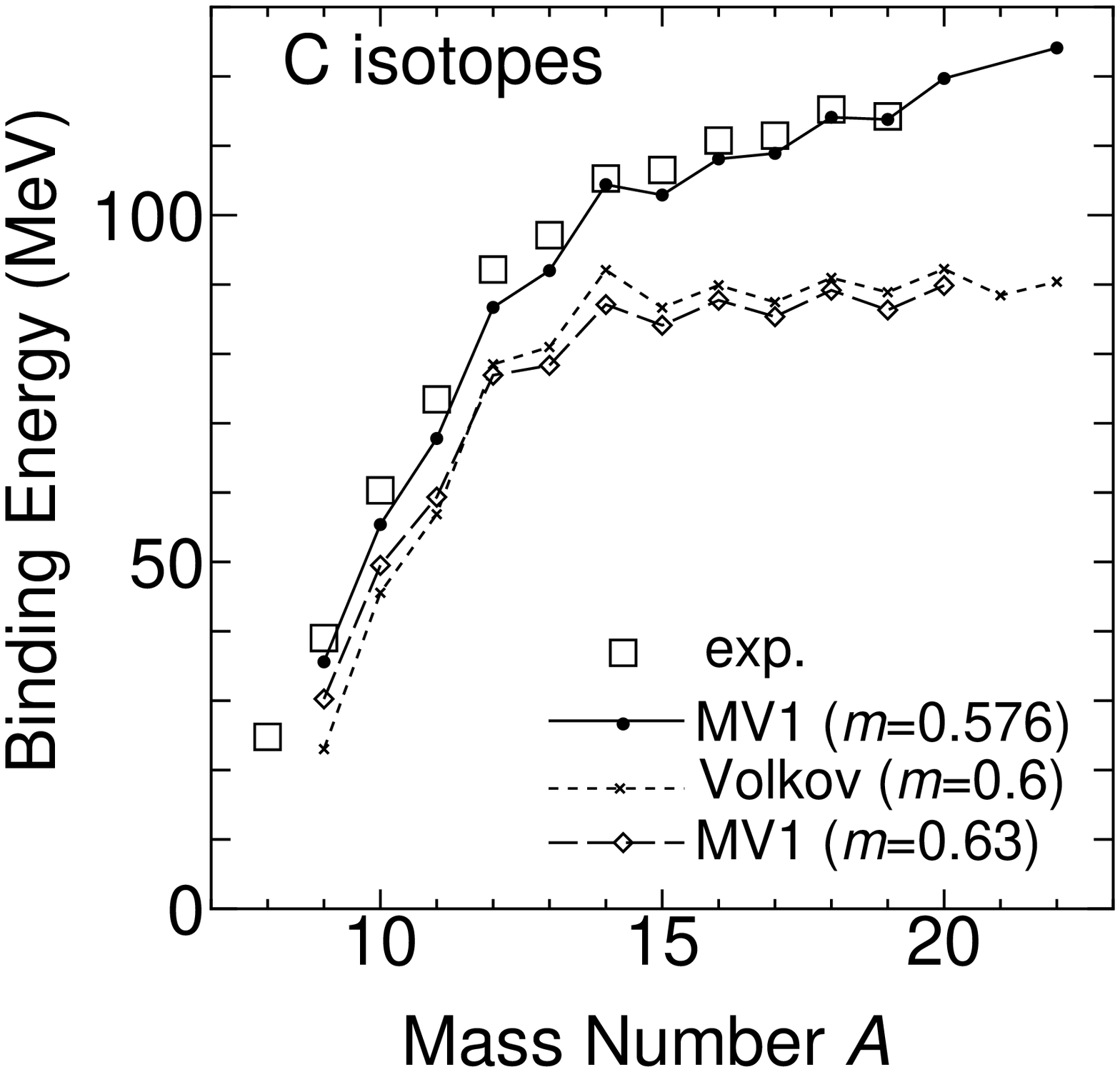}
}
\end{figure}
%%%%%%%%%%%% FIGURE %%%%%%%%%%%%%%%%%%%%%%%%%

The binding energies of B and C isotopes are presented in 
Fig.\ref{fig:bene} and Fig.\ref{fig:cene}, respectively.
For B isotopes, the figure shows theoretical results with the MV1 force
with $m=0.576$, $m=0.63$, and $m=0.65$. In Fig. \ref{fig:cene},
results of C isotopes with Volkov($m=0.60$) and MV1 force with $m=0.576$ and
$m=0.63$ are shown. For $^{15}$C the energies of $5/2^+$ states are shown, 
although the ground state should be $1/2^+$.
In both B and C isotopes, MV1 force with $m=0.576$ reproduces well the 
experimental data.

%%%%%%%%%%%% FIGURE %%%%%%%%%%%%%%%%%%%%%%%%%
\begin{figure}
\caption{\label{fig:libespe}
The energy levels of Li and Be isotopes.
The calculations with the interaction (a) 
Volkov No.1 force ($m=0.56$) and 
(b) MV1 force ($m=0.576$)
are compared with the experimental data.
}
\centerline{
\epsfxsize=1.0\textwidth
\epsffile{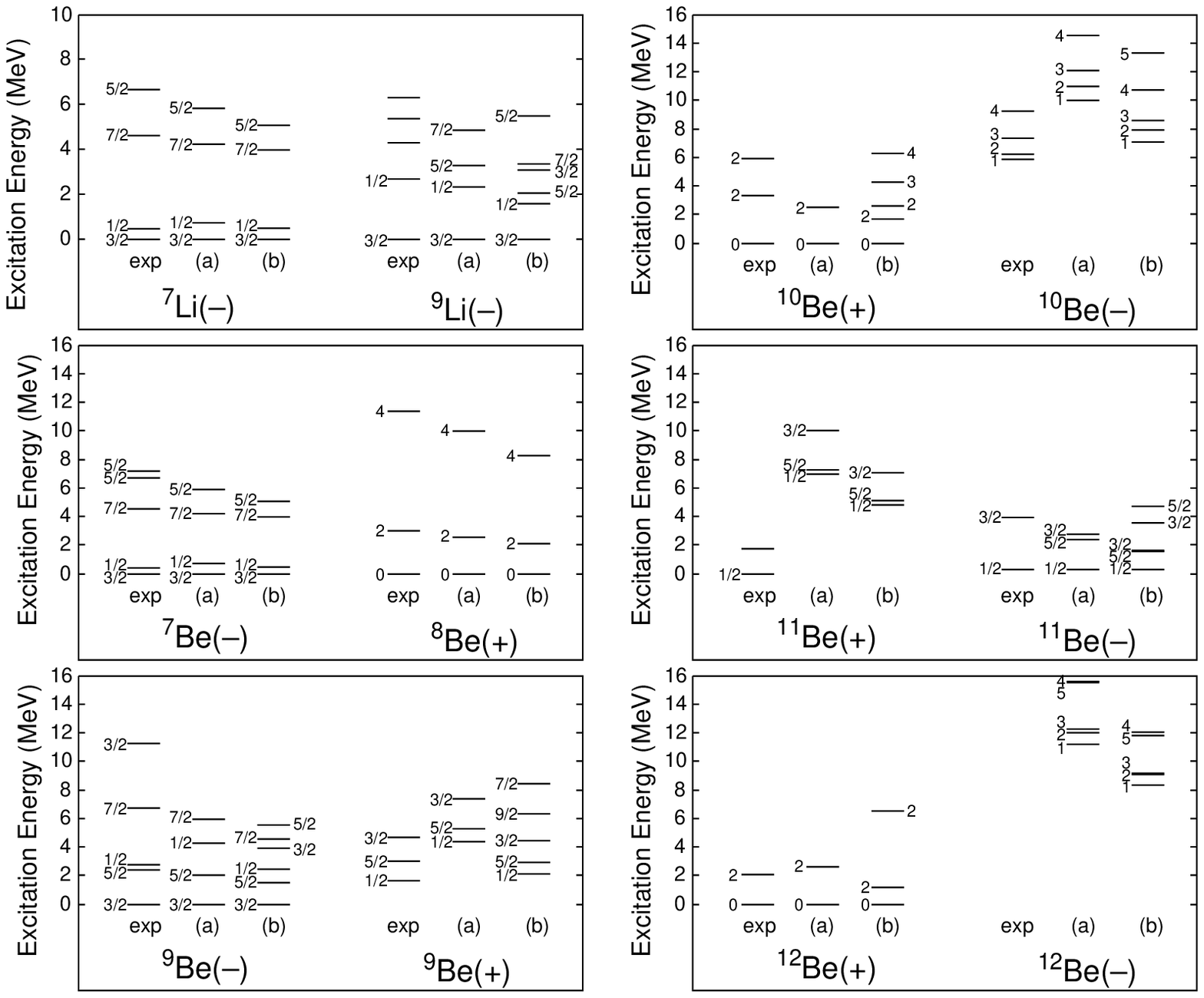}
}
\end{figure}
%%%%%%%%%%%% FIGURE %%%%%%%%%%%%%%%%%%%%%%%%%

%%%%%%%%%%%% FIGURE %%%%%%%%%%%%%%%%%%%%%%%%%
\begin{figure}
\caption{\label{fig:bspe}
The energy levels of B isotopes.
The adopted interactions are the case (b) 
MV1 force ($m=0.576$). The results for $^{11}$B with 
the MV1 ($m=0.56$) and the spin-orbit force 
($u_{I}=-u_{II}$=1500MeV) are also shown (f).}
\centerline{
\epsfxsize=1.0\textwidth
\epsffile{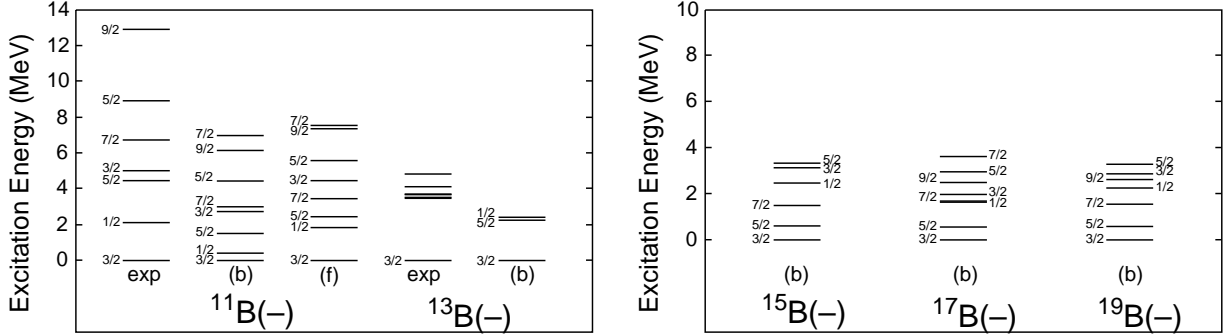}
}
\end{figure}
%%%%%%%%%%%% FIGURE %%%%%%%%%%%%%%%%%%%%%%%%%

%%%%%%%%%%%% FIGURE %%%%%%%%%%%%%%%%%%%%%%%%%
\begin{figure}
\caption{\label{fig:cspe}
The energy levels of C isotopes.
The theoretical values are obtained with the 
Volkov force ($m=0.6$) (f) and with the interaction 
(b) MV1 force ($m=0.576$).
For the positive parity states of $^{13}$C and $^{15}$C,
the results with the interaction (e) are presented 
instead of the ones with the interaction (b).}
\centerline{
\epsfxsize=1.0\textwidth
\epsffile{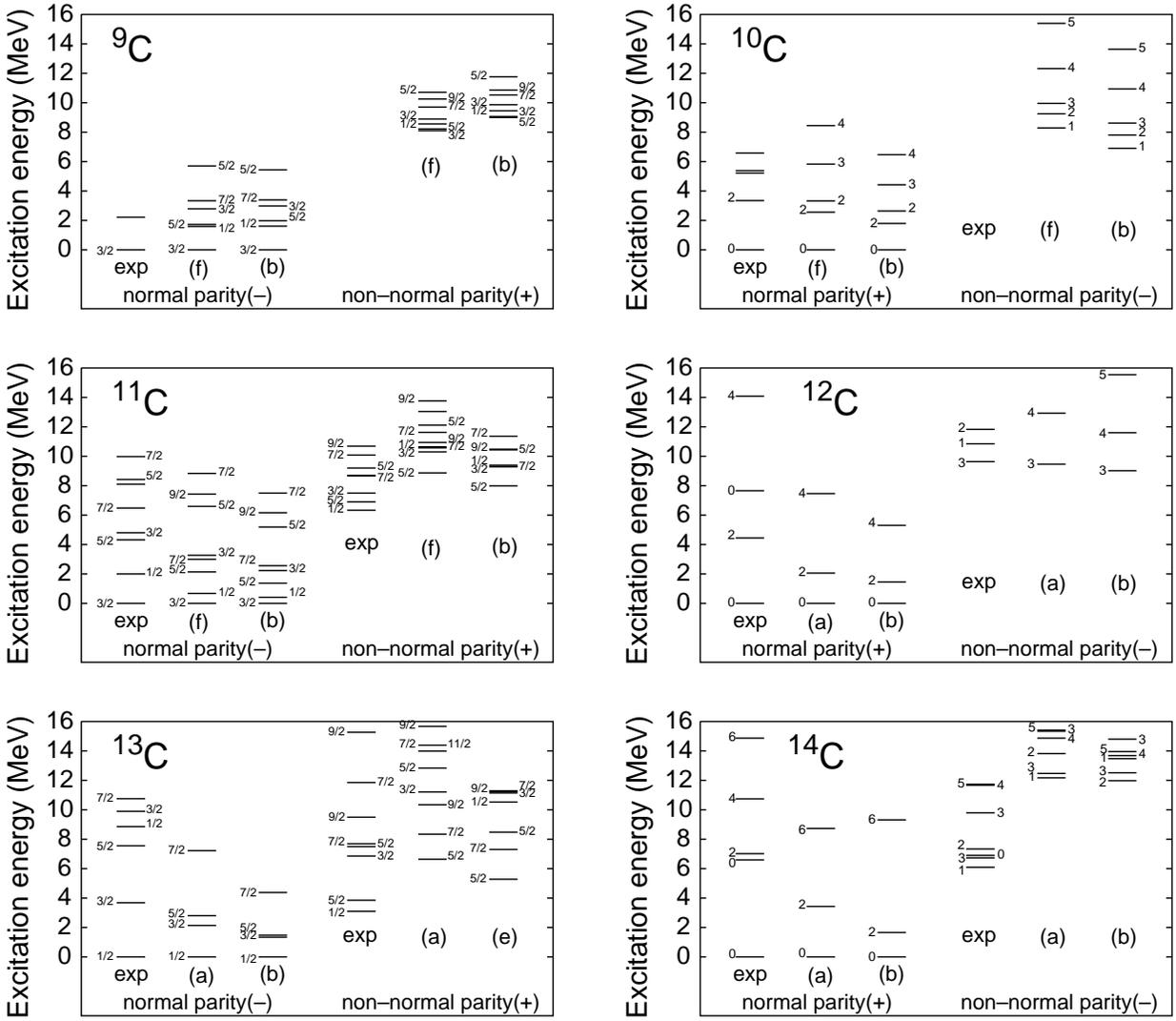}
}
\end{figure}
%%%%%%%%%%%% FIGURE %%%%%%%%%%%%%%%%%%%%%%%%%
%%%%%%%%%%%% FIGURE %%%%%%%%%%%%%%%%%%%%%%%%%
\begin{figure}
\caption{\label{fig:cspe2}
The energy levels of C isotopes. (See the figure caption in 
Fig.\protect\ref{fig:cspe})}
\centerline{
\epsfxsize=1.0\textwidth
\epsffile{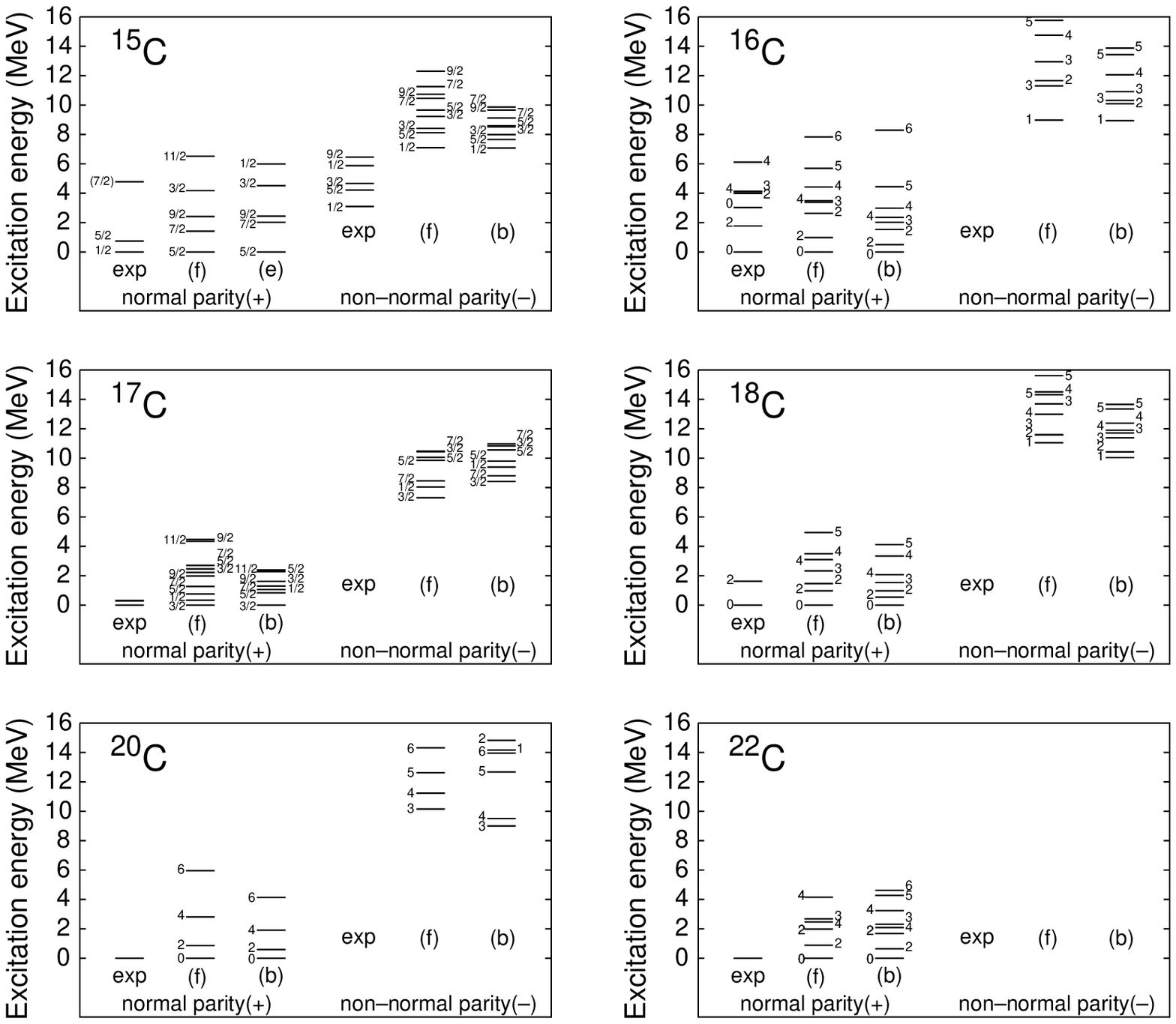}
}
\end{figure}
%%%%%%%%%%%% FIGURE %%%%%%%%%%%%%%%%%%%%%%%%%

The energy levels of Li, Be, B, and C  isotopes 
are displayed in Fig.\ref{fig:libespe},
Fig.\ref{fig:bspe}, Fig.\ref{fig:cspe}, and Fig.\ref{fig:cspe2}. 
The adopted interactions are explained in the figure captions.
The second $J^\pm$ states are obtained by diagonalizing
Hamiltonian with respect to   
$K$ quantum numbers. In many nuclei, the theoretical values of 
rotational bands such as excitation energies and spin sequence 
in low energy region well correspond 
to the experimental data.
It means that many low-lying levels are approximately described as   
the rotated states of the intrinsic states obtained by 
the simplest AMD calculations which are VAP for parity 
projection but VBP for total-spin
projection in the framework of AMD. 
Comparing the level spacing calculated with and without the three-body
force, it is found that the states obtained with the three-body force 
have smaller level spacing, that is, larger moment of inertia. 
In many cases we found that the VBP calculation gives smaller level spacing
than the VAP calculation. 

The energy levels depend on the parameter $m$ of the Majorana 
exchange term and the presence of the three-body force.
In general the results without the three-body 
force overestimate the excitation 
energies of non-normal parity states in most nuclei. 
The differences of the excitation energies 
between the normal parity and the non-normal parity 
states are improved in the results with the three-body force. 
In general the non-normal parity states have wider extension of the density
distribution and therefore they feel relatively weaker repulsive force 
due to the three-body terms. It is the reason why the calculations with 
the three-body force give smaller excitation energies for the 
 non-normal parity states. 
Even in the results with the three-body force, however,
the non-normal parity $1/2^+$ state is still higher than the $1/2^-$ states
in $^{11}$Be which has been known to have abnormal parity of the 
ground state $1/2^+$.
Careful choice of interactions
and the improvement of the wave function is necessary to reproduce this 
feature of the parity inversion. For instance, VAP calculations with 
appropriate interaction parameters succeed to obtain the lower
energy of $1/2^+$ state than the one of $1/2^-$ state. 
The detail will be mentioned later.

The energy levels of odd-even and even-odd nuclei are also sensitive 
to the strength of the spin-orbit force. 
The other set of interaction parameters (f) explained in the figure
caption of Fig.\ref{fig:bspe} give rather good results 
for energy levels of $^{11}$B.

In $^{13}$C and $^{15}$C the lowest positive parity states are known to be
$1/2^+$ states. However, the calculated intrinsic states 
of positive parity with MV1($m=0.576$) and Volkov force($m=0.60$) 
contain little component of $1/2^+$ states. 
Adopting other interaction parameter (e) in Table \ref{tbl:intparam}:
 MV1 force with $m=0.336$, Bartlett and Heisenberg components 
$b=-0.2$, $h=0.4$ and the slightly stronger spin-orbit force with the 
magnitude of $u_I=-u_{II}=1500$MeV, we obtained the $1/2^+$ component
whose energy is still rather higher than the one of $5/2^+$ state.
In the result with every interaction, the main component of 
the intrinsic state obtained with 
VBP calculation is $5/2^+$, therefore, the variation affects to minimize 
the energy of $5/2^+$ state. 
For the lowest $1/2^+$ states of $^{13}$C and $^{15}$C,
variations after spin-parity projection are useful instead of 
VBP calculations. 

%\subsubsection{Low lying energy levels}

%\subsubsection{single particle energies}

\subsubsection{Radii}
\begin{figure}
\caption{\label{fig:liberadii}
The root-mean-square radii of Li and Be isotopes.
The solid lines and the dotted lines are 
the AMD results calculated with 
the interaction (a) Volkov force with $m=0.56$, and 
(b) MV1 force with $m=0.576$.
The square points indicate the experimental data
deduced from the interaction cross sections
\protect\cite{TANIHATAb}.}
\centerline{
\epsfxsize=1.2\textwidth
\epsffile{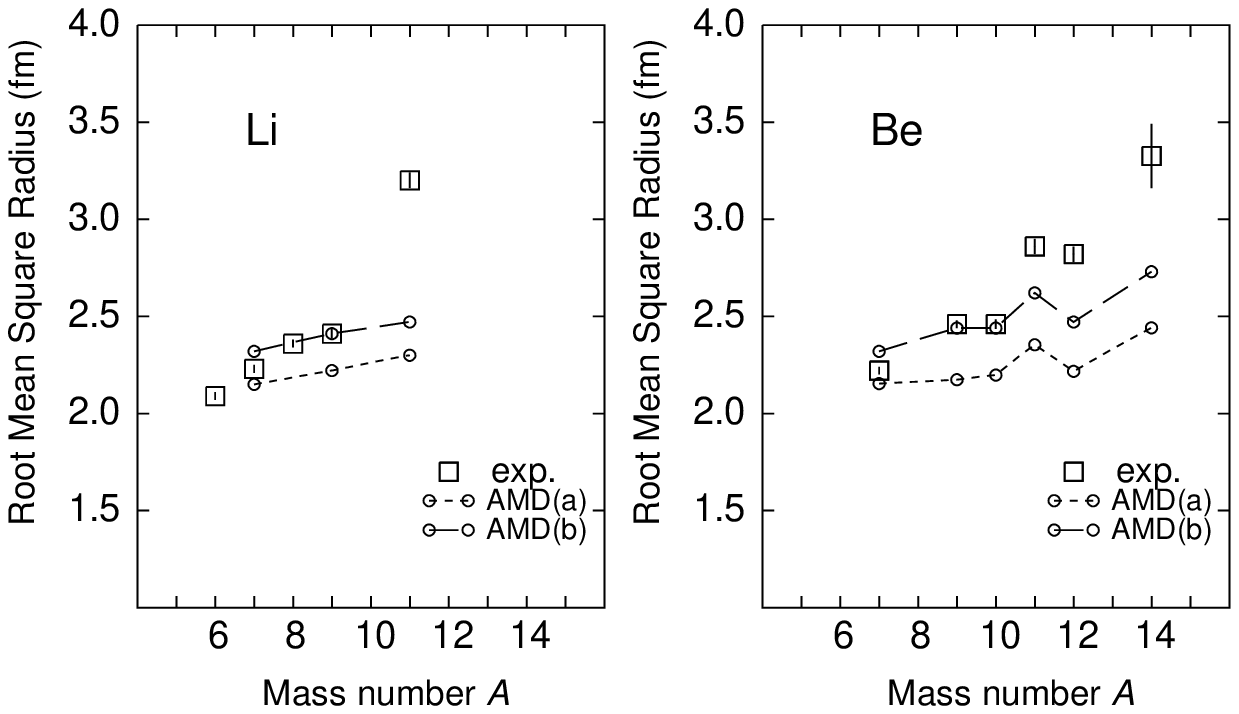}
}
\end{figure}

Figure \ref{fig:liberadii} shows the radii of Li and  Be
isotopes. 
Dashed lines are the theoretical root-mean-square radii 
by AMD with the three-body force 
and dotted lines show results without three-body force. 
Square points indicate the experimental radii reduced from the
experimental data of the interaction cross section.
The AMD calculations with the three-body force seem to qualitatively 
agree with the observed radii except for the very neutron-rich 
nuclei. The theory underestimates the extremely large radii 
of $^{11}$Li, $^{11}$Be and $^{14}$Be which are considered to have 
the neutron halo structures.  
For reproduction of such large radii due to the halo structures, 
improvement of the wave function and 
careful choice of interactions should be important.
The radius of the positive parity state of $^{12}$Be 
calculated with the present simple AMD calculation is smaller than
the experimental datum because the obtained state has 
the closed neutron-$p$-shell structure. 
However VAP calculation with a set of interaction parameters 
which reproduces the parity inversion of the $^{11}$Be ground state
gives a ground 
state of $^{12}$Be as $2\hbar\omega$ state with 2
particles in $sd$ shell and 2 holes in $p$ shell in neutron configuration,
 whose radius is as large as the experimental data.

%%%%%%%%%%%% FIGURE %%%%%%%%%%%%%%%%%%%%%%%%%
%%%%%%%%%%%% FIGURE %%%%%%%%%%%%%%%%%%%%%%%%%

%%%%%%%%%%%% FIGURE %%%%%%%%%%%%%%%%%%%%%%%%%
\begin{figure}
\caption{\label{fig:bradii}
The root-mean-square radii of B isotopes.
The triangles connected with the dotted line indicate the AMD
results by the use of interaction (b) the fixed Majorana parameter 
$m=0.576$.
The solid line shows the results by the interaction (c) with
a mass-dependent Majorana 
parameter.
The square points indicate the experimental data
deduced from the interaction cross sections
\protect\cite{TANIHATAb}.}
\centerline{
  \epsfxsize=0.6\textwidth
\epsffile{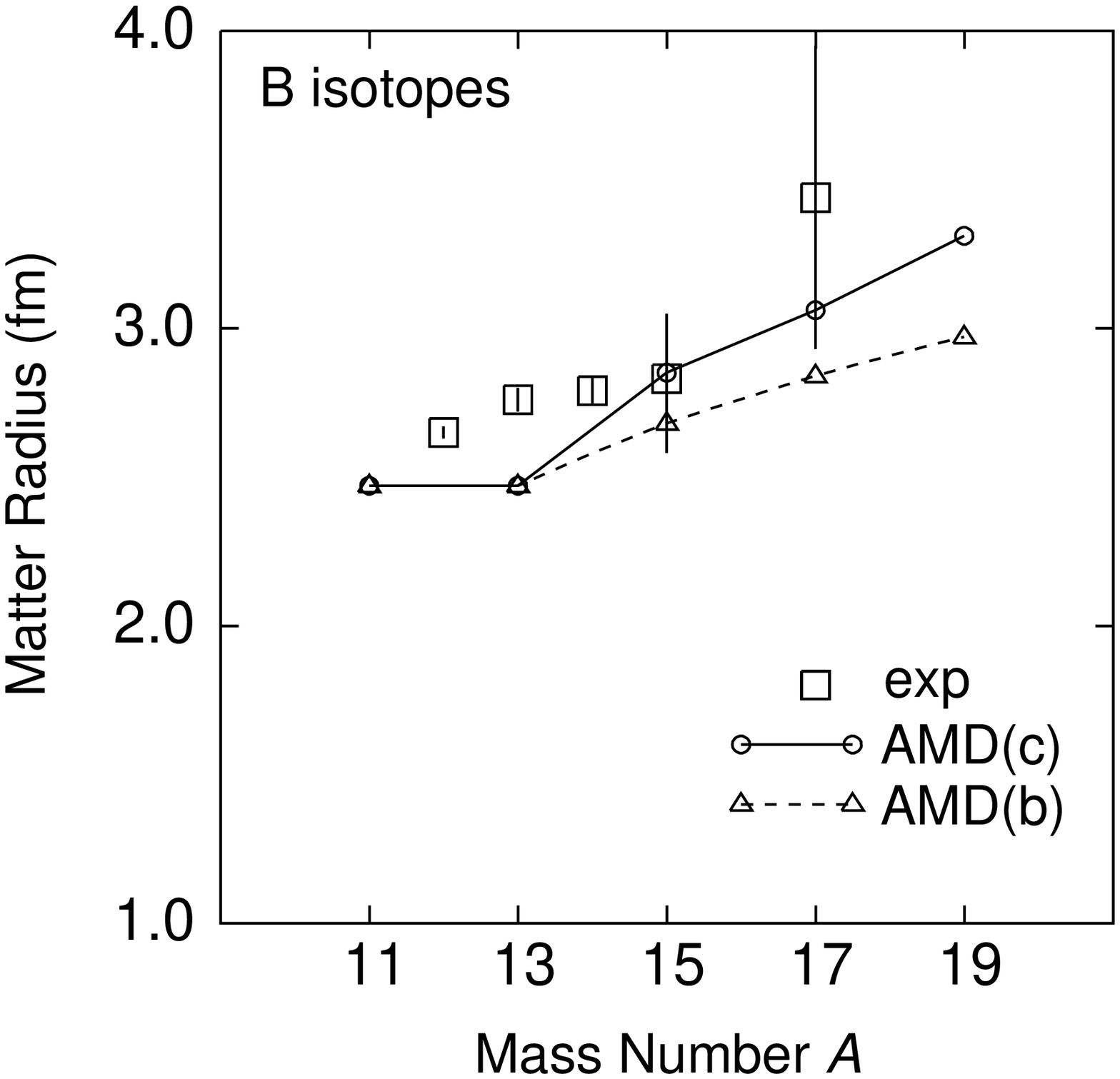}
}
\end{figure}
%%%%%%%%%%%% FIGURE %%%%%%%%%%%%%%%%%%%%%%%%%

In Fig. \ref{fig:bradii}, theoretical radii of B isotopes 
are compared with the experimental interaction radii.
The triangles connected with the dotted line indicate the AMD
results by the use of the fixed Majorana parameter $m=0.576$
(interaction(b)).
The solid line shows the results by the interaction (c) with
a mass-dependent Majorana 
parameter; $m=$0.576, 0.576, 0.63, 0.65, and 0.65 for $^{11}$B,
$^{13}$B, $^{15}$B, $^{17}$B, and $^{19}$B, respectively. 
Considering that the use of larger $m$ value for the heavier nuclear
system is generally reasonable, it is not unnatural to adopt the 
mass-dependent $m$ values adopted here. The results with mass-dependent
Majorana parameter reasonably fit to the experimental data.

%%%%%%%%%%%% FIGURE %%%%%%%%%%%%%%%%%%%%%%%%%
\begin{figure}
\caption{\label{fig:cradii}
The root-mean-square matter radii of C isotopes.
The solid and dashed lines are the theoretical results
by the use of interaction (b) $m=0.576$ and (d) $m=0.63$. 
The experimental data
deduced from the interaction cross sections are displayed with the square 
points
\protect\cite{OZAWA}.}
\centerline{
\epsfxsize=0.7\textwidth
\epsffile{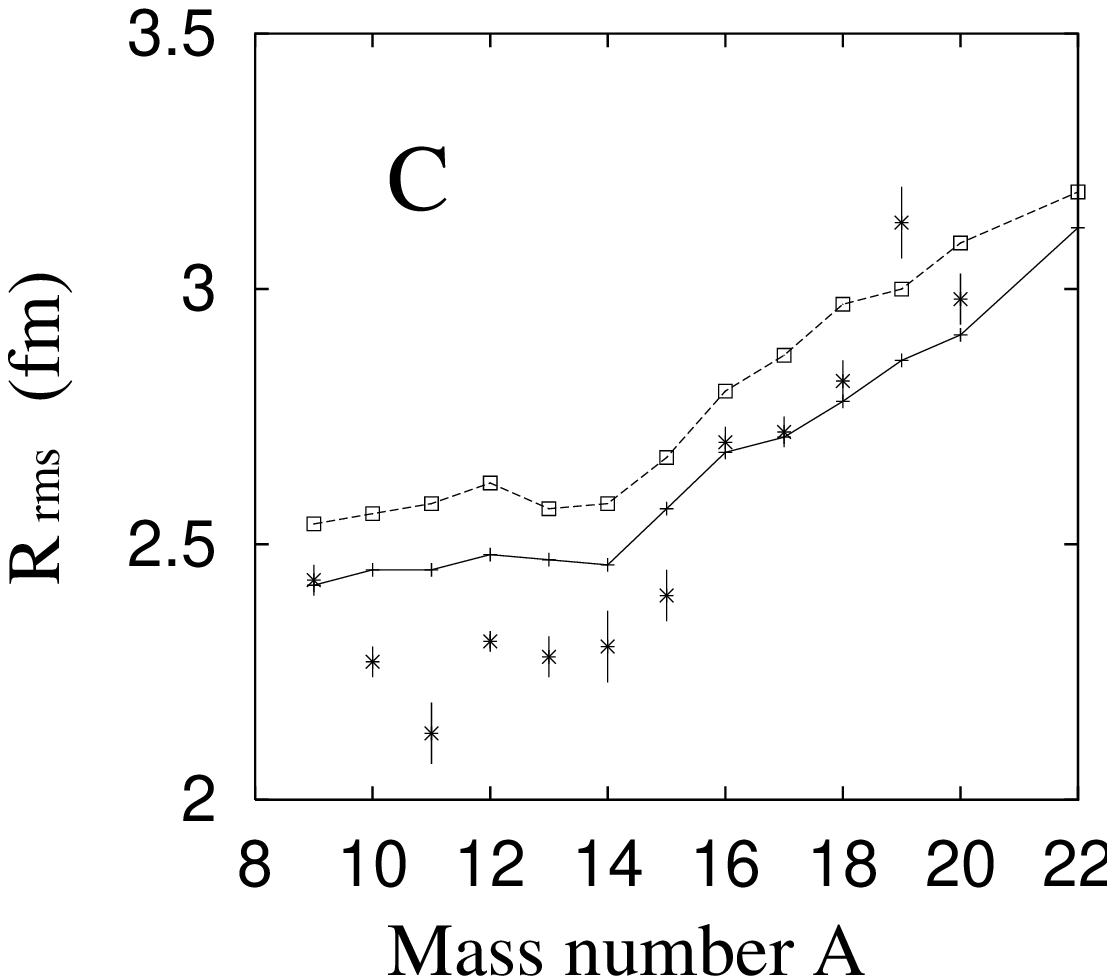}
}
\end{figure}
%%%%%%%%%%%% FIGURE %%%%%%%%%%%%%%%%%%%%%%%%%

The radii of C isotopes are presented
in Fig.\ref{fig:cradii}. 
The theoretical results are calculated 
by the use of the MV1 force (the solid line for 
$m=0.576$ and the dotted line for $m=0.63$). 
As seen in the figure, the recent experimental data
of the interaction radii of
C isotopes \cite{OZAWA} are found to be 
consistent with our theoretical predictions \cite{ENYOdoc}.
The radii of C isotopes have a kink at $^{14}$C and 
increase as the neutron number becomes larger
in the neutron-rich region $N > 8$. 
It is easy to quantitatively fit the theoretical values to 
the experimental ones except for the valley at $^{11}$C 
by using mass-dependent Majorana parameter
in a similar way to the case of B isotopes.
In AMD results, we did not find any reason for the small radius of 
$^{11}$C.
A detailed discussion of the radii of the neutron-rich 
C isotopes will be given later 
in subsection \ref{subsub:nskin} about the neutron-skin structure. 

\subsubsection{Magnetic moments}

\begin{figure}
\caption{\label{fig:libemu}
Magnetic dipole moments $\mu$ of even-odd Be and odd-even 
Li and B isotopes.
The calculations with the interaction (b) are compared with the experimental
data.}
\centerline{
\epsfxsize=0.55\textwidth
\epsffile{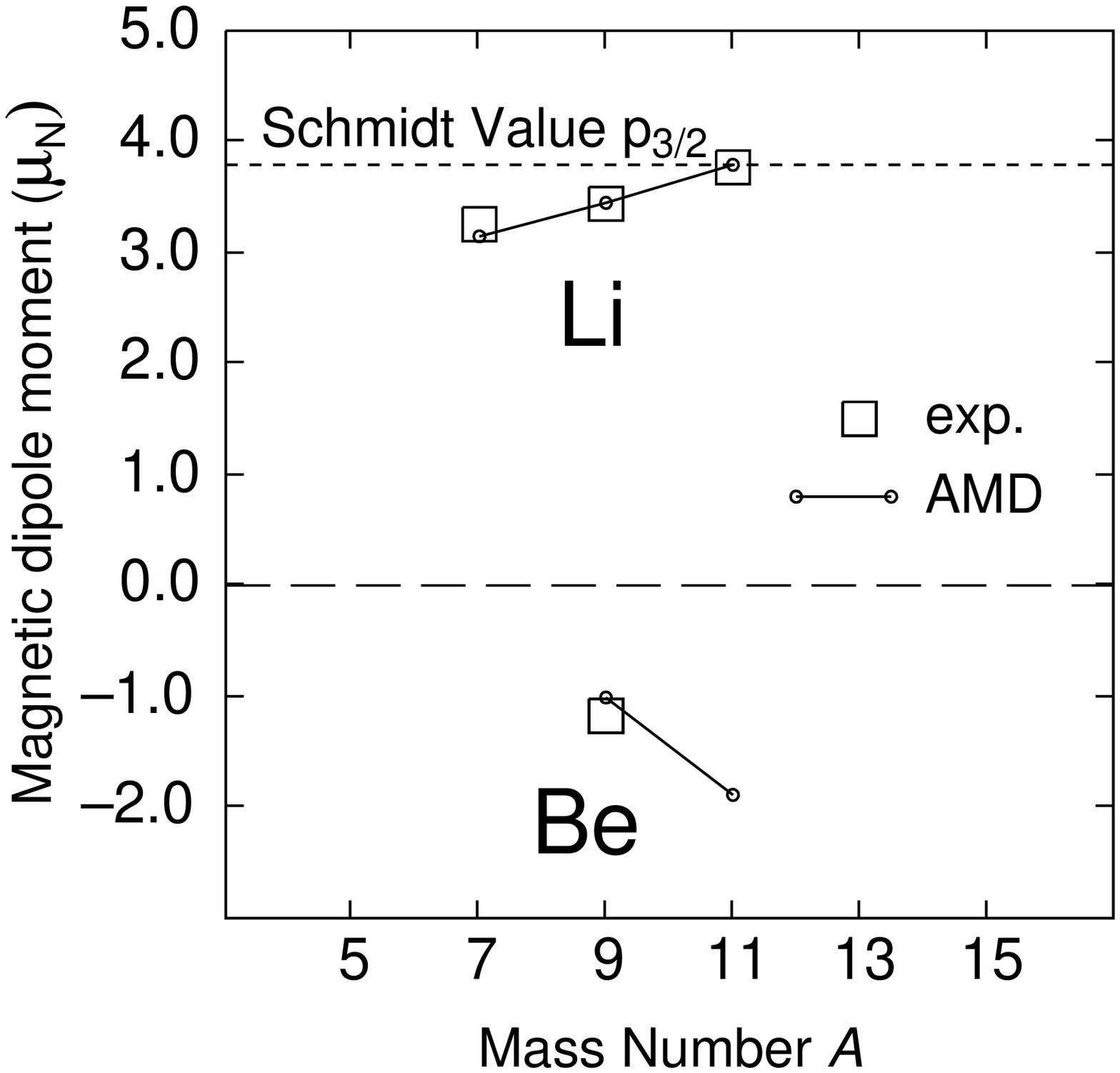}
\epsfxsize=0.55\textwidth
\epsffile{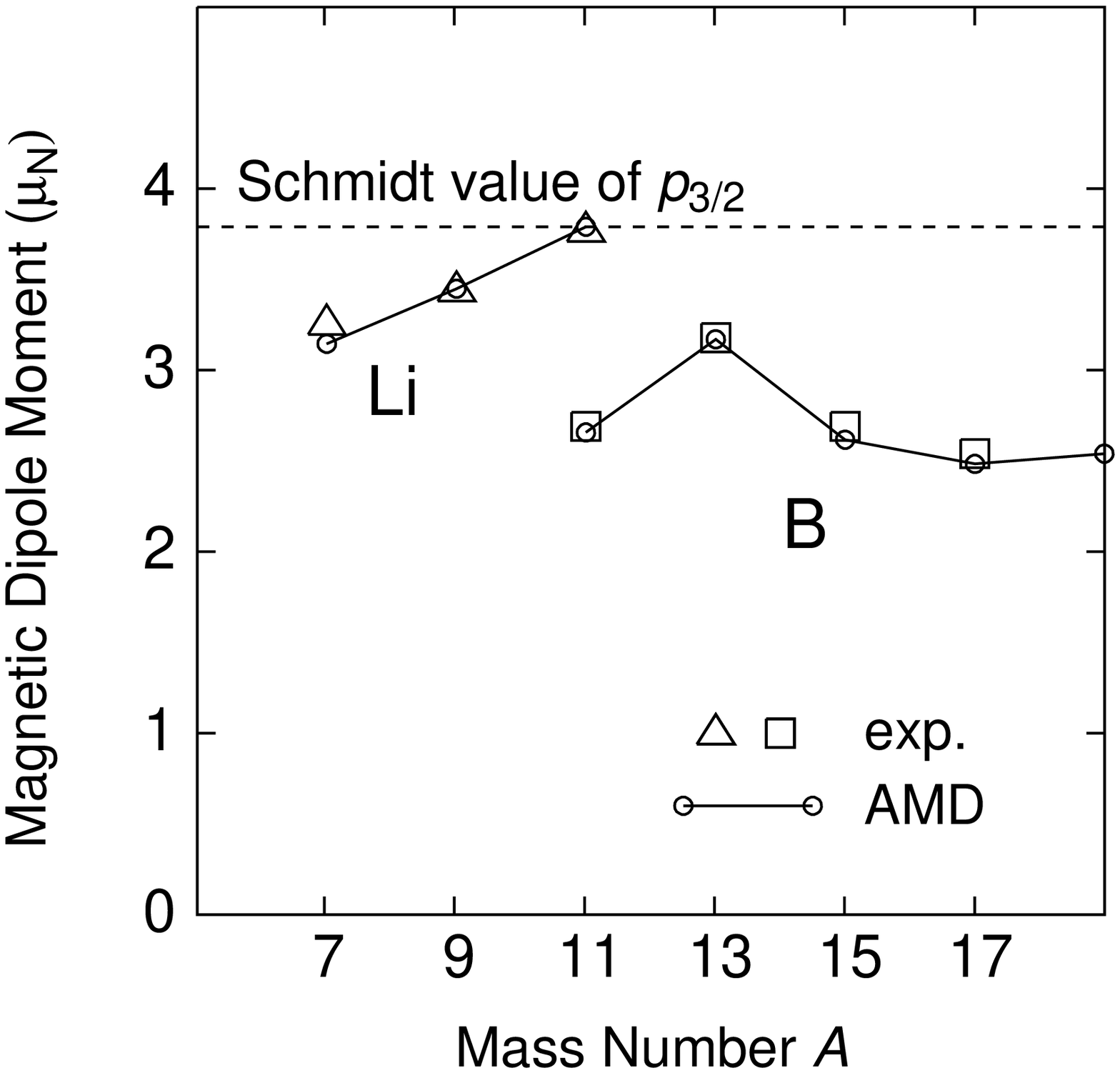}
}
\end{figure}

Magnetic dipole moments by the of AMD wave function are obtained 
by numerically calculating the expectation values of the magnetic dipole 
operator $\hat \mu_z$ by the spin-parity projected states with the 
highest $z$-component of the spin,
\begin{equation}
\frac{
\langle P^{J\pm}_{JK}\Phi|\hat\mu_z| P^{J\pm}_{JK}\Phi\rangle
}
{
\langle P^{J\pm}_{JK}\Phi|P^{J\pm}_{JK}\Phi\rangle
},
\end{equation}
where $\hat \mu_z$ is the $z$ component of 
the operator of the magnetic dipole moment
$\hat {\mbox{\boldmath $\mu$}} =g_{sn}{\bf s}_n+g_{sp}{\bf s}_p+g_l{\bf l}_p$ 
with the bare $g$-factor $g_{sp}=5.58$, $g_l=1.0$
for protons and $g_{sn}=-3.82$ for neutrons.
We choose an appropriate $K$ quantum number 
which minimizes the energy of the state $P^{J\pm}_{JK}\Phi$.
Figure \ref{fig:libemu} shows the magnetic dipole moments
$\mu$ of odd-even Li isotopes, even-odd Be isotopes, and odd-even B isotopes.
The theoretical results of AMD calculations agree with the experimental
data for many nuclei very well.
It should be emphasized that the AMD method
is the first framework which has succeeded in reproducing the magnetic 
dipole moments for these isotopes systematically.

%%%%%%%%%%%% FIGURE %%%%%%%%%%%%%%%%%%%%%%%%%
%%%%%%%%%%%% FIGURE %%%%%%%%%%%%%%%%%%%%%%%%%

The dependence of the $\mu$ moment on the neutron number in Li isotopes and 
in B isotopes is closely related to the structure change with the increase 
of the neutron number. Therefore these data of $\mu$ moments 
should carry important informations about the nuclear structures. 
We will give detailed discussions about the correlation
of the nuclear structure and the observed electromagnetic properties
in later section.

The theoretical value of the $\mu$ moment of $^{11}$Be(1/2$^+$)
 is sensitive to the strength of the spin-orbit force.
In the case of MV1 force with $m=0.576$, the $\mu$ moment
 of $^{11}$Be(1/2$^+$) calculated with the strength of 
the spin-orbit force $u_I=u_{II}=900$MeV is $-1.9$ $\mu_N$  which is 
as much as the Schmidt value, while with 
the stronger spin-orbit force $u_I=u_{II}=1200$MeV the $\mu$ moment
is $-1.2$ $\mu_N$. 
In the case of 1/2$^+$ state of $^{11}$Be, the $\mu$ moment directly 
reflects the spin configuration of the last $7$-th 
neutron. Since the strength of the spin-orbit force affects
the orbit of the last neutron in $sd$ shell
and also the core excited component of the neutron $p_{3/2}$ closed shell, 
it is natural that the the $\mu$ moment
of $^{11}$Be depends on the strength of the spin-orbit force.

On the other hand, in the case of $\mu$ moments of odd-even
 Li and B isotopes 
the calculated results do not depend on the interaction parameters 
so much as the one of $^{11}$Be because
the main contribution to $\mu$ moments originates from the 
the spin configuration of the valence proton in the $p_{3/2}$ orbit.
A slight dependence of $\mu$ moments of $^9$Li and the mirror nucleus 
$^9$C on the spin-orbit force 
has been discussed in Ref.\cite{ENYOd}.

In Table \ref{tbl:cmumom} the $\mu$ moments of even-odd C isotopes are
presented. The theoretical results are calculated with the interaction
 (b) MV1 force with 
$m=0.576$ and the spin-orbit force with $u_{I}=-u_{II}=900$MeV
except for positive parity states of $^{13,15}$C. The positive parity states
of $^{13,15}$C are calculated with the interaction (e) MV1 force with $m=0.336$,
$b=-0.2$ and  $h=0.4$ and the spin-orbit force with the 
magnitude of $u_I=-u_{II}=1500$MeV which is the same interaction
as mentioned in \ref{subsub:energy}. The theoretical results well agree 
to the experimental data.

\begin{table}[tbh]

\caption{\label{tbl:cmumom}
Magnetic dipole moments and electric quadrupole moments of C isotopes.
We adopt the interaction (b) except for positive parity states
of $^{13}$C and $^{15}$C. The interaction (e) is used 
for positive parity states of  $^{13}$C and $^{15}$C.
The experimental data are quoted from Ref.\protect\cite{MATSUTAa}$^{\rm a}$, 
and Ref.\protect\cite{NTABLE}$^{\rm b}$.
}
\begin{center}
\begin{tabular}{lccccc}
 & \multicolumn{2}{c}{$\mu (\mu_N$)}&\multicolumn{2}{c}{$Q$(e mb)} \\
& exp. &model &exp.&model\\
\hline
                &       &       &               &       &\\
$^9$C(3/2$^-$)  &$|1.39|^a$     &$-$1.53& $-$           &$-$27&\\
\hline
                &       &       &               &       &\\
$^{10}$C(2$^+$) &$-$    &0.70   & $-$           &$-$38&\\
\hline
                &       &       &               &       &\\
$^{11}$C(3/2$^-$)       &$-$0.96$^b$&$-$0.90& 34.3$^b$          &20     &\\
\hline
                &       &       &               &       &\\
$^{12}$C(2$^+$) &$-$    &1.01   & 60$\pm$30$^b$ &51     &\\
\hline
                &       &       &               &       &\\
$^{13}$C(1/2$^-$)       &0.70$^b$       &0.99   & $-$           &$-$    &\\
$^{13}$C(1/2$^+$)       &$-$    &$-$1.90& $-$           &$-$    &\\
$^{13}$C(5/2$^+$)       &$|1.40|^b$     &$-$1.52& $-$           &$-$45&\\
\hline
                &       &       &               &       &\\
$^{14}$C(2$^+$) &$-$    &3.11   & $-$           &36     &\\
\hline
                &       &       &               &       &\\
$^{15}$C(1/2$^+$)       &$|1.32|^b$     &$-$1.26& $-$           &$-$    &\\
$^{15}$C(5/2$^+$)       &$-$1.76$^b$&$-$1.64& $-$               &2      &\\
\hline
                &       &       &               &       &\\
$^{17}$C(3/2$^+$)       &$-$    &$-$1.05& $-$           &26     &\\
\end{tabular}
\end{center}
\end{table}

\subsubsection{Electric quadrupole moments and B(E2)}
It should be pointed out that we can describe electric 
properties such as quadrupole moments and $E2$ transition strength $B(E2)$
by using not effective charges but the bare charges for protons and neutrons
 in the AMD framework. 
It is because the drastic changes of proton and 
neutron structures are directly treated in the framework.
Here, we just present the theoretical results of electric quadrupole moments
and $B(E2)$ comparing with the experimental data.
We will give detailed discussions 
on the relation between observable electric properties and the intrinsic
structures later in \ref{subsec:amddis}.

\begin{figure}
\caption{\label{fig:libebqmom}
Electric quadrupole moments $Q$ of Li, Be, and B isotopes.
The theoretical values and the experimental data for 
the ground $3/2^-$ states of odd-even Li and B isotopes,
and $^9$Be are shown. Solid lines (A) and triangle points indicate 
the calculated results using the MV1 force with $m=0.576$ 
(interaction (b)) for Li and Be, and
using the mass-dependent $m$ parameters (interaction (c))
 for B isotopes. The dashed line (B) indicates the results of 
B isotopes with Bartlett and Heisenberg terms described in the text.
The point (C) is the calculation of $^7$Li with 
the wave function improved by taking account for 
the long tail of the relative motion between clusters $\alpha$ and $t$.
The experimental data are quoted from Refs.
\protect\cite{ARNOLDb,ASAHIa,NTABLE}
}
\centerline{
\epsfxsize=.7\textwidth
\epsffile{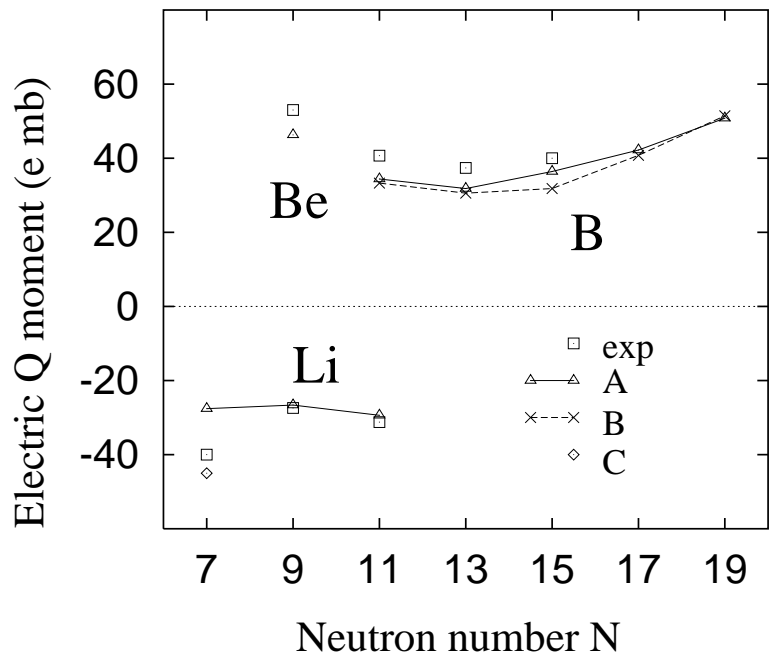}
}
\end{figure}

Although the intrinsic structures of many nuclei are qualitatively not so much
sensitive to the adopted interaction,
the calculated $Q$ moments depend on the adopted interaction 
with or without a density-dependent term and also depend 
on the Majorana parameter $m$,
because the radii are sensitive to these interaction parameters.
Figure \ref{fig:libebqmom} shows the electric quadrupole moments of Li,
Be, and B isotopes. The solid lines and the triangle points (A) 
indicate the AMD calculation of Li and Be isotopes 
by the use of MV1 force with $m=0.576$ 
(interaction (b)), and that of B isotopes with the mass-dependent 
$m$ parameters (interaction (c)) which reproduce the observed 
radii of B isotopes.
The calculations agree to the experimental $Q$ moments systematically .

In order to check the interaction dependence of 
$Q$ moments we also show the theoretical values for B isotopes 
calculated by adopting the 
interaction with additional Bartlett and Heisenberg terms 
as $b=-0.2$ and $h=0.4$ in Fig.\ref{fig:libebqmom} (dashed line).
Here the Majorana parameters are changed from $m$ into
$m=M+{2\over 5}(b-h)$ so as to give as the same binding
energies for $\alpha$ nuclei as the ones obtained with the interaction (c)
of no Bartlett and Heisenberg terms.
Comparing the solid line and the dashed line for B isotopes,
it is found that the additional components do not give significant effects
on $Q$ moments. 

Calculations with the interaction (b) fit well
to the experimental $Q$ moments 
of Li and Be isotopes except for $^7$Li. 
The $Q$ moment of $^7$Li is underestimated by theory.  
By improving the wave function of $^7$Li in the following way,
we have obtained the large $Q$ moment of $^7$Li indicated by 
a point (C) in Fig. \ref{fig:libebqmom}
which is as much as the experimental data.
As we explain the calculated intrinsic structure later in detail, 
the AMD wave function of $^7$Li has proved 
to have the well-developed cluster structure of $\alpha+t$. 
When the clustering is well developed, the relative wave function
between clusters spreads out toward the outer spatial region 
resulting in a long tail. 
However, 
since the single nucleon wave function of AMD is a Gaussian wave packet,
the relative wave function between clusters is also 
necessarily a Gaussian wave packet. Because of 
the lack of the outer tail part of the relative wave function
between $\alpha$ and $t$ ,
the quadrupole moment of $^7$Li is underestimated by 
the simplest AMD wave function. Therefore we have improved the inter-cluster
relative wave function of the AMD by
superposing several AMD wave functions which are written as  
$\alpha+t$ clustering states 
with different distances between the centers of two clusters. 
Superposition of the spin-parity eigen states projected from these 
wave functions has been made by diagonalizing the total Hamiltonian.
The improved wave function has proved to reproduce the electric 
quadrupole moment well as seen in Fig.\ref{fig:libebqmom}.

Also in the case of C, 
the calculations with the interaction (b) reasonably agree with the observed
$Q$ moments of $^{11}$C and $^{12}$C.
As seen in the radii of unstable nuclei with neutron halo and 
also in the $Q$ moments of $^7$Li,
the simple AMD wave function is not sufficient to describe
the outer tail of wave function because of the Gaussian form.
If the proton-rich nuclei such as $^9$C
have the outer tail of the valence protons, the theoretical predictions
may underestimate the $Q$ moments of these nuclei.

%%%%%%%%%%%%%%%%%%%%%%%%%%%%%%%%%%%%%%%%%%%%%%%%%%
\begin{figure}
\caption{\label{fig:be2}
$E2$ transition strength 
in Li, Be, B, and C isotopes.
Square points indicate the experimental data
and triangle points are the AMD calculations with
the interaction (b) MV1 $m=0.576$.
The initial and final states are the lowest $J^\pi$ levels  
except for the levels with subscripts in $^{11}$B.
The cross point for $B(E2:1/2\rightarrow 3/2)$ of $^7$Li 
is the calculation with 
the wave function improved by taking into account
the long tail of the relative motion between $\alpha$ and $t$ clusters.
}
\centerline{
\epsfxsize=0.7\textwidth
\epsffile{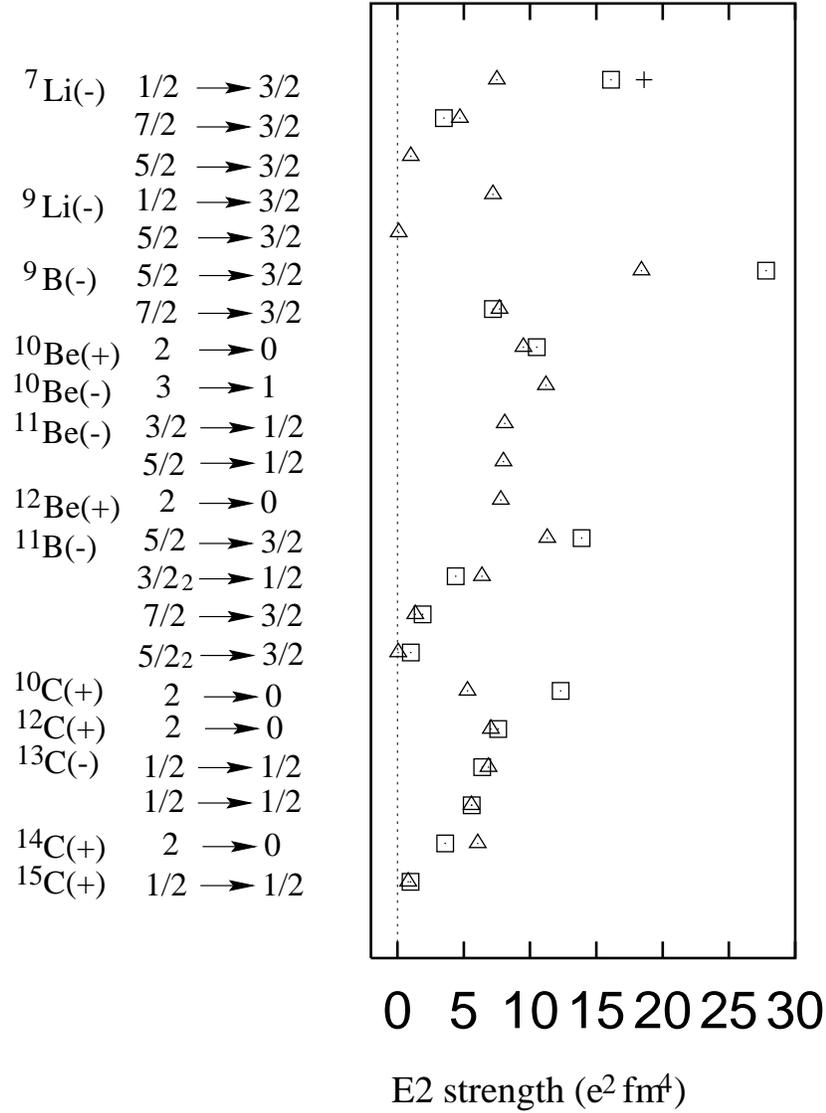}
}
\end{figure}
%%%%%%%%%%%%%%%%%%%%%%%%%%%%%%%%%%%%%%%%%%%%%%%%%%

Figure \ref{fig:be2} shows the $E2$ transition strength. 
The theoretical values are calculated by using the interaction (b),
MV1 force with $m=0.576$, except for the data of the positive 
parity states of $^{15}$C.
Theoretical values well agree with the experimental data.
$B(E2; 1/2^-\rightarrow 3/2^-)$ 
in $^7$Li is underestimated by simple AMD wave function because of 
the lack of outer tail of the inter-cluster ($\alpha$-$t$) 
relative wave function.
The strength can be reproduced by the theoretical results 
with the improved wave function 
described above.
In the simple AMD results 
the strength $B(E2; 2^+\rightarrow 0^+)$ of $^{12}$Be is calculated to be
smaller than the one of $^{10}$Be. However a larger strength
of $B(E2; 2^+\rightarrow 0^+)$ of $^{12}$Be than the one of $^{10}$Be 
is predicted by 
the VAP calculations with a set of interactions 
which gives a largely deformed ground state with $2p-2h$ 
in the neutron orbits.

\subsection{Discussion}\label{subsec:amddis}
In this section we make systematic study of the intrinsic structures.
After describing the deformations and clustering structures of Li,
Be, B and C isotopes, we discuss the effect of the intrinsic structure
on the observable quantities to deduce the informations of the 
nuclear structure
from the experimental data. We notice some interesting features
such as the effect of clustering structures on the 
electromagnetic properties, the opposite deformation between protons
and neutrons, and the neutron skin structures. 

\subsubsection{Shapes and clustering structure}
\begin{figure}
\caption{\label{fig:liden}
The density distributions of the normal-parity states of Li isotopes.
The density of the intrinsic wave function before the 
parity projection is integrated along the line parallel to 
an appropriate principal axis as explained in the text. 
The AMD wave functions are obtained with the interaction (b).
}
\centerline{
\epsfxsize=0.5\textwidth
\epsffile{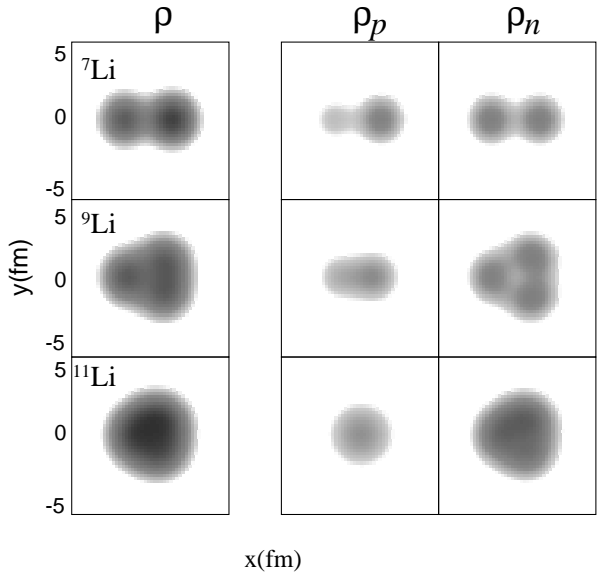}
}
\end{figure}

\begin{figure}
\caption{\label{fig:beden}
The density distributions of the normal-parity states of Be isotopes.
The adopted interaction is the case (b).
}
\epsfxsize=0.49\textwidth
\noindent
\begin{minipage}[t]{0.49\textwidth}
\epsfxsize=1.0\textwidth
\centerline{\epsffile{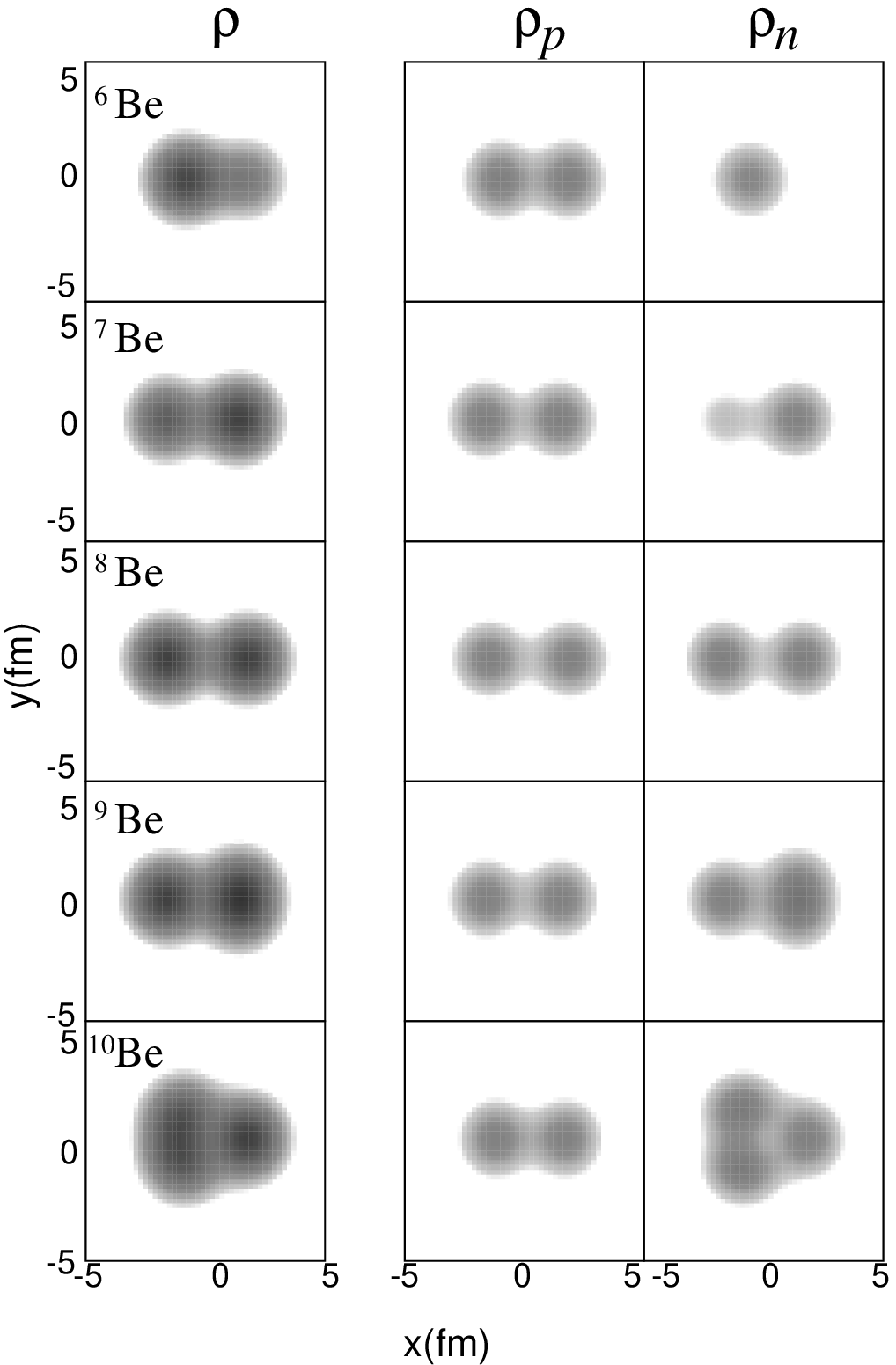}}
\end{minipage}
\hskip 0.01\textwidth
\begin{minipage}[t]{0.49\textwidth}
\epsfxsize=1.0\textwidth
\centerline{\epsffile{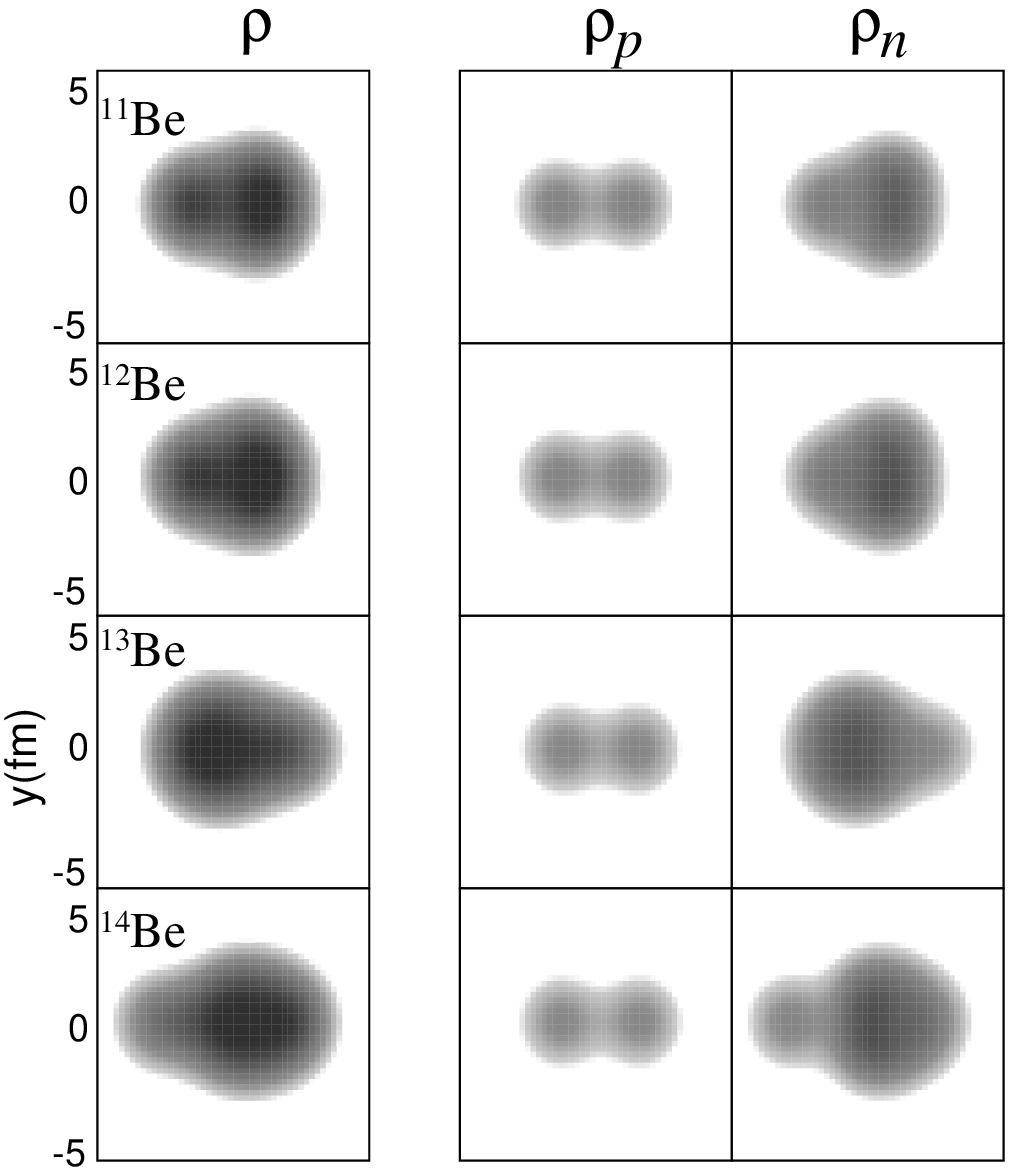}}
\end{minipage}
\end{figure}

\begin{figure}
\caption{\label{fig:bedenn}
The density distributions of the non-normal-parity states of Be isotopes.
The adopted interaction is the case (b).
}
\epsfxsize=0.49\textwidth
\centerline{\epsffile{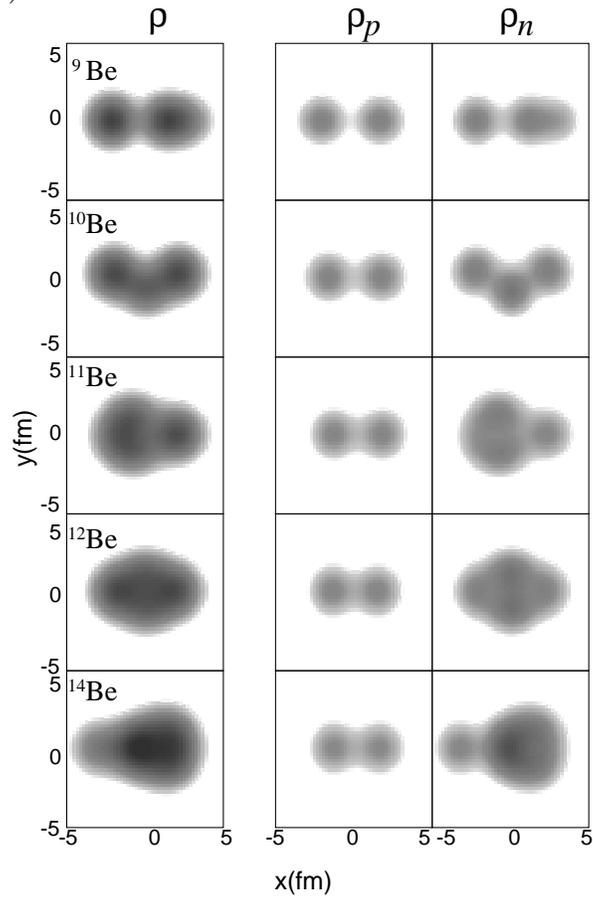}}
\end{figure}

\begin{figure}
\caption{\label{fig:bden}
The density distributions of the normal-parity states of B isotopes.
The intrinsic wave functions are calculated by the use of the 
interaction (c) with the mass-dependent Majorana parameter.
}
\centerline{
\epsfxsize=0.5\textwidth
\epsffile{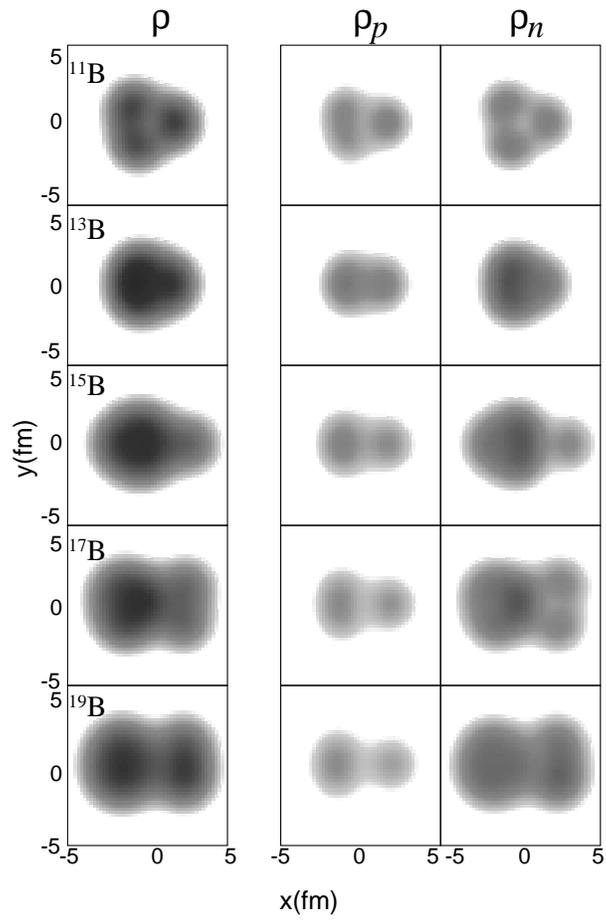}
}
\end{figure}

\begin{figure}
\caption{\label{fig:cden}
The density distributions of the normal-parity states of C isotopes.
The intrinsic wave functions are calculated with the
interaction (b) except for $^{15}$C.
For $^{15}$C, we obtain the intrinsic state with
the interaction (e).
}
%\epsfxsize=0.49\textwidth
%\begin{minipage}[b]{0.49\textwidth}
%\epsfxsize=1.0\textwidth
\centerline{\epsfxsize=0.48\textwidth
\epsffile{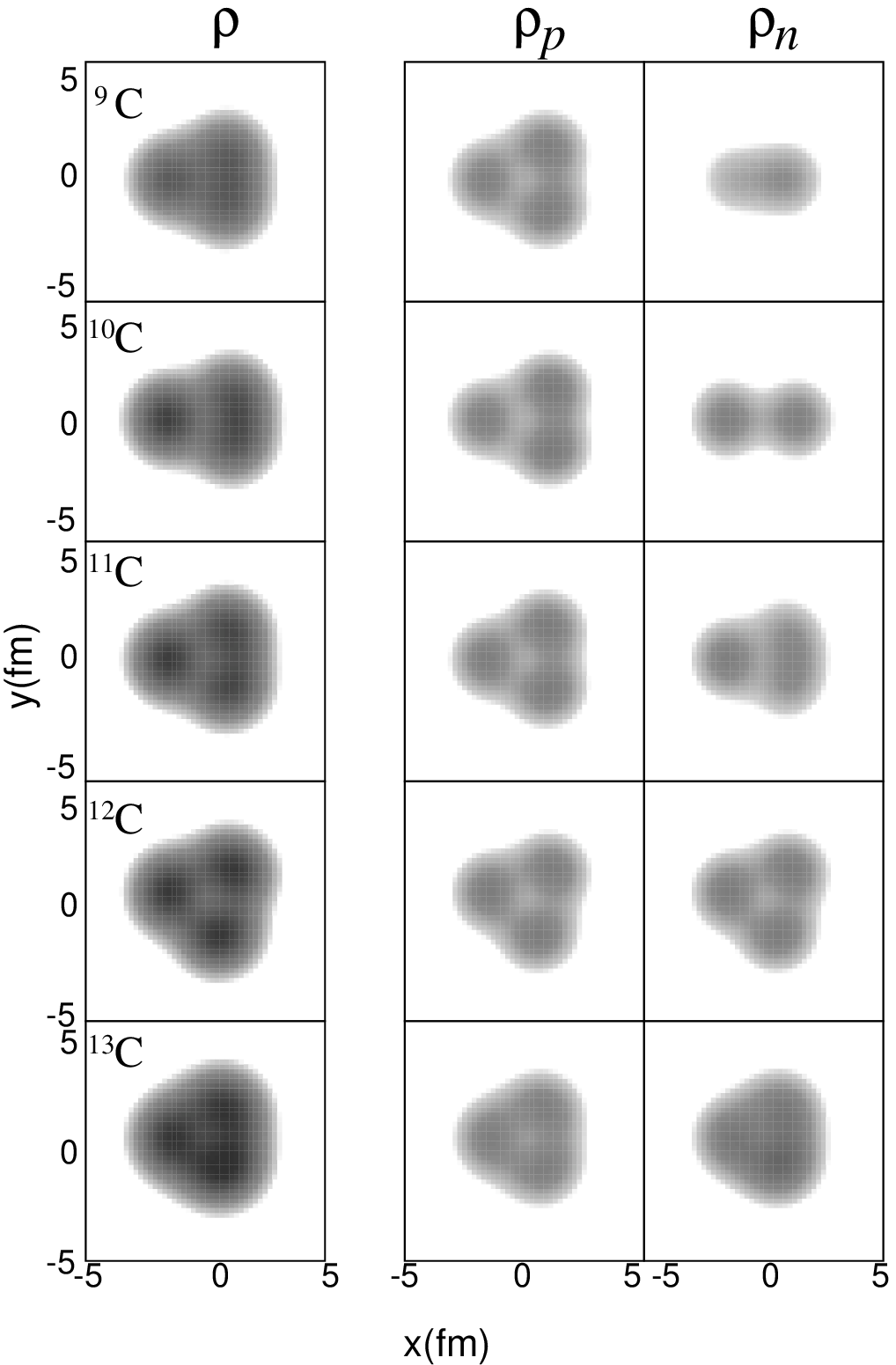}
\begin{minipage}[b]{0.5\textwidth}
\epsfxsize=0.96\textwidth
\epsffile{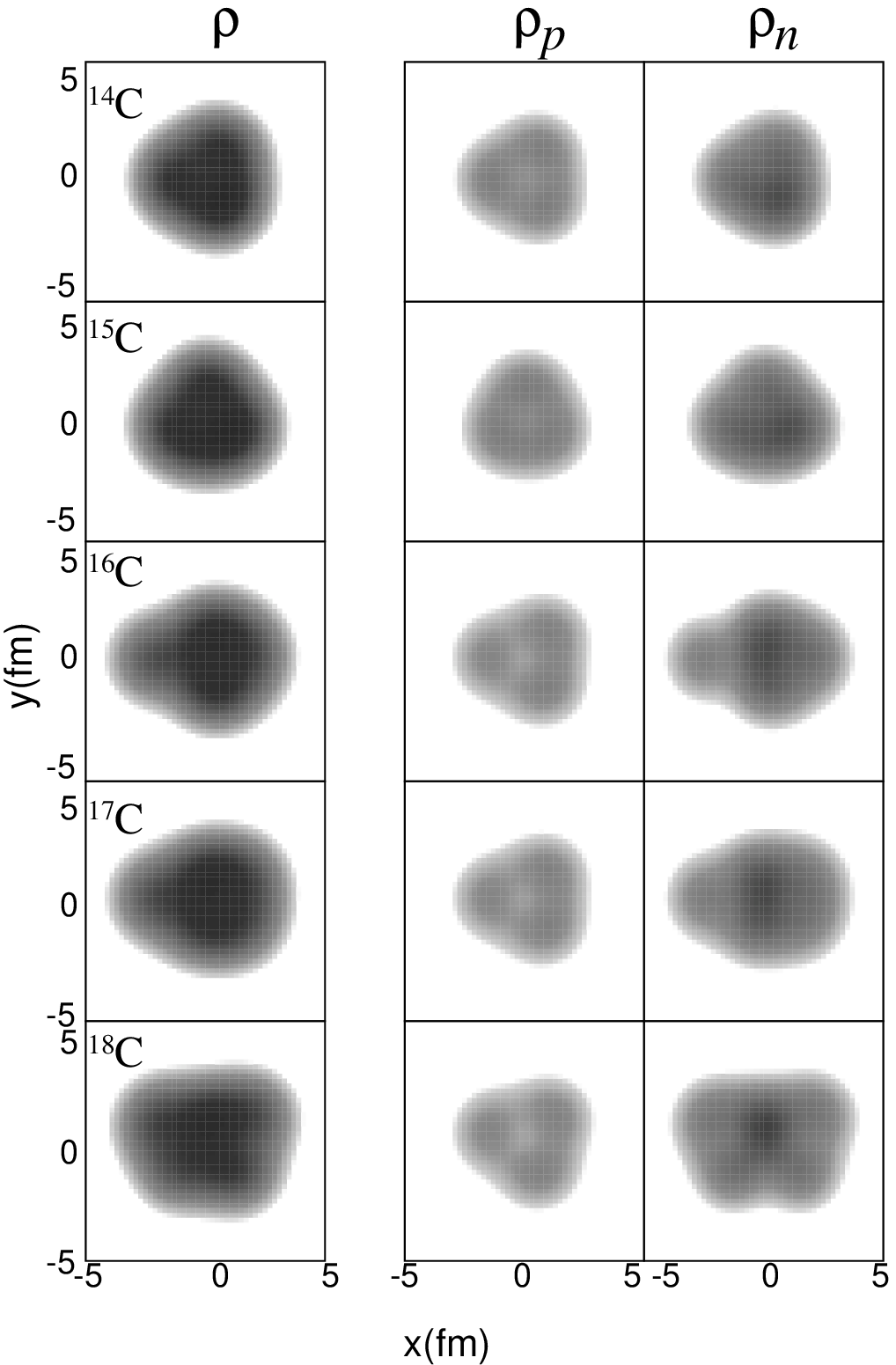}
\end{minipage}}
%}
%\end{minipage}
%\hskip 0.01\textwidth
%\centerline{
%}
%\end{minipage}
\hskip 0.01\textwidth
\begin{minipage}[b]{0.5\textwidth}
\epsfxsize=0.96\textwidth
\centerline{\epsffile{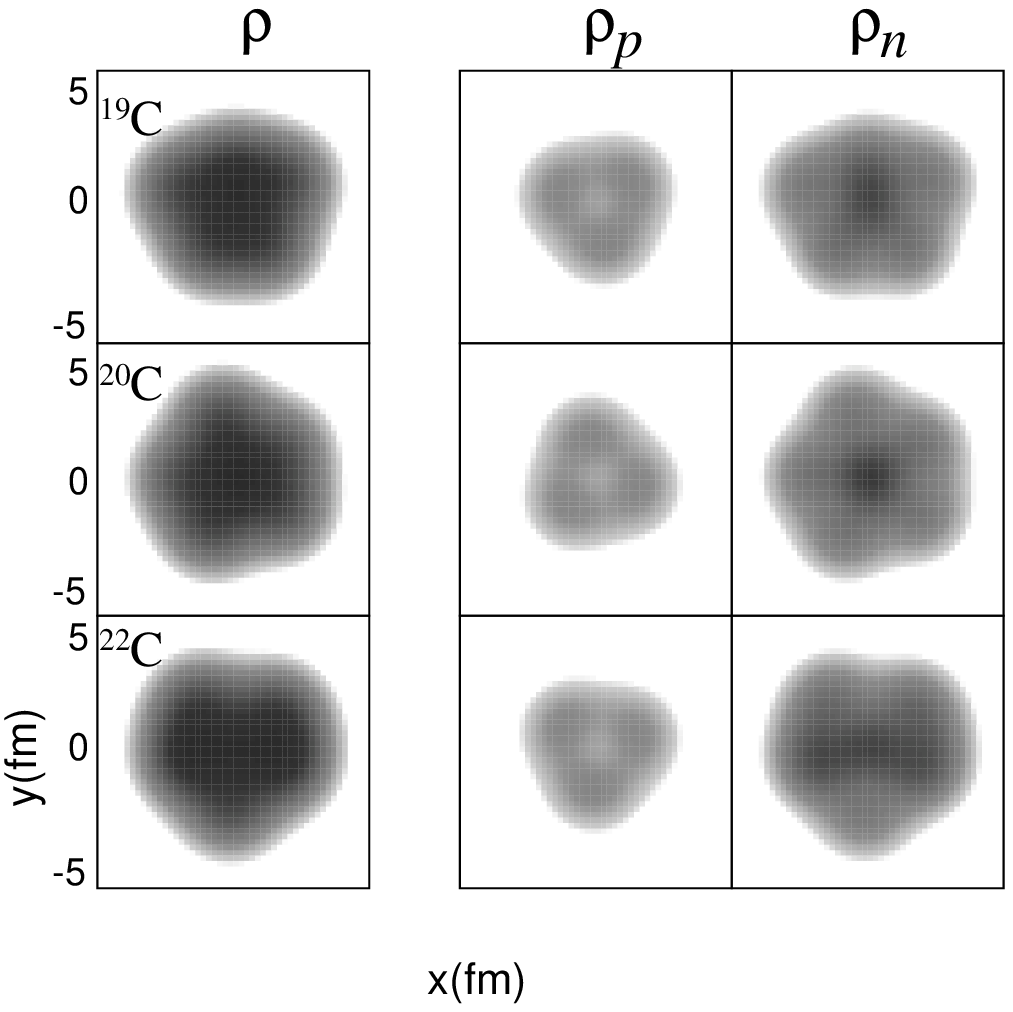}}
\end{minipage}
\end{figure}

By analyzing the intrinsic wave function, we discuss 
the shapes, deformations and clustering aspects of
Li, Be, B and C isotopes. First we show the density distributions of the 
intrinsic states of Li, Be, B and C  isotopes
in the Figs. \ref{fig:liden}, \ref{fig:beden}, \ref{fig:bedenn},
 \ref{fig:bden},
and \ref{fig:cden}.
In drawing the figures, the density of each intrinsic wave function before
parity projection is
projected onto an $X$-$Y$ plane by integrating out along the line 
parallel to the $Z$ axis.
Here $X$, $Y$, $Z$ axes are chosen so as to be 
$\langle \sum_i x_i^2 \rangle \ge \langle \sum_i y_i^2
 \rangle \ge \langle \sum_i z_i^2 \rangle $
and $\langle \sum_i x_iy_i \rangle=\langle \sum_i y_iz_i
 \rangle=\langle \sum_i z_ix_i \rangle=0$.
We see the drastic structure change along the increase of the neutron number
in each series of isotopes.

%%%%%%%%%%%%%%%%%%%%%%%%%%%%%%%%%%%%%%%%%%%%%%%%%%
%%%%%%%%%%%%%%%%%%%%%%%%%%%%%%%%%%%%%%%%%%%%%%%%%%

In the results of Li isotopes (Fig. \ref{fig:liden}), 
it is easily seen that the $^7$Li system has the largest deformation
with the $\alpha$+$t$ clustering structure. $^9$Li also has a deformed
shape, though the deformation is not as large as the one in
$^7$Li. The ground state of $^{11}$Li has an almost spherical structure
that can be expressed by a
shell model wave function with the closed neutron $p$ shell.

%%%%%%%%%%%%%%%%%%%%%%%%%%%%%%%%%%%%%%%%%%%%%%%%%%
%%%%%%%%%%%%%%%%%%%%%%%%%%%%%%%%%%%%%%%%%%%%%%%%%%

%%%%%%%%%%%%%%%%%%%%%%%%%%%%%%%%%%%%%%%%%%%%%%%%%%
%%%%%%%%%%%%%%%%%%%%%%%%%%%%%%%%%%%%%%%%%%%%%%%%%%

\begin{figure}
\caption{\label{fig:berpp}
The distance $R_{pp}$ between two pairs of protons in the intrinsic states
 of the normal parity states (solid line)
and the non-normal parity states (dotted line)  of  
Be isotopes. In Be nuclei with $N \ge 4$ the distance 
$R_{pp}$ is considered to be the inter-cluster distance between
$2\alpha$.}
\centerline{
\epsfxsize=0.7\textwidth
\epsffile{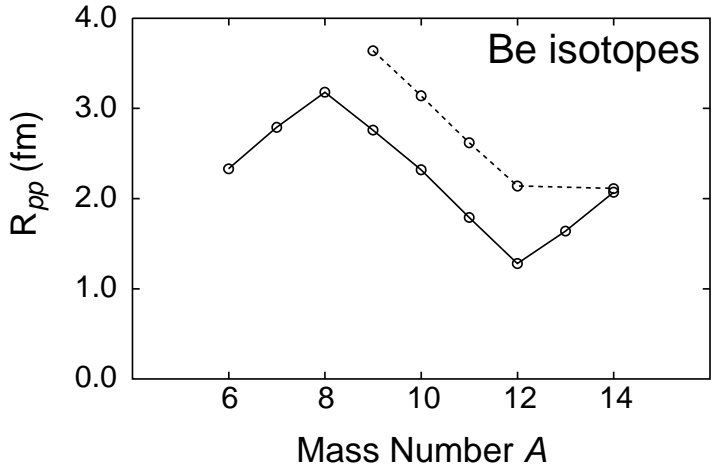}
}
\end{figure}

As for the Be isotopes (Fig. \ref{fig:beden} and \ref{fig:bedenn}), 
separated two pairs of protons are found in the 
proton density. It means that $2\alpha$ core exists in 
all the heavier Be nuclei than $^7$Be.
The development of clustering in Be isotopes is estimated 
by the relative distance between two pairs of protons, which we show
in Fig. \ref{fig:berpp}. 
In the non-normal-parity states we find that 
the $2\alpha$ clustering is largest in $^9$Be as already well
known. In the normal-parity states of Be isotopes, the clustering
becomes weaker and weaker with the increase of neutron number up to 
$^{12}$Be with neutron magic number $N=8$. In $^{14}$Be
the clustering develops again and the deformation becomes larger
than $^{12}$Be.
In the non-normal-parity states of Be isotopes,
there are many exotic structures 
with developed clustering structures and larger deformations
than the normal parity states (Fig.\ref{fig:bedenn}). 
These largely deformed states give rise to the rotational bands
which well agree to the experimental data of energy levels of the
non-normal-parity states in $^9$Be and $^{10}$Be.
The ground state of $^{11}$Be is known to
be a non-normal state with $1/2^+$. The calculated positive-parity state of 
$^{11}$Be which corresponds to the ground $1/2^+$ state has the developed 
prolate deformation as large as the normal-parity state of $^9$Be
(see Fig. \ref{fig:berpp}).
The deformation in the positive parity state of $^{11}$Be
is considered to be one of the reasons for the parity inversion of the 
ground state.

%%%%%%%%%%%%%%%%%%%%%%%%%%%%%%%%%%%%%%%%%%%%%%%%%%
%%%%%%%%%%%%%%%%%%%%%%%%%%%%%%%%%%%%%%%%%%%%%%%%%%

Here we stress again the possibility of the 
abnormally deformed ground state of $^{12}$Be. 
With the simple AMD calculations, the obtained normal-parity state of 
$^{12}$Be has the closed $p$-shell structure with 
a spherical shape of neutrons.
However, the VAP calculations with the set of interaction which 
reproduces the abnormal spin parity $1/2^+$ of the ground state
of $^{11}$Be suggest that the ground state of  $^{12}$Be is 
a $2\hbar\omega$ state with 2 neutrons in $sd$ shell 
in the language of a simple shell-model. In that case, 
the ground state of $^{12}$Be has a large prolate deformation 
with a developed clustering structure, and instead, the spherical $p$-shell
closed-shell state is found in the second $0^+_2$ state.
In the later section on VAP, we will mention the details 
about the structure of excited states of 
neutron-rich Be isotopes.
Although the $1\hbar\omega$ and $2\hbar\omega$ states are suggested
to be the ground states of Be isotopes in $N\ge 7$ region, 
in this section based on the simple AMD results
we discuss the so-called $0\hbar\omega$ 
states of normal-parity states and $1\hbar\omega$ 
states of the non-normal parity states
which are expected 
to be the ground or low excited states.

%%%%%%%%%%%%%%%%%%%%%%%%%%%%%%%%%%%%%%%%%%%%%%%%%%
%%%%%%%%%%%%%%%%%%%%%%%%%%%%%%%%%%%%%%%%%%%%%%%%%%

%%%%%%%%%%%%%%%%%%%%%%%%%%%%%%%%%%%%%%%%%%%%%%%%%%
%%%%%%%%%%%%%%%%%%%%%%%%%%%%%%%%%%%%%%%%%%%%%%%%%%

Also in the AMD results of B isotopes shown in Fig. \ref{fig:bden},
the drastic structure change with the increase of the neutron number
is found.
The total matter density $\rho$ indicates the deformed state with 
a three-center-like clustering in $^{11}$B, while the nucleus $^{13}$B
with a neutron magic number $N=8$ has the most
spherical shape among B isotopes. It is very interesting that in the
neutron-richer region, $^{15}$B, $^{17}$B and $^{19}$B,  
the clustering structure with prolate deformation develops again.
In the right column for neutron density of Fig. \ref{fig:bden}
we can see the neutron structures. 
In $^{11}$B, six neutrons have an oblate-deformed distribution,
while eight neutrons in $^{13}$B constitute the closed shell.
On the other hands, ten neutrons in $^{15}$B posses a large
prolate deformation. It is consistent with the familiar features
that the ordinary nucleus $^{12}$C with $N=6$ has an oblate shape
, the nucleus $^{16}$O with $N=8$ is spherical and $^{20}$Ne with $N=10$
posses a large prolate deformation.
The neutron densities in $^{17}$B and $^{19}$B are 
found to have largely prolately deformed 
structures. The prolate deformation in the system with $N=14$ is not
an obvious feature, but is a characteristic seen in neutron-rich B 
isotopes in which five protons prefer a prolate shape.
Generally speaking proton structure follows the change of neutron deformation
in B isotopes.
It means that in the neutron-rich region
the proton density with two clusters stretches outward as the neutron 
number goes up toward the neutron-drip line.
As a result we find that the
 two-center-like clustering develops more largely in $^{17}$B 
and most largely in $^{19}$B. It is interesting that the theory predicts 
the development of the clustering in B isotopes near the neutron-drip-line 
as well as Be isotopes.
The present results for the development of 
clustering are consistent with the previous work by 
Seya et al. \cite{SEYA}  where they calculated
B isotopes by assuming the existence of a $2\alpha$ core.
It should be pointed out that the present work by AMD 
is the first calculation which predicts the 
clustering structure in neutron-rich Be and B isotopes
without a priori assumption of the cluster cores as far as we know. 
We consider that these clustering structures 
in the unstable nuclei near the neutron drip-line consist of cluster cores and
surrounding neutron cloud because the valence neutrons
are weakly bound in these nuclei. In that sense the clustering structures 
seen in neutron-rich region should be different from
the well-known clustering structure in ordinary nuclei
without valence nucleons around clusters.

In contrast to Be and B isotopes, clustering structures are not found in the 
normal-parity states of neutron-rich C isotopes. 
The results show the general tendency of the oblately deformed proton 
density in the C isotopes.
According to the AMD calculations, 
the well-known $3\alpha$ clustering in $^{12}$C has been checked 
with the framework free from the assumption of the existence of clusters.
The figure for neutron density (right column in Fig.\ref{fig:cden})
presents the drastic change of neutron structure with the increase of 
the neutron number. In the neutron-rich region the neutron density
stretches widely in outer region. However proton density does not 
change so much in spite of the drastic change of neutron structure
and remains in the inner compact region. The stability of the proton
structure is a characteristic of the neutron-rich C with six protons.
As a result the neutron skin structure may appear in the neutron-rich 
C. We will make more quantitative discussions of the neutron skin in C
later.  In the non-normal parity states of the proton-rich C isotopes,
there are many exotic shapes with the large deformation. 
The well-developed $3\alpha$ clustering of $^{12}$C
seen in the negative parity states constructs a rotational $K=3^-$ band which 
well corresponds to the lowest negative-parity band 
observed in the experimental levels.

%%%%%%%%%%%%%%%%%%%%%%%%%%%%%%%%%%%%%%%%%%%%%%%%%%
\begin{figure}
\caption{\label{fig:defopara}
The deformation parameters of the intrinsic states of odd-even
Li and B isotopes and even-even Be and C isotopes.
The definitions $\beta$ and $\gamma$ are described in the text.
The adopted interactions are the force (b) MV1 with $m=0.576$ 
for the nuclei $N\le 8$ and the MV1 force with $m=0.63$ 
for the heavier nuclei $N > 8$.  
}
\centerline{
\epsfxsize=.8\textwidth
\epsffile{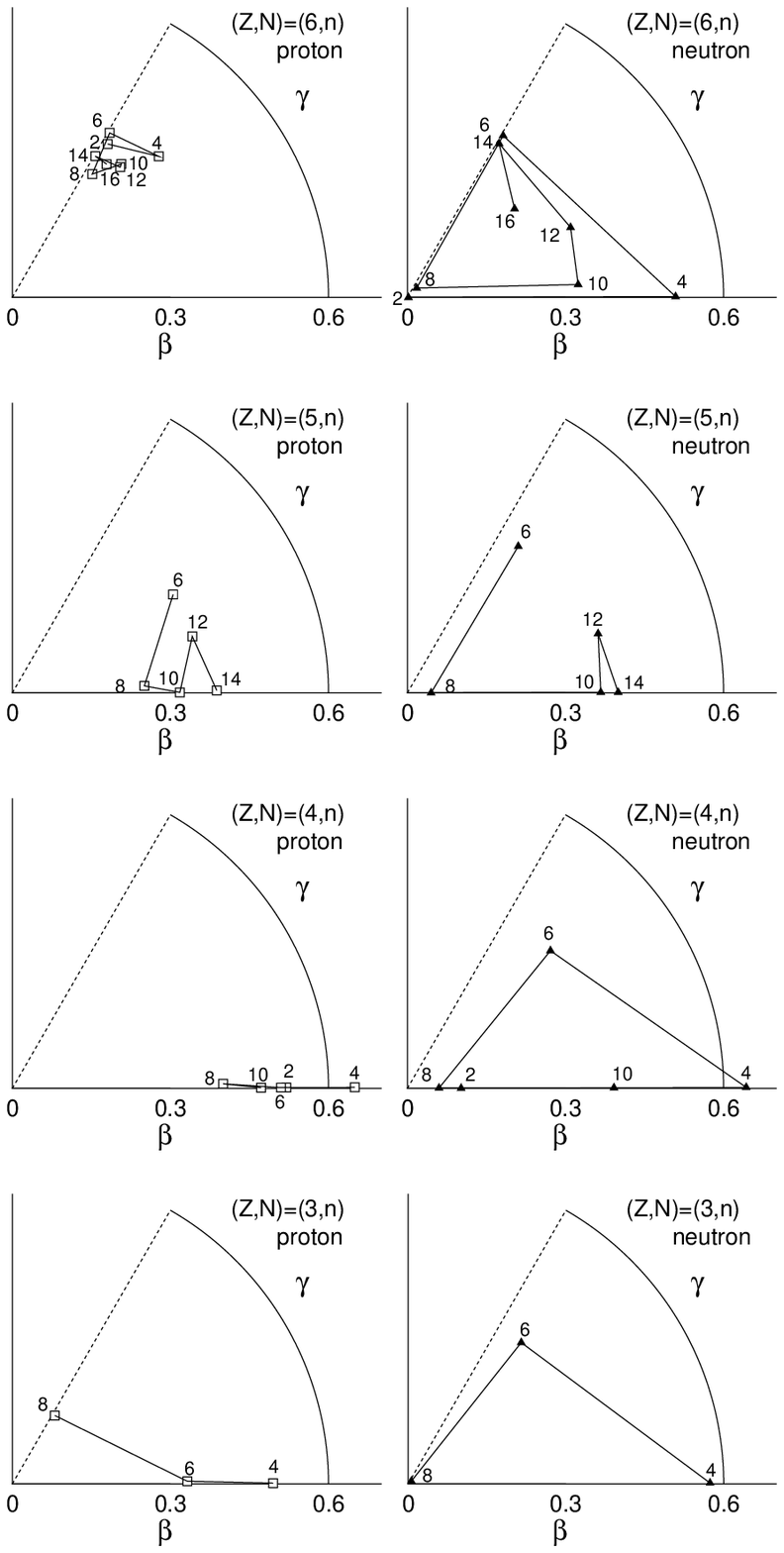}
}
\end{figure}
%%%%%%%%%%%%%%%%%%%%%%%%%%%%%%%%%%%%%%%%%%%%%%%%%%

For the sake of the systematic study of deformations of proton density and 
neutron density, here we explain
the deformation parameters defined by the moments 
$\langle x^2\rangle$,
$\langle y^2\rangle$, and
$\langle z^2\rangle$
for the intrinsic AMD wave functions
as follows;
\begin{eqnarray}
{\langle x^2 \rangle^{1\over 2} \over
\left(\langle x^2 \rangle
\langle y^2 \rangle
\langle z^2 \rangle\right)^{1\over 6}} & \equiv&
{\rm exp}\left(\delta_1\right)
={\rm exp}\left[
\sqrt{5\over 4\pi}\beta {\rm cos}\left(\gamma+{2\pi\over 3}\right)
\right],\nonumber\\
{\langle y^2 \rangle^{1\over 2} \over
\left(\langle x^2 \rangle
\langle y^2 \rangle
\langle z^2 \rangle\right)^{1\over 6}} & \equiv&
{\rm exp}\left(\delta_2\right)
={\rm exp}\left[
\sqrt{5\over 4\pi}\beta {\rm cos}\left(\gamma-{2\pi\over 3}\right)
\right],\nonumber\\
{\langle z^2 \rangle^{1\over 2} \over
\left(\langle x^2 \rangle
\langle y^2 \rangle
\langle z^2 \rangle\right)^{1\over 6}} & \equiv&
{\rm exp}\left(\delta_3\right)
={\rm exp}\left[
\sqrt{5\over 4\pi}\beta {\rm cos}\gamma
\right].\label{eqn:defomoment}
\end{eqnarray}
Here the directions $x$, $y$, and $z$ are chosen so as to satisfy the
relation 
$\langle x^2\rangle \le
\langle y^2\rangle \le
\langle z^2\rangle$. The deformation parameters $\beta$ $\gamma$ 
calculated for protons 
and for neutrons are displayed in Fig. \ref{fig:defopara}.
The Figures are for the normal-parity states of 
Li, Be, B and C isotopes with the even neutron number obtained 
with the interaction (b) $m=0.576$ for the nuclei with $N\le 8$
and $m=0.63$ for the nuclei with $N > 8$.
The behavior of the deformation parameters is not so sensitive
to the Majorana parameter.
It is found that the neutron shape changes rapidly with the increase of 
neutron number in all the series of isotopes.
In the region $N\le 12$, the main feature of deformations of the
neutrons, prolate or oblate shape, is dominated 
only by the neutron number. It means that the neutron shape 
is not sensitive to the proton number in the light region.
The neutron density in the system with $N=2$ has a spherical shape.
In $N=4$ system the neutron density deforms prolately, and
in the case of $N=6$ the neutrons prefer oblate deformation.
In $N=8$ the neutron shape becomes almost spherical due to 
the closed neutron $p$ shell.
When the neutron number becomes 10, the prolate neutron deformation
appears again.
As for the proton deformation, it depends on the proton number.
In the system with $Z=2$, Be isotopes, the proton density prefers the 
prolate deformation. The degree of prolate deformation 
of Be isotopes changes following the drastic change of the neutron deformation.
Especially the prolate deformation of protons is well developed in the 
system with the prolately deformed neutron density.
Contrary to Be isotopes, in the case of C isotope with $Z=6$,
the proton density always prefers the oblate deformation.
The deformation parameter for proton is stable in spite of the 
change between prolate and oblate shapes of the neutron density.
In the case of $Z=3$ and $Z=5$, Li and B isotopes, the 
proton shape depends on the 
neutron number so as to follow the deformed mean field given by the 
neutron density.
In the system with the heavier neutron number such as $N=14$,
the neutron shape possesses both characters, which are seen in  
the oblate neutron shape in $^{20}$C and in the prolate deformation 
in $^{19}$B.
In such a case with a few choice of the neutron shape, 
the neutron shape is determined by the  proton number.

As mentioned above, in the very light region
the neutron deformation is dominated by the neutron number, and the
proton shape is  basically determined by the proton number.
The dependence of neutron(proton) shape 
on the neutron(proton) number is consistent with the 
ordinary understanding of the shell effect for the stable nuclei. 
That is to say 
the spherical shape is seen 
when the neutron number equals the magic number $M$, 
the prolate deformation at $M+2$,  and oblate one at $M-2$.
However one of the new features found in this study of the light 
unstable nuclei is that the proton shape and the neutron shape do
 not necessarily correlate together in light region. As a result 
interesting phenomena such as the opposite deformation 
between protons and neutrons may occur in unstable nuclei.
For instance, C nuclei prefer oblate deformation of protons 
while neutrons tend to deform prolately when the neutron number 
equals to 4 and 10. Therefore $^{10}$C and $^{16}$C may have the 
opposite deformation between protons and neutrons.
The detail of this problems in proton-rich C will be discussed later. 
The other interesting feature is the large deformation in the 
nuclei with prolate protons and prolate neutrons. 
The developments of deformations in 
$^8$Be, $^{14}$Be, $^{15}$B, $^{17}$B, and $^{19}$Be are
 understood as follows.
Once a prolate shape of protons is chosen by the proton number, 
the prolate deformation is enhanced and the clustering of protons is developed
by the neutron deformations if the neutrons deform prolately.

%%%%%%%%%%%%%%%%%%%%%%%%%%%%%%%%%%%%%%%%%%%%%%%%%%
\begin{figure}
\caption{\label{fig:quant}
The deviation of the proton orbits and neutron orbits
 from the harmonic oscillator shell-model wave functions
 in the normal-parity states obtained with AMD calculations. 
The expectation values of the total number of the oscillator quanta
are calculated with the spin-parity eigen states projected from 
the intrinsic AMD wave functions. The detail is mentioned in the text.
The interaction (b) MV1 with $m=0.576$ is used 
in the region $N\le 8$, and the MV1 force with $m=0.63$ is adopted
in the region $N > 8$.
}
\centerline{
\epsfxsize=1.1\textwidth
\epsffile{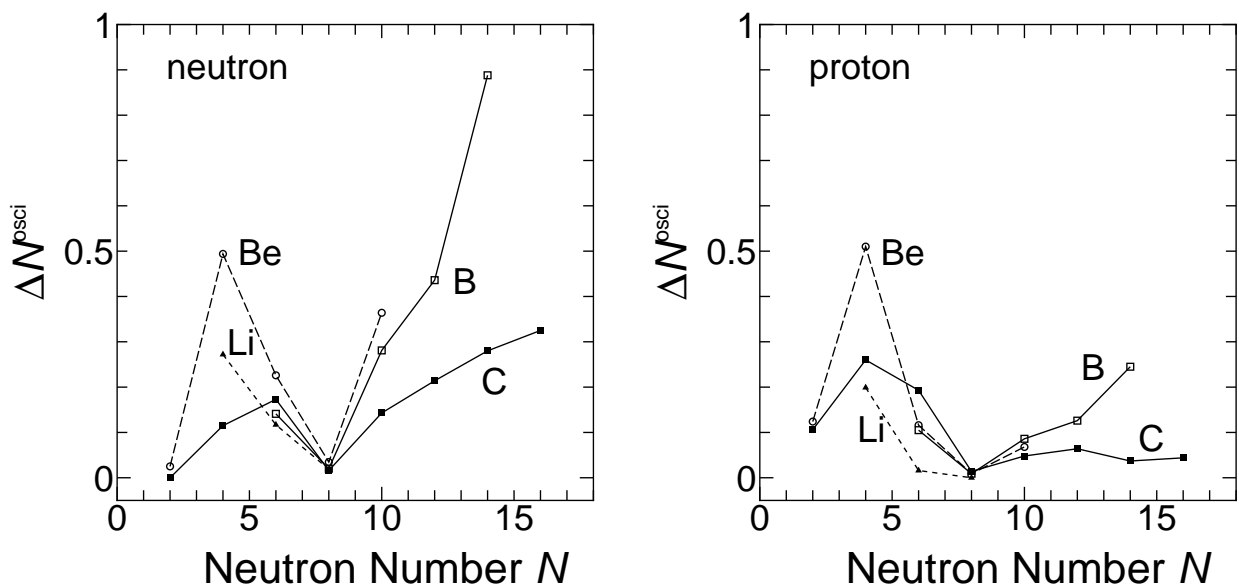}
}
\end{figure}
%%%%%%%%%%%%%%%%%%%%%%%%%%%%%%%%%%%%%%%%%%%%%%%%%%

In order to analyze clustering development quantitatively,
we calculate the expectation values for the total number of the oscillator 
quanta. 
In general if the clustering structure develops 
the wave function of the system contains the components of 
a large amount of the orbits in higher shells, therefore the 
the expectation values of the oscillator quanta become larger. 
On the other hands a small value of the oscillator quantum indicates
that the state is almost written by the shell-model states
in the $0\hbar\omega$ configurations
but the spatial clustering does not exist.
We introduce the value $\Delta N_p$ ($\Delta N_n$) 
which stands for the deviation of the proton (neutron)
orbits in the AMD wave function from those in the $0\hbar\omega$ states 
of the harmonic oscillator shell model basis,
\begin{eqnarray}
&\Delta N_p \equiv 
\frac
{\langle P^{J\pm}_{MK}\Phi_{AMD}|{N}^{op}_p|
P^{J\pm}_{MK}\Phi_{AMD}\rangle}
{\langle P^{J\pm}_{MK}\Phi_{AMD}|
P^{J\pm}_{MK}\Phi_{AMD}\rangle}
-N^{min}_p
,  \\
&
\Delta N_n \equiv 
\frac
{\langle P^{J\pm}_{MK}\Phi_{AMD}|{N}^{op}_n|
P^{J\pm}_{MK}\Phi_{AMD}\rangle}
{\langle P^{J\pm}_{MK}\Phi_{AMD}|
P^{J\pm}_{MK}\Phi_{AMD}\rangle}
-N^{min}_n,  \\
\end{eqnarray}
where $N^{op}_p$ and $N^{op}_n$ are the oscillator quantum number
operators and $N^{min}_p$ ($N^{min}_n$) are the minimum values of 
oscillator quantum numbers for protons (neutrons) 
given by the $0\hbar\omega$ state. 
We choose the same width parameters for $N^{op}_p$ and $N^{op}_n$ 
as the width of the Gaussians in the AMD wave functions for simplicity. 
In Fig. \ref{fig:quant}, $\Delta N_p$ and
$\Delta N_n$ for the normal parity states 
are displayed as a function of neutron number. 
We can estimate the clustering development by 
the deviation of proton orbits $\Delta N_p$ from the $0\hbar\omega$ state.
As is expected, the dependence of $\Delta N_p$ of Be isotopes on the 
neutron number is found to be
very similar to the that of the inter-cluster distance 
between $2\alpha$
which has been already shown in Fig.\ref{fig:berpp}. 
In the Li, Be, and B isotopes the neutron number 
dependence of $\Delta N_p$ is qualitatively similar to 
the behavior of $\Delta N_n$ for neutrons. 
It is because the development of the clustering structure in these isotopes
is sensitive to the neutron structure determined by the neutron number.
In the Be and B isotopes, the increase of $\Delta N_p$ in 
the region $N > 8$ indicates that the clustering develops as the neutron 
number increases toward the neutron-drip line.
In all the nuclei with $N=8$, the $\Delta N_n$ and also $\Delta N_p$ 
are almost zero, which stands for that the states are  approximately 
same with the shell-model states and can be  
written by the major shell orbits. 
We stress again that we analyzed the structure change of 
the normal-parity states where the main components are $0\hbar\omega$ 
states even if they are not necessarily ground states. 
The effect of the neutron magic number $N=8$ is clearly seen
in the valley at $N=8$ in the variation of $\Delta N_p$.
In the case of the C isotopes, the large $\Delta N_p$
at $N=6$ indicates the developed clustering structure 
in $^{12}$C, while the small values of $\Delta N_p$ in the region $N\ge 8$
are because of the disappearance of clustering in the neutron-rich C
as already mentioned.

\subsubsection{Effects of intrinsic structures on 
electromagnetic properties}
  
Here we 
consider the effect of clustering structures 
on the observable electric and magnetic moments by
analyzing $\mu$ and $Q$ moments of the ground states in Li and B isotopes.
As shown above 
the simple AMD calculations well agree to the experimental data of  
the electromagnetic properties such as the
magnetic dipole moments $\mu$ and the electric quadrupole 
moments $Q$, the strength of transitions. 
Roughly speaking, the electromagnetic properties of the odd-even nucleus
reflect the orbit of the last valence proton.
In that sense the last
proton in $p$-shell may dominate the moments in the odd-even 
Li and B isotopes, 
however, $\mu$ and $Q$ moments in these nuclei
shift as functions of the neutron number (Fig. \ref{fig:libemu}
 and \ref{fig:libebqmom}).
The $N$ dependence of the experimental data of the 
electric and the magnetic properties 
can be explained in relation to the drastic change between the cluster
and shell-model-like structures. 

In the following discussion, 
we consider two kinds of the fundamental effects of the cluster
structures on the properties such as the electric and magnetic moments. 
One is caused by the spatial relative distance between clusters
(spatial clustering effects), and the other 
is concerned with the angular momentum coupling correlation of nucleons. 
As a typical example of the latter effect, we recall the so-called 
shell-model cluster in the SU$^3$ coupling shell-model 
\cite{BRINK,BAYBOHR} configuration.
As we show below, it is found that 
the effect of the cluster structure on the magnetic dipole 
moments of Li and B isotopes is given only by the cluster coupling 
of angular momenta, while the quadrupole moments are effected by 
the spatial cluster as well as the cluster coupling 
of angular momenta.
 In order to extract the effect of the cluster coupling of angular 
momenta from the AMD wave functions, we have artificially made the 
inter-cluster relative distances in the AMD wave functions very small
so as to obtain the states in the shell-model limit.
In the obtained shell-model limit states, the spatial cluster is not
recognizable any more but only the effects of the cluster coupling 
of angular momenta persist. 

\begin{table}%[tbh]
\caption{\label{tbl:libqmusl}
The expectation values of the magnetic dipole operator $\mu$,
the electric quadrupole operator $Q$,  
the square of total orbital-angular-momenta
of protons $\langle {\bf L}^2_p \rangle$,
the square of total orbital-angular-momenta
of neutrons  $\langle {\bf L}^2_n \rangle$,
the square of total orbital-angular-momenta of the system
$\langle {\bf L}^2 \rangle$,
the total intrinsic spin of protons
$\langle {\bf S}^2_p \rangle$,
the total intrinsic spin of neutrons
$\langle {\bf S}^2_n \rangle$, and
the total intrinsic spin of the system
$\langle {\bf S}^2 \rangle$
for the highest states $|{3\over 2}{3\over 2}\rangle $ 
in the simplest AMD calculations of Li and B isotopes. 
Those in the shell-model limit are also listed.}
\begin{center}
\begin{tabular}{cccccccccc}
\hline\hline
 & & {\hfill$\mu$\hfill} & {\hfill$Q$\hfill}  
 & $\langle{\bf L}^2_n\rangle$ 
 & $\langle{\bf L}^2_p\rangle$  
 & $\langle {\bf L}^2 \rangle$
& $\langle{\bf S}^2_n\rangle$ 
& $\langle{\bf S}^2_p\rangle$ 
& $\langle{\bf S}^2\rangle$ \\
 & & ($\mu_N$) & (e$\cdot$mb) & & & & & & \\
\hline
 & EXP.& 3.27 & $-$40(3) 
& & & & & & \\
 $^{{7}}$ Li & AMD & 3.15 & $-$27.6 & 2.8 & 2.2 &2.0
& 0.00 & 0.75 & 0.75\\
 & SML & 3.14 &$-$15.1 & 2.6 & 2.0 & 2.0 & 0.00 & 0.75 & 0.75\\
\hline
 & EXP.& 3.44 & $-$27(1) 
& & & & & & \\
 $^{{9}}$ Li & AMD & 3.42 & $-$27.0 & 1.1 & 2.0 &2.1 
& 0.13 & 0.75 & 0.90\\
 & SML & 3.44 & $-$23.2 & 1.1 & 2.0 & 2.0 & 0.13 & 0.75 & 0.90\\
\hline
 & EXP.& 3.76 & $-$31(5) 
& & & & & & \\
 $^{{1}{1}}$ Li & AMD & 3.79 & $-$29.4 & 0.0 & 2.0 &2.0
& 0.00 & 0.75 & 0.75\\
 & SML & 3.79 & $-$29.4 & 0.0 & 2.0 & 2.0 & 0.00 & 0.75 & 0.75\\
\hline
\hline
 & EXP.& 2.69 & 40 
& & & & & & \\
 $^{11}$ B & AMD & 2.65 & 34.0 & 2.5  & 3.6 & 2.8 & 0.04 & 0.75
& 0.78 \\
 & SML & 2.66 & 25.9 &  2.3 & 3.4 &2.8& 0.04& 0.75 & 0.78\\
\hline
 & EXP.& 3.17 & 37(4) 
& & & & & & \\
 $^{13}$ B & AMD & 3.17 & 31.7 &  0.0 & 2.7 & 2.7&0.00 & 0.75
& 0.75 \\
 & SML & 3.18 & 28.6 &  0.0 & 2.7 & 2.7&0.00& 0.75  & 0.75\\
\hline
 & EXP.& 2.66 & 38(1) 
& & & & & & \\
 $^{{1}{5}}$ B & AMD & 2.63 & 34.3 & 3.7 & 3.8 &2.8 &0.00 & 0.75
& 0.75 \\
 & SML & 2.64 & 22.5 &  3.5 & 3.7 &2.8& 0.00 & 0.75 & 0.75 \\
\hline
 & EXP.& $2.54$ & $-$
& & & & & & \\
 $^{{1}{7}}$ B & AMD & 2.49 & 42.2 & 4.4 & 4.1 &2.9& 0.07 & 0.75&
 0.81\\
 & SML & 2.50 & 22.6 &  4.0 & 3.7 & 2.9&0.33 & 0.75 & 0.77\\
\hline
 & EXP.& $-$ & $-$ 
& & & & & & \\
 $^{{1}{9}}$ B & AMD & 2.53 & 50.8 & 4.3 & 4.2 & 2.90&0.00 & 0.75
& 0.75 \\
 & SML & 2.55 & 24.9 &  3.9 & 3.8 & 2.9&0.00& 0.75& 0.75\\
\hline\hline
\end{tabular}
\end{center}
\end{table}

\begin{table}%[tbh]
\caption{
\label{tbl:libqmusl2}
The expectation values of the 
$z$-components of the operators which effect on the $\mu$ moments;
the total orbital-angular-momenta
of protons  $\langle L_{pz} \rangle$,
the total orbital-angular-momenta
of neutrons $\langle L_{nz} \rangle$,
the total intrinsic spin
of protons  $\langle S_{pz} \rangle$, and 
the total intrinsic spin
of neutrons  $\langle S_{nz} \rangle$,
for the highest states $|{3\over 2}{3\over 2}\rangle$ 
in the simplest AMD calculations and in the shell-model limit
for Li and B isotopes.
}
\begin{center}
\begin{tabular}{cccccc}
\hline\hline
&&&&&\\
 & 
& $\langle S_{pz} \rangle$
& $\langle S_{nz} \rangle$ 
& $\langle L_{pz} \rangle$ 
& $\langle L_{nz} \rangle$\\ 
\hline
&&&&&\\
 $^{{7}}$ Li & AMD & 0.50 & 0.00 & 0.36 & 0.64 \\
 & SML & 0.50 & 0.00& 0.35 & 0.65 \\
\hline
&&&&&\\
 $^{{9}}$ Li & AMD & 0.50 & 0.02 & 0.71 & 0.27 \\
 & SML & 0.50 & 0.02 & 0.72 & 0.26 \\
\hline
&&&&&\\
 $^{{1}{1}}$ Li & AMD & 0.50 & 0.00& 1.00 & 0.00\\
 & SML & 0.50 & 0.00& 1.00 & 0.00\\
\hline
\hline
&&&&&\\
 $^{11}$ B & AMD & 0.34 & 0.00 & 0.74 & 0.41 \\
 & SML & 0.34& 0.00 & 0.74 & 0.41\\
\hline
&&&&&\\
 $^{13}$ B & AMD & 0.37& 0.00& 1.13 & 0.00\\
 & SML & 0.37& 0.00&1.13 & 0.00\\
\hline
&&&&&\\
 $^{{1}{5}}$ B & AMD & 0.33& 0.00& 0.77& 0.40\\
 & SML & 0.34 &0.00 & 0.77 & 0.40 \\
\hline
&&&&&\\
 $^{{1}{7}}$ B & AMD & 0.33& 0.00 & 0.68&
 0.49 \\
 & SML & 0.33 & 0.00 & 0.68 & 0.49\\
\hline
&&&&&\\
 $^{{1}{9}}$ B & AMD &0.33 & 0.00 & 0.73
& 0.45 \\
 & SML & 0.33&0.00 & 0.73& 0.45\\
\hline\hline
\end{tabular}
\end{center}

\end{table}

\begin{figure}
\caption{\label{fig:libmusl}
The effects due to the intrinsic spins $\mu_s$ and of 
the orbital-angular momentum $\mu_l$ on the total 
magnetic dipole moments $\mu$ of Li and B isotopes.
The experimental data are also shown with square points.
}
\centerline{
\epsfxsize=0.7\textwidth
\epsffile{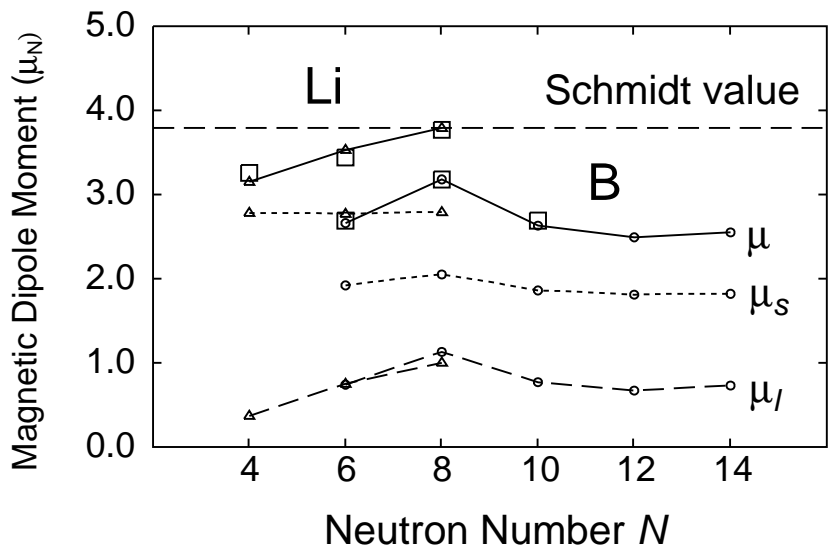}
}
\end{figure}

Table \ref{tbl:libqmusl} shows the results 
of $\mu$ and $Q$ moments calculated with 
the spin-parity projected states from the shell-model limit wave functions,
 which are compared with the original AMD results.
In the table we also list the expectation values of the squared 
total-angular momenta ${\bf J}_p$, ${\bf J}_n$, ${\bf J}$, 
the orbital-angular momenta ${\bf L}_p$, ${\bf L}_n$, ${\bf L}$
and the intrinsic spins ${\bf S}_p$, ${\bf S}_n$, ${\bf S}$
for protons, neutrons and for the total system, 
which have close relations with the spin configurations. 
We also calculate the $z$-components of the 
orbital-angular momenta and the intrinsic spins of protons and neutrons
in the highest $M$ states $|JM\rangle=|\frac{3}{2}\frac{3}{2}\rangle$.
Since the $\mu$ moments in the shell-model limit 
are found to be almost the same as 
those of the original AMD, it is confirmed that the magnetic dipole 
moments do not depend on the spatial clustering but are effected only by
the cluster coupling of angular momenta. It is easily understood 
because the expectation values of linear terms of operators, 
{\bf J}, {\bf L} and
{\bf S} are mainly determined by the cluster coupling of angular momenta. 
As shown in the Table \ref{tbl:libqmusl} 
the magnitude of the total neutron intrinsic spin almost equals 
to 0 because the intrinsic spins of the even neutrons are
almost paired off.
It means that the direct contribution to the $\mu$ moments from 
the neutrons is little. 
Then the $\mu$ moments of odd-even Li and B 
isotopes approximately consist of two
terms from $z$-components of ${\bf S}_p$ and ${\bf L}_p$, 
\begin{eqnarray}
& \mu\approx \mu_s+\mu_l\\
& \mu_s\equiv 5.58 
\langle{3\over 2}{3\over 2}|S_{pz}|{3\over 2}{3\over 2}\rangle \mu_N, 
 \mu_l\equiv 
\langle{3\over 2}{3\over 2}|L_{pz}|{3\over 2}{3\over 2}\rangle \mu_N,
\end{eqnarray}
where $\mu_N$ stands for the nuclear magneton.
Figure \ref{fig:libmusl} presents the components from the 
two terms $\mu_s$ and $\mu_l$ in the 
total $\mu$ moments. 

%%%%%%%%%%%%%%%%%%%%%%%%%%%%%%%%%%%%%%%%%%%%%%%%%%
%%%%%%%%%%%%%%%%%%%%%%%%%%%%%%%%%%%%%%%%%%%%%%%%%%
%%%%%%%%%%%%%%%%%%%%%%%%%%%%%%%%%%%%%%%%%%%%%%%%%%
\begin{figure}
\caption{\label{fig:qmomsl}
The $Q$ moments of Li and B isotopes for the original AMD wave functions
and the shell-model limit described in the text.
Squares are the experimental data.
The spatial clustering effects can be estimated by subtracting
the $Q$ moments (cross points )
in the shell-model limit from the original results (open circles).
}
\centerline{
\epsfxsize=0.9\textwidth
\epsffile{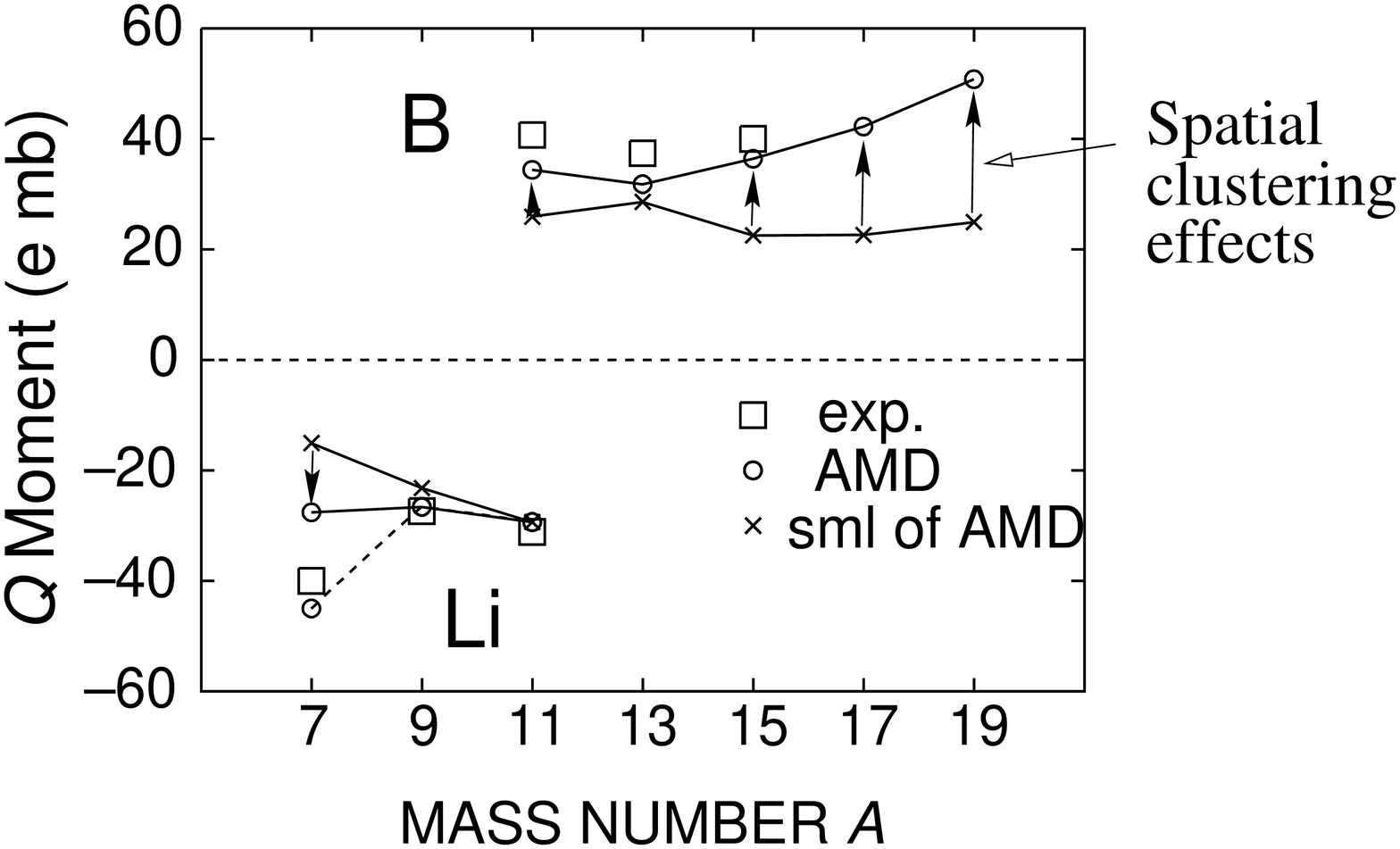}
}
\end{figure}
%%%%%%%%%%%%%%%%%%%%%%%%%%%%%%%%%%%%%%%%%%%%%%%%%%

In the Li isotopes, $\mu_s$ is almost constant 
because $|{\bf S}_p|={1\over 2}$ and $S_{pz}=0.5$ 
(see Table \ref{tbl:libqmusl2}).
The deviation of $\mu$ from the Schmidt value and 
its $N$ dependence originate from the latter terms
$\mu_l$ due to the orbital-angular momentum of protons. 
The reason of the $N$ dependence of $\mu_l$ can be understood as follows.
The total angular momenta $|{\bf L}|$ are always 1 
and $L_{z}=1.0$ in Li isotopes. However the state with the cluster
coupling contains the components of $|{\bf L}_n|\ne 0$ states with 
non-zero total-orbital-angular momentum of neutrons,
which makes $L_{pz}$ decrease in the highest 
$|{3\over 2}{3 \over 2}\rangle$ states. 
That is why $^{7}$Li with the clustering structure has the smallest 
$\mu$ moments in odd-even Li isotopes.
In other words, the $N$ dependence
of $\mu$ moments of Li isotopes is described by the 
non-zero ${\bf L}_n$ of neutrons due to the 
clustering coupling effects. 

In the case of B isotopes, the deviations of the $\mu$ moments from the 
Schmidt value 3.79$\mu_N$ for the $p_{3/2}$ proton orbit are not 
as small as the case of Li isotopes. 
In B isotopes the $z$-component $\langle S_{pz} \rangle$ has proved to be
about 0.35 which gives a smaller magnitude $\mu_s\approx 1.8\mu_N$
than the case of $\langle S_{pz} \rangle=0.5$.
In the results in Table \ref{tbl:libqmusl}, it is found that  
the calculated values of $\langle {\bf L}^2\rangle$ 
are almost constantly 2.8 in B isotopes and 2.0 in Li isotopes.
It implies that 
the wave functions of Li isotopes are the almost pure $L=1$ states, 
while those of  B isotopes contains $L=1$ states and $L=2$ whose components 
are easily estimated to be about 20\%. 
The mixing of the $L=2$ states in B isotopes makes ${\bf S}_p$ tilted from the
$z$ direction in the $|JM\rangle =|{3\over 2}{3\over 2}\rangle$ 
states as to $S_{pz}\approx 0.35$.
The reason of the pure $L=1$ state in Li and the mixing ratio 
of the $L=2$ states in B are described in Ref.\cite{ENYOc} in more detail.
The second term $\mu_l$ causes the $N$ dependence of $\mu$ moments of 
B isotopes as similar way with Li isotopes.
Even though the magnitude $|{\bf L}|$ and 
the $z$-component $\langle L_z \rangle$ 
of the total-orbital-angular momentum 
 is approximately constant in the B isotopes,
the clustering gives
the components of ${\bf L}_{n}\ne 0$ states which reduce 
the magnitude of $\langle L_{pz} \rangle$.
Because of such a effect of the clustering coupling on the $\mu_l$ term, 
the magnetic moments of $^{11}$B, $^{15}$B, $^{17}$B and $^{19}$B
are smaller than the one of $^{13}$B 
which has the shell-model like structure.
Since the closed neutron shell in the nuclei $^{13}$B with  
$N=8$ make ${\bf S}_n={\bf L}_n={\bf J}_n=0$, the $\mu$ moments of 
$^{13}$B is the largest in B isotopes.
Similarly to Li isotopes, it is concluded that 
the $N$ dependence of $\mu$ moments of B isotopes
is understood by the cluster coupling effect.

In contrast to the magnetic dipole moments, 
the electric quadrupole moments are sensitive to the relative distance
between clusters. For the sake of estimating the spatial cluster 
effects on $Q$ moments, we try to decompose 
the calculated $Q$ moments into two components: 
The first component is originated by the spatial clustering, 
and the other is due to other properties including 
the cluster coupling of angular momenta.
We consider that the second components are given by the $Q$ moments
calculated by the shell-model-limit wave function where 
the spatial clustering has been already removed.
They are shown in Table\ \ref{tbl:libqmusl} together with the $Q$-moments
of the AMD calculations. We display the two components in $Q$ moments
in the Fig. \ref{fig:qmomsl}, which is helpful to analyze the $N$ dependence
of the each component.
As for the B isotopes the calculated $Q$ moments 
in the shell model limit are 25.9, 28.6, 22.5, 22.6 and 24.9
e$\cdot$mb for $^{11}$B, $^{13}$B, $^{15}$B,
$^{17}$B and $^{19}$B, respectively.
It shows that the second components in these nuclei except for $^{13}$B
are smaller than the one in $^{13}$B with $N=8$.
With similar argument as for $\mu$-moments, we expect that
the reduction of the $Q$-moments in other B isotopes than $^{13}$B
should be explained by the mixing of the component of the
non-zero neutron orbital-angular momentum.
By subtracting these second components from the 
total $Q$-moments ( namely the $Q$-moments of the AMD calculations ), 
we can estimate the contribution of the first component 
due to the spatial clustering
as 8.1, 3.1, 11.8, 19.6 and 25.9
e$\cdot$mb 
for $^{11}$B, $^{13}$B, $^{15}$B, $^{17}$B and $^{19}$B, respectively.
This component is smallest in $^{13}$B and becomes larger
as the neutron number increases toward the neutron drip-line.
Such dependence on $N$ of the first component is indeed consistent
with the clustering development mentioned 
in the previous subsection.
In the case of Li isotopes the first component is largest in $^7$Li 
and decreases toward $^{11}$Li as the clustering structure weakens.

Thus it is proved that 
the systematic $N$-dependent features of 
experimental data for the electric magnetic properties 
are quantitatively explained  
by the structure change given by our AMD results. 
The reader is referred to Ref.\cite{ENYOc} for the detailed discussions.

\subsubsection{Opposite deformation between protons and neutrons}

\begin{figure}%----------------------------------------

\caption{\label{fig:cpdefo}
Deformation parameters $(\beta,\gamma)$ for proton and neutron
densities in the intrinsic states of
proton-rich C isotopes with $A=9\sim 11$ are given by triangles and
squares, respectively. 
The mass number  $A$ is indicated beside each point 
in the figure.
The interaction (b) is used.
}
\epsfxsize=0.6\textwidth
%\begin{minipage}[b]{0.45\textwidth}
\centerline{\epsffile{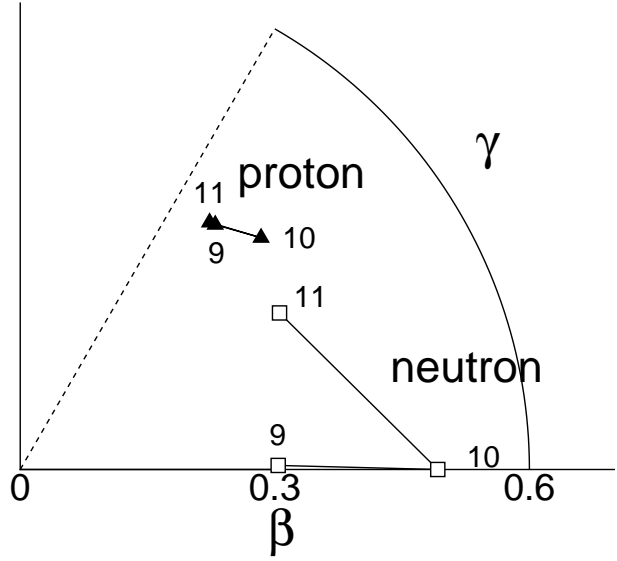}}
%\end{minipage} \ \hskip 0.05\textwidth \
%
\end{figure}

As mentioned above, AMD results suggest that 
the opposite deformations between protons and neutrons may be found in 
the proton-rich C isotopes.
In Fig. \ref{fig:cpdefo}, we illustrate the deformation parameters 
of protons and neutrons in proton-rich C isotopes. 
It is notable that the neutrons prefer the prolate and the triaxial deformations
rather than the oblate shape in these nuclei
because of the neutron number $N=3,4,5$.
On the other hands, the protons prefer oblate shape in C isotopes.
As a result the disagreement between the proton
and the neutron deformations is found in proton-rich C.
Our purpose here is to confirm
the disagreement between the proton and the neutron shapes
by the help of the electric quadrupole moments and
transitions in C and the ones in the mirror nuclei.

\begin{figure}%----------------------------------------

\noindent
\caption{\label{fig:obpro}
Schematic figures of the nucleus with oblate proton and prolate neutron
deformations. 
In the body-fixed flame $x,y,z$ is chosen so that 
moments of inertia obey the relation 
${\cal I}_{zz}\le {\cal I}_{yy} \le {\cal I}_{xx}$.
}
%\end{minipage} \ \hskip 0.05\textwidth \
%
\epsfxsize=0.7\textwidth
%\begin{minipage}[b]{0.45\textwidth}
\centerline{\epsffile{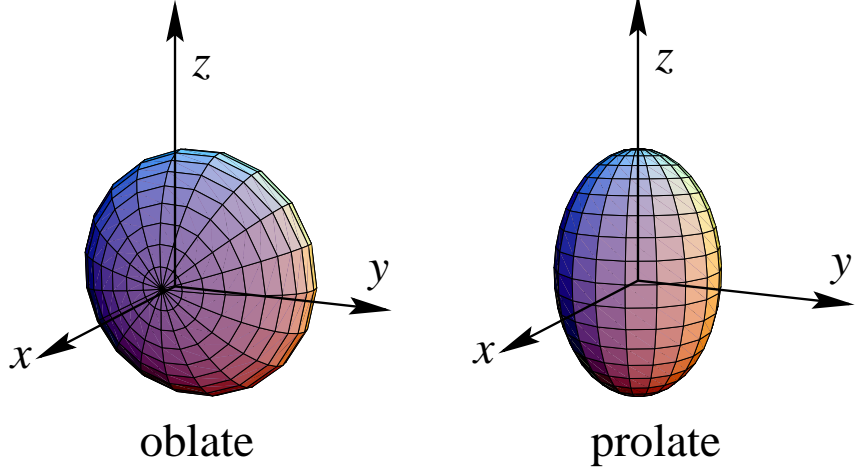}}
\end{figure}

First we discuss $Q$ moments of $^{11}$C and the mirror 
nucleus $^{11}$B.
Based on mirror symmetry for proton and neutron deformations, 
we compare the proton and the
neutron deformations 
by analyzing the ratio of the electric quadrupole moment 
$Q$ in $^{11}$C to the one in $^{11}$B.
We introduce the well-known approximate relation between
the electric quadrupole moment $Q$ in the laboratory frame 
and the intrinsic quadrupole moment $Q_0$;
\begin{equation}
Q=Q_0{3 K^2 -J(J+1)\over (2 J+3)(J+1)}.
\end{equation}
By using Eq. \ref{eqn:defomoment} we can express
the intrinsic electric 
quadrupole moment $Q_0$ as follows in the first order of the deformation
parameter $\beta_p$,
\begin{equation}
Q_0=\sqrt{16\pi\over 5}{3\over 4\pi} Ze\beta_p {\rm cos}\gamma_p R_e^2,
\label{eqn:q0}
\end{equation}
where $\beta_p$ and $\gamma_p$ are the deformation parameters for
the proton density, and 
$Z$ and $R_e$ are the proton number and the charge radius, respectively.
Instead of the usual deformation parameter 
$\beta_p$ in the usual equation 
$Q_0=\sqrt{16\pi\over 5}{3\over 4\pi} Ze\beta_p R_e^2$,
we introduce the effective deformation parameter 
$\beta_p{\rm cos}\gamma_p$ in Eq.\ref{eqn:q0} 
which is necessary for the system with 
different proton and neutron shapes as described below.
We explain the appropriate principal axes 
in the nucleus where the different proton and neutron shapes
coexist as shown in schematic Figure \ref{fig:obpro}.
For example in the nucleus with oblate proton and prolate neutron
deformations, the approximate symmetry axis $x$ for protons usually differs
from the approximate symmetry axis $z$ for neutrons 
so as to make the largest overlap between the proton density and
the neutron density. 
In many cases, it has been found that a symmetry axis $z$ for the prolate
neutron density should be chosen as the principal axis $Z$ of 
the total intrinsic system for the total-angular momentum projection
in generating the lowest $J^\pm$ state with an approximately 
good $K$ quantum.
In such cases, the usual formula for $Q_0$ should be
modified by using the effective deformation to $z$-axis, 
$\beta_p$cos$\gamma_p$ instead of $\beta_p$. 
In other words, an oblate deformation gives a smaller contribution 
to the intrinsic quadrupole moment $Q_0$ than is expected.

%%%%%%%%%%%%%%%%%%%% BEGIN OF FIGURE %%%%%%%%%%%%%%%%%%%%%%%%
%%%%%%%%%%%%%%%%%%%%%%%%%%%%%%%%%%%%
%%%%%%%%%%%%%%%%%%%% BEGIN OF FIGURE %%%%%%%%%%%%%%%%%%%%%%%%
%%%%%%%%%%%%%%%%%%%%%%%%%%%%%%%%%%%%

Assuming these simple approximations
the ratio of the $Q$ moment in $^{11}$C to that in $^{11}$B 
is represented by the product of three terms, the 
ratio of proton numbers, that of the
proton deformation parameters, and the one of the charge radii.
When we assume the mirror symmetry for the deformation parameters
and replace the deformation parameter $\beta_p{\rm cos}\gamma_p
(^{11}$B) 
of the proton density in $^{11}$B
by $\beta_n{\rm cos}\gamma_n(^{11}$C) 
of the neutron density in $^{11}$C, the ratio of $Q$
is written as,
\begin{equation}
{Q(^{11}{\rm C})\over Q(^{11}{\rm B})} =
\left({Z=6\over Z= 5}\right)\times 
\left({\beta_p{\rm cos}\gamma_p\left(^{11}{\rm C}\right) \over
\beta_n{\rm cos}\gamma_n\left(^{11}{\rm C}\right) }
\right)\times \left( {R_e(^{11}{\rm C})
\over R_e(^{11}{\rm B})} \right)^2.
\label{eqn:ratiomm}
\end{equation}

We take the third term for the charge radii to be unity 
since $^{11}$C and $^{11}$B are the nuclei close to the stability line.
If the neutron deformation 
agrees to the proton deformation
in $^{11}$C,
the second term gives no contribution to the ratio of $Q$ moments and
the ratio can be explained only by
the charge ratio 1.2. 
However the ratio $0.88$ deduced from the experimental data is
less than a unity and is inconsistent with the $Z$ ratio
because the experimental data of $Q(^{11}$C)=34.8 e$\cdot$mb 
 is smaller than $Q(^{11}$B)=40.7 e$\cdot$mb.     
According to AMD calculations, this problem can be resolved by
taking into account the difference
between the intrinsic deformations of proton and neutron densities
in $^{11}$C.
Through the second term in Eq.(\ref{eqn:ratiomm}), the ratio of $Q$ moments
reflects the difference of intrinsic 
shapes between proton and neutron densities.
As shown above in Fig.\ref{fig:cpdefo} 
the proton deformation is oblate while the 
neutron density becomes a triaxial shape in $^{11}$C.
Since $\beta_p{\rm cos}\gamma_p(^{11}$C) is smaller than 
$\beta_n{\rm cos}\gamma_n(^{11}$C),
the second term in Eq. \ref{eqn:ratiomm} becomes
less than unity, which cancels the effect 
of the first term of the $Z$ ratio.

We make further quantitative discussion by the use of theoretical 
values of 
deformation parameters in the intrinsic states obtained with AMD.
In the present results,
the ground state of $^{11}$C with $J=3/2$ 
is obtained by a total-angular-momentum projection 
on a state $|J=3/2,K=3/2\rangle$ with respects to the principal $z$
axis with the minimum moment of inertia.
Using the theoretical values of $\beta_p{\rm cos}\gamma_p$ 
and $\beta_n{\rm cos}\gamma_n$ shown in
Fig.\ref{fig:cpdefo} we can estimate the ratio of $Q$ moments 
with the first and the second terms in Eq.\ref{eqn:ratiomm}.
The estimated ratio is found to be 0.87
which is as small as the value of 0.88
deduced from the experimental data.
In fact the theoretical results of 
electric $Q$ moments for total-angular
momentum projected states are 20 e$\cdot$mb for $^{11}$C and
34 e$\cdot$mb for $^{11}$B, which are consistent with the experimental data 
of $Q(^{11}$C)$<Q(^{11}$B) (Table.\ref{tbl:cqmom}).

\begin{table}[tbh]

\caption { \label{tbl:cqmom}
Electric quadrupole moments and transitions of proton-rich C isotopes
and the mirror nuclei.
Calculations are with MV1 force ($m=0.576$)
and the experimental data are taken from Ref.\protect\cite{NTABLE}.
}
\begin{center}
\begin{tabular}{cc|ccc}
& &\multicolumn{2}{c}{electric $Q$ moments}& \\
nucleus & level & exp. & theory & \\
\hline 
$^{11}$C & $3/2^-$ &  34.3 e$\cdot$mb & 20 e$\cdot$mb &\\
$^{11}$B & $3/2^-$ &  40.7(3) e$\cdot$mb & 34 e$\cdot$mb &\\
\hline
$^{10}$C & $2^+$ &   $-$ & $-38$ e$\cdot$mb &\\
$^{10}$Be & $2^+$ &   $-$ & $-65$ e$\cdot$mb &\\
\hline
$^{9}$C & $3/2^-$ &   $-$ & $-28$ e$\cdot$mb &\\
$^{9}$Li & $3/2^-$ &  $-27.8$ e$\cdot$mb & $-27$ e$\cdot$mb &\\
\hline
\hline
& &\multicolumn{2}{c}{$E2$ transition strength} & \\
nucleus & level & exp. & theory & \\
\hline
$^{11}$C & $5/2^-\rightarrow 3/2^- $ &  $-$
 & 6.8 e$^2{\rm fm}^4$ &\\
$^{11}$B & $5/2^-\rightarrow 3/2^- $ &   13.9(3.4) e$^2{\rm fm}^4$
 & 11.3 e$^2{\rm fm}^4$ &\\
\hline
$^{10}$C & $2^+\rightarrow 0^+ $ &  12.3(2.0) e$^2{\rm fm}^4$
 & 5.3 e$^2{\rm fm}^4$ &\\
$^{10}$Be & $2^+\rightarrow 0^+ $ &   10.5(1.0) e$^2{\rm fm}^4$
 & 9.5 e$^2{\rm fm}^4$ &\\
\hline
$^{9}$C & $1/2^-\rightarrow 3/2^- $ &  $-$ 
 & 5.7 e$^2{\rm fm}^4$ &\\
$^{9}$Li & $1/2^-\rightarrow 3/2- $ &   $-$
 & 7.2 e$^2{\rm fm}^4$ &
\end{tabular}
\end{center}
\end{table}

It is concluded that the difference of 
the intrinsic deformation of the proton density
in $^{11}$C from that in $^{11}$B
(an oblate shape in $^{11}$C and a triaxial shape in $^{11}$B)
makes significant effects to the ratio of 
the $Q$ moment of $^{11}$C to that of $^{11}$B.
By assuming the mirror symmetry, it is theoretically 
suggested that the disagreement between proton and neutron deformations 
in $^{11}$C is supported by the experimental fact of the 
ratio $Q(^{11}$C)/$Q(^{11}$B)/ less than a unity.

Next we make similar analysis of the deformations for $^{10}$C.
 We find that the difference
between proton and neutron deformations in $^{10}$C is important 
to understand the ratio of the $E2$ transition 
strength $B(E2;2^+\rightarrow 0^+)$ in $^{10}$C to 
that in the mirror nucleus $^{10}$Be.
Assuming mirror symmetry, 
the ratio of $B(E2)$ is approximated similarly to Eq.\ref{eqn:ratiomm}
as,
\begin{equation}
{B(E2;^{10}{\rm C})\over B(E2;^{10}{\rm Be})} =
\left({Z=6\over Z= 4}\right)^2\times 
\left({\beta_p{\rm cos}\gamma_p\left(^{10}{\rm C}\right) \over
\beta_n{\rm cos}\gamma_n\left(^{10}{\rm C}\right) }
\right)^2\times 
\left( {R_e(^{10}{\rm C})\over R_e(^{10}{\rm Be})} \right)^4.
\label{eqn:ratiobe2}
\end{equation}
The first term of the charge ratio $(6/4)^2$=2.25 is much larger than the
ratio of experimental values 12.3(2.0)e$^2$fm$^4$/10.5(1.0)e$^2$fm$^4$
=1.2(0.3). 
The reason why the square of the charge ratio fails to reproduce
the ratio of $B(E2)$ in the mirror nuclei $^{10}$C and $^{10}$Be 
is because of the disagreement between proton and neutron deformations
in $^{10}$C. 

In the intrinsic state of $^{10}$C,
the proton density deforms oblately with $\beta_p{\rm
cos}\gamma_p=0.28$
 while the neutron
 deformation is prolate with a larger value of the effective deformation 
parameter $\beta_n{\rm cos}\gamma_n$=0.49
(Fig.\ref{fig:cpdefo}), which makes the second
term in Eq.\ref{eqn:ratiobe2} less than unity.
If the third term is omitted, the ratio is roughly
estimated as,
\begin{equation}
{B(E2;^{10}{\rm C})\over B(E2;^{10}{\rm Be})}= 
 2.25\times (0.28/0.49)^2 \sim 0.75.
\end{equation}
Although the ratio estimated above
is smaller than the ratio 1.2 deduced from the experimental data,
it is found that the reduction of the ratio of the proton numbers 
is made by the ratio of the deformation parameters 
The microscopic calculations of $Q$ moments with AMD are
shown in Table \ref{tbl:cqmom} and are
compared with the experimental data. Since the calculations underestimate the
value of $B(E2;^{10}$C), the ratio
$B(E2;^{10}$C)$/B(E2;^{10}$Be) is smaller than a unity.
In the results with VAP calculation which will be mentioned later,
the theoretical value of $B(E2;^{10}$C) is improved.
  
As for the $^9$C and the mirror nucleus $^9$Li, 
the present prediction is that the
$Q$ moments of $^9$C is smaller to
the one of $^9$Li. We should note that the present results do not include 
the effect of the long tail of valence nucleons.
However, if the proton halo of $^9$C exists 
because the separation energy of protons is small,
the orbits of the valence protons may give the large effect on 
the $Q$ moments.

\subsubsection{Neutron skin and halo \label{subsub:nskin}}

The presence of a neutron skin structure has been discussed
for a long time. 
Recently thick neutron skins have been reported
in He isotopes \cite{HESKIN} and in $^{20}$N \cite{NSKIN} by the help of 
the experimental data of interaction cross sections.
The appearance of the neutron skin has been also shown
in the comparison of neutron radii with proton radii
along a chain of Na isotopes by
combining the data of the isotope-shift for charge radii with 
those of the matter radii deduced from interaction cross sections
\cite{NASKIN}. 
Also the theoretical studies of skin structures have been tried
in unstable nuclei \cite{RING,ENYOc,ENYOdoc,FUKUNISHIa}.

The present results suggest that in the neutron-rich nuclei of 
B and C isotopes the density of neutrons stretches far toward the outer 
region. The simple AMD calculations predict the presence of 
``neutron skin structure'', which is the surface region with rather high
neutron density but low proton density.
In particular, C isotopes are expected to have thicker skins than 
those of B isotopes because 
the neutron-rich C have no developed 
clustering structure as is seen in the proton density of $^{20}$C
which remains in the inner region as compact as that 
of stable C nuclei.
One of characteristics of C isotopes is 
the stationary structure of protons
in spite of the drastic change of neutron structure 
along the increase of neutron number.
It is in contrast with neutron-rich B isotopes which are predicted to have 
the clustering structure. According to simple AMD calculations,
in a series of $N=14$ isotones 
the neutron skin which is developed in $^{20}$C weakens with the increase of 
the proton number because the mean-field for the neutrons
given by proton density becomes deeper to decrease the neutron radii.

%%%%%%%%%%%%%%%%%%%% BEGIN OF FIGURE %%%%%%%%%%%%%%%%%%%%%%%%
\begin{figure}%----------------------------------------

\caption{\label{fig:crden}
The densities $\rho_p$ of protons and $\rho_n$ of neutrons 
as a function of radius in C isotopes.
The densities of the $0^+$ states are obtained with AMD calculations
by using MV1 force with $m=0.63$.
The proton density of $^{12}$C is written by the dotted lines. 
The densities of in $^{19}$B are also shown 
together with the proton density of 
$^{13}$B.
}
\epsfxsize=0.9\textwidth
%\begin{minipage}[b]{0.9\textwidth}
\centerline{\epsffile{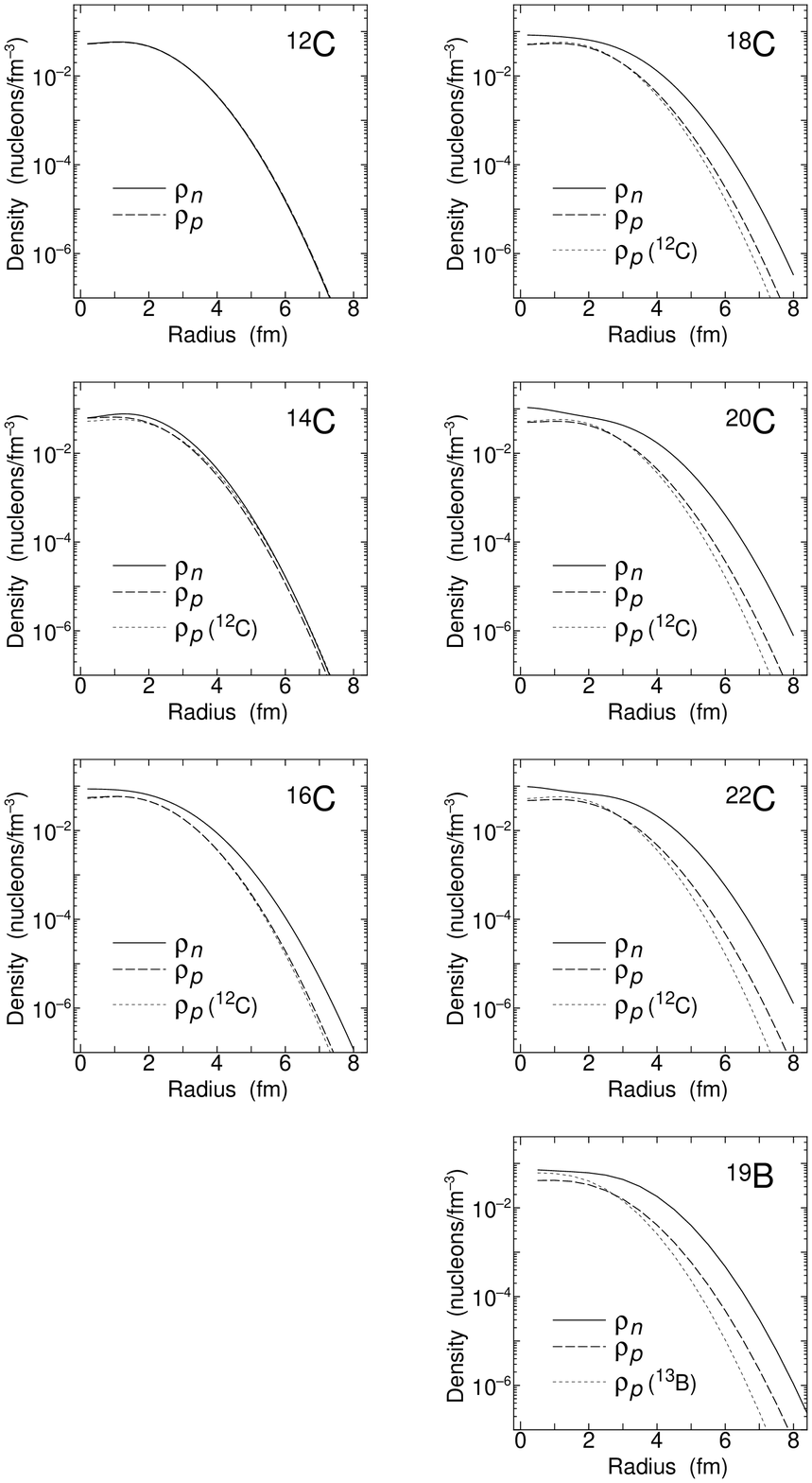}}
%\end{minipage} \ \hskip 0.05\textwidth \
%
\end{figure}
%%%%%%%%%%%%%%%%%%%%%%%%%%%%%%%%%%%%

Figure \ref{fig:crden} shows the proton and the neutron densities
of C isotopes as a function of radius.
The difference between proton and neutron densities 
in the surface region around $r\approx 4$ fm enhances 
as the neutron number increases from $N=10$ 
toward the neutron-drip line.
In the surface region of $^{20}$C,
the line for the neutron density (solid) seems to be shifted outward 
by about 1 fm compared with the line
for proton density (dashed).

%%%%%%%%%%%%%%%%%%%% BEGIN OF FIGURE %%%%%%%%%%%%%%%%%%%%%%%%
\begin{figure}%----------------------------------------

\caption{\label{fig:cpnradii}
Theoretical values of root-mean-square radii of protons (dotted lines)
and neutrons (solid lines). Calculations are the results
with the interaction (b) MV1 ($m=0.576$) and 
with the MV1 force with $m=0.63$.
}
\epsfxsize=0.6\textwidth
\centerline{\epsffile{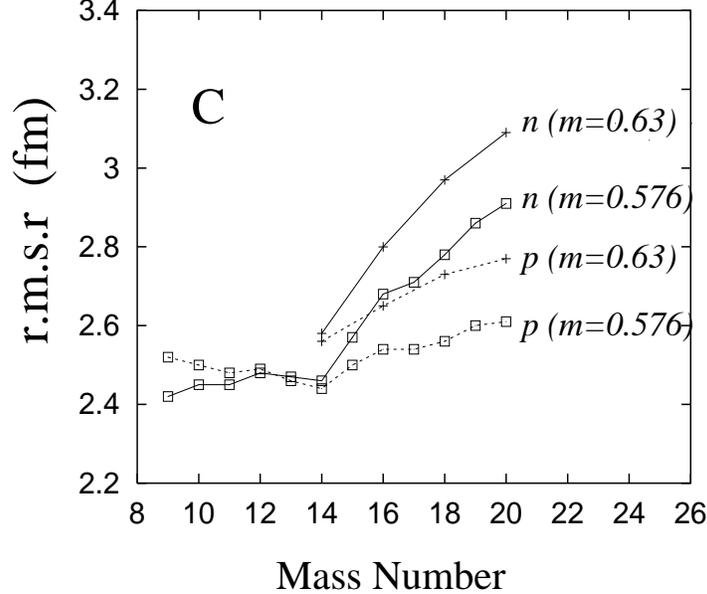}}
\end{figure}
%%%%%%%%%%%%%%%%%%%%%%%%%%%%%%%%%%%%

As already mentioned
the recently measured radii of C isotopes \cite{OZAWA} agree 
well systematically
with the simple AMD calculations 
except for $^{19}$C which is expected to have a neutron halo structure
(Fig. \ref{fig:cradii}). 
We show the root-mean-square radii of neutrons and protons
separately in Fig.\ref{fig:cpnradii}. It is found that 
the neutron radii become larger and larger in the region heavier than
$^{14}$C while the proton radii are rather stable with the 
increase of the neutron number. Although the 
radii depend on the adopted interaction parameters, in both calculations
with Majorana parameter $m=0.576$ and $m=0.63$
the difference between the proton radius and the neutron radius
in $^{20}$C is more than 0.3 fm.
According to the present results,
the increase of the matter radii in the neutron-rich C isotopes is 
mainly due to the development of the neutron skin structure. 

\begin{figure}%----------------------------------------

\caption{\label{fig:chfe}
The single-particle energies of the intrinsic states for 
the normal parity states of C isotopes.
The method to calculate the single-particle energies is explained
in the text.
The energies of proton(neutron) orbits are presented on the left(right).
The MV1 force with $m=0.63$ is used.
}
\epsfxsize=1.1\textwidth
\centerline{\epsffile{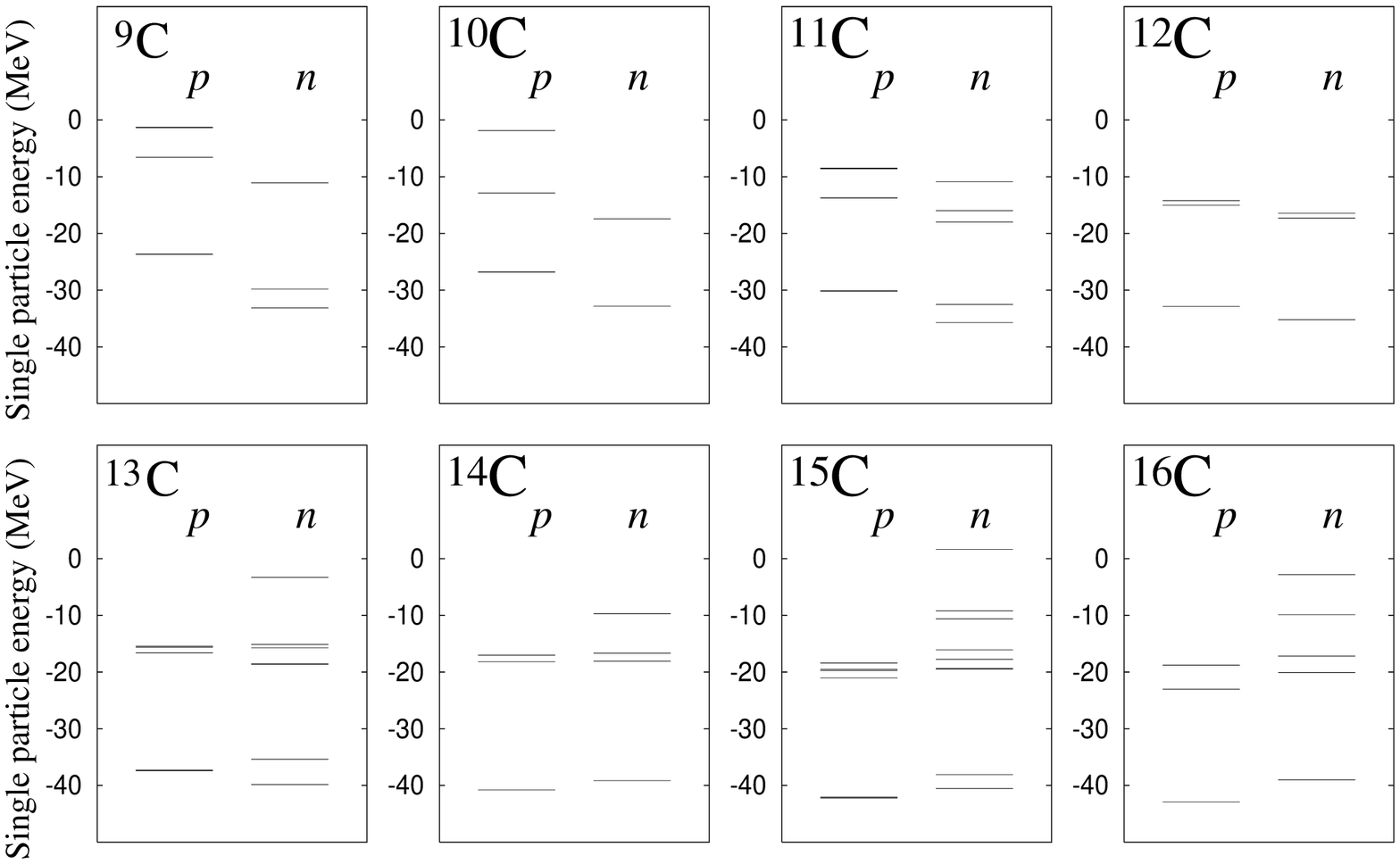}}
\epsfxsize=1.1\textwidth
\centerline{\epsffile{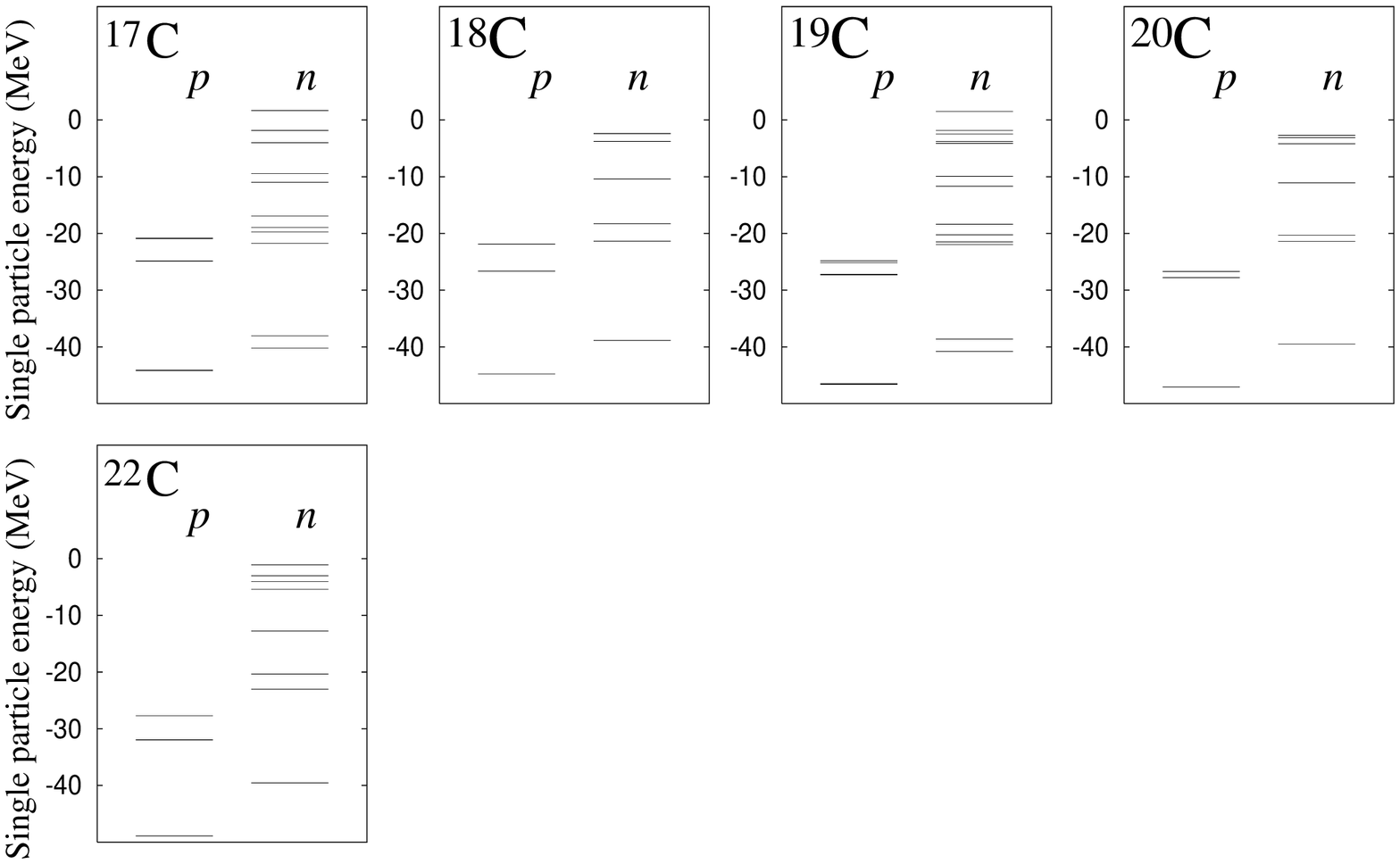}}
\end{figure}

The radial behavior of the densities of protons and 
neutrons is related closely 
with the single-particle energies. 
We calculate the 
single-particle energies and the single-particle wave functions 
by diagonalizing the single-particle
Hamiltonian with the analogy to Hartree-Fock theory.
First we transform the set of single-particle wave functions $\varphi_i$ 
of the solved of AMD wave function into an orthonormal 
base $\tilde\varphi_\alpha$.
The single-particle Hamiltonian
can be constructed by the use of the orthonormal base as follows
\cite{ENYOdoc,DOTE,ENYOg};
\begin{eqnarray}
h_{\alpha\beta}&=&
\langle\tilde\varphi_\alpha|\hat t|\tilde\varphi_\beta\rangle
+\sum^A_{\gamma}\langle\tilde\varphi_\alpha\tilde\varphi_\gamma|\hat v|
\tilde\varphi_\beta\tilde\varphi_\gamma-
\tilde\varphi_\gamma\tilde\varphi_\beta\rangle \nonumber\\
&+&{1\over 2}\sum^A_{\gamma,\delta}
\langle \tilde\varphi_\alpha\tilde\varphi_\gamma\tilde\varphi_\delta
|\hat v_3|\tilde\varphi_\beta\tilde\varphi_\gamma\tilde\varphi_\delta
+\tilde\varphi_\delta\tilde\varphi_\beta\tilde\varphi_\gamma
+\tilde\varphi_\gamma\tilde\varphi_\delta\tilde\varphi_\beta
-\tilde\varphi_\beta\tilde\varphi_\delta\tilde\varphi_\gamma
-\tilde\varphi_\gamma\tilde\varphi_\beta\tilde\varphi_\delta
-\tilde\varphi_\delta\tilde\varphi_\gamma\tilde\varphi_\beta\rangle,
\end{eqnarray}
where the Hamiltonian operator is
 written by a sum of the kinetic term, the two-body interaction term 
and the three-body interaction term; $H=\sum_{i}\hat t+\sum_{i<j} \hat v_2
+\sum_{i<j<k} \hat v_3$.
We note that the single-particle energies defined as above can be positive
because the model space for single particle 
wave functions is restricted to  the Gaussian wave packet
giving rise to an artificial wall due to the zero-point kinetic energy of the 
packet which prevents the single-particle 
wave function from escaping out of the nucleus.
In Fig.\ref{fig:chfe} we present single-particle energies 
in C isotopes which are calculated from the 
obtained intrinsic AMD wave functions. 
It is found that in the neutron-rich C isotopes heavier than $^{14}$C
the Fermi energy is at most a few MeV. 
In the nuclei near the neutron-drip line, 
many valence neutrons occupy the higher orbits a few MeV below
the zero energy which correspond to $sd$ orbits. 
These weakly bound neutrons build the neutron skin structure.
In the even-odd neutron-rich C isotopes such as $^{19}$C, 
the energy of the last valence neutron
is about zero energy.
The possible existence of halo structure in $^{19}$C is
suggested experimentally by the measurements of
the longitudinal-momentum distribution 
of $^{18}$C after the one-neutron breakup of $^{19}$C \cite{C19} 
and also by the interaction cross sections
\cite{OZAWA}.
Although the 
halo structure may not be seen in the present results
because the AMD wavefunction is not sufficient to describe a 
long tail of the halo tail,
considering the small binding energy of valence neutrons 
we naturally expect that the halo structure may appear
if the single-particle wave function of such a loosely bound 
valence neutron is described more precisely than the present model space.

%%%%%%%%%%%%%%%%%%%% BEGIN OF FIGURE %%%%%%%%%%%%%%%%%%%%%%%%
%%%%%%%%%%%%%%%%%%%%%%%%%%%%%%%%%%%%

On the other hands, the protons are  deeply bound in the
neutron-rich C. The binding energies of protons grow
rapidly from $^9$C to $^{14}$C because the number of neutrons in
$p$ shell increases. In the heavier region from $^{14}$C to
the neutron-drip line, the proton energy becomes deeper and deeper 
gradually. Taking into account the 
potential depth of protons and the enlargement of 
the neutron density distribution, 
the kink of radii at $^{14}$C 
seems to be a natural phenomenon which reflects the shell
effect of neutrons at $N=8$.

\section{Study with VAP} \label{sec:vap}

Recently the structures of 
the excited states as well as the ground states are very attractive in the
study of unstable nuclei.
It is natural that various molecule-like 
states may appear
in the excited states of light unstable nuclei
because the excitation due to the relative motion between 
clusters is important in the light nuclear region.
We apply the method of the 
variational after spin-parity projection in the framework 
of AMD to the light unstable nuclei 
for the aim to make systematic study of the structure change 
with the increase of the excitation energy.
The formulation has been already explained in the section \ref{sec:formura}.
The applicability of the framework for the study of the excited states
in the stable nuclei has been confirmed in Ref.\cite{ENYOe} on
the structure of $^{12}$C.
Here we study the structures of the 
excited states of $^{10}$Be and $^{12}$Be.

\subsection{$^{10}$Be \label{subsec:be10}}
$^{10}$Be, one of the challenges in the study of 
light unstable nuclei, has been investigated experimentally 
not only by use of the unstable nuclear beams but also  
in such experiments as the transfer and pick-up reactions.
The recent experiments of the charge exchange 
reactions $^{10}$B($^3$He,$t$)$^{10}$Be \cite{FUJIWARA}
let us know the strength of the Gamow-Teller transitions to 
the excited states of $^{10}$Be. These new data of $\beta$ transition
strength which are deduced from the cross sections at the forward angle
are very helpful to study the structure of the excited states.
The structures of $^{10}$Be have been studied hard also theoretically
by microscopic calculations, for example, shell models
\cite{BROWN,WOLTERS}, 
cluster models \cite{OGAWA,ITAGAKI,SEYA}, Hartree-Fock \cite{TAKAMI},
and antisymmetrized molecular dynamics \cite{ENYOb,ENYOc,DOTE}.
We study the structure of the excited states of $^{10}$Be
with the VAP calculations in the framework of AMD.

\subsubsection{Results}
In this subsection we display the theoretical results of the excitation 
energies, $E2$, $E1$, and $\beta$ transitions
which can be directly compared with the experimental data.
The detail of the structures is discussed in the next subsection. 

The adopted interaction for the central force is the case 3 of 
MV1 force \cite{TOHSAKI}.
The adopted parameters are 
$m=0.62$, $b=h=0$ for the Majorana,
Bertlett and Heisenberg 
terms of the central force and the strength of the spin-orbit force
$u_I=-u_{II}=3000$ MeV (interaction (g)).
Trying another set (h) of parameters  with $m=0.65$, $b=h=0$ and 
$u_I=-u_{II}=3700$ MeV, we did not find significant differences
 in the results. The set of parameters of case(g) is the one
 adopted in the work on $^{12}$C \cite{ENYOe}. On the other hands, 
the VAP calculations with the set of interactions 
case(h) reproduce the abnormal parity of the ground state 
of $^{11}$Be.
The optimum width parameters $\nu$ of wave packets are chosen to be 
0.17 fm$^{-2}$ for case(g) and 0.19 fm $^{-2}$ for case(h) 
which give the minimum energies in VBP calculations of $^{10}$Be.
The resonance states are treated within a bound state approximation 
by situating an artificial barrier out of the surface.

The lowest $J^\pm$ states are obtained by 
VAP calculations for $P^{J\pm}_{MK'}\Phi_{AMD}$ with $(J^\pm,K')$=
$(0^+,0)$, $(2^+,0)$, $(3^+,+2)$, $(4^+,0)$, $(1^-,-1)$, $(2^-,-1)$,  
$(3^-,-1)$, $(4^-,-1)$. 
Considering $0^+_2$ state to be $0^+$ state in the second 
$K^\pi=0^+_2$ band, 
the $0^+_2$ state is calculated by VAP as the higher excited state orthogonal
to the lowest $0^+_1$ state as explained 
in the subsection \ref{subsec:excited}. That is to say that 
the $0^+_2$ state is obtained by VAP for $\Phi({\bf Z})$ 
in Eq. \ref{eqn:excite} with 
$(J^\pm,K',n)=(0^+,0,2)$.
In the case of higher $2^+$ states, 
we impose the constraint that 
the approximately principal $z$-axis of the intrinsic 
deformation equals to the 3-axis of the Euler angle in the 
total-spin projection.
According to VBP calculations 
the second $2^+$ state is described 
as the band head of the lowest $K^\pi=2^+$ band.
Therefore we construct the state $2^+_2$ 
by choosing $(J^\pm,K')$ of  $P^{J\pm}_{MK'}\Phi_{AMD}$
as $(J^\pm,K')=(2^+,+2)$ under the constraint on the principal $z$-axis.
Because of the constraint and the choosing of the $K'$ quantum number
it keeps the approximately orthogonality to the 
lowest $2^+$ state with $(J^\pm,K')=(2^+,0)$.
The third $2^+_3$ state is easily conjectured to 
be $2^+$ state in the second $K^\pi=0^+_2$ band like $0^+_2$ state.
We obtain the $2^+_3$ state by VAP for $\Phi({\bf Z})$ in Eq.\ref{eqn:excite}
with $(J^\pm,K',n)=(2^+,0,2)$
by imposing the constraint so as to make 
the principal $z$-axis equal with the $3$-axis in spin projection.
It means that the orthogonal condition of $2^+_3$ to $2^+_1$ 
is kept by superposing two 
wave functions 
as described in the subsection \ref{subsec:excited},  
while the orthogonality to $2^+_2$ ($K^\pi=2^+$) is taken into account
by choosing the different $K$ quantum $K'=0$.

\begin{figure}%----------------------------------------
\noindent
\caption{\label{fig:be10sped}
Excitation energies of the levels in $^{10}$Be.
Theoretical results are calculated by the diagonalization of the 
states obtained with VAP by using the interaction case (g).
}
\epsfxsize=1.0\textwidth
\centerline{\epsffile{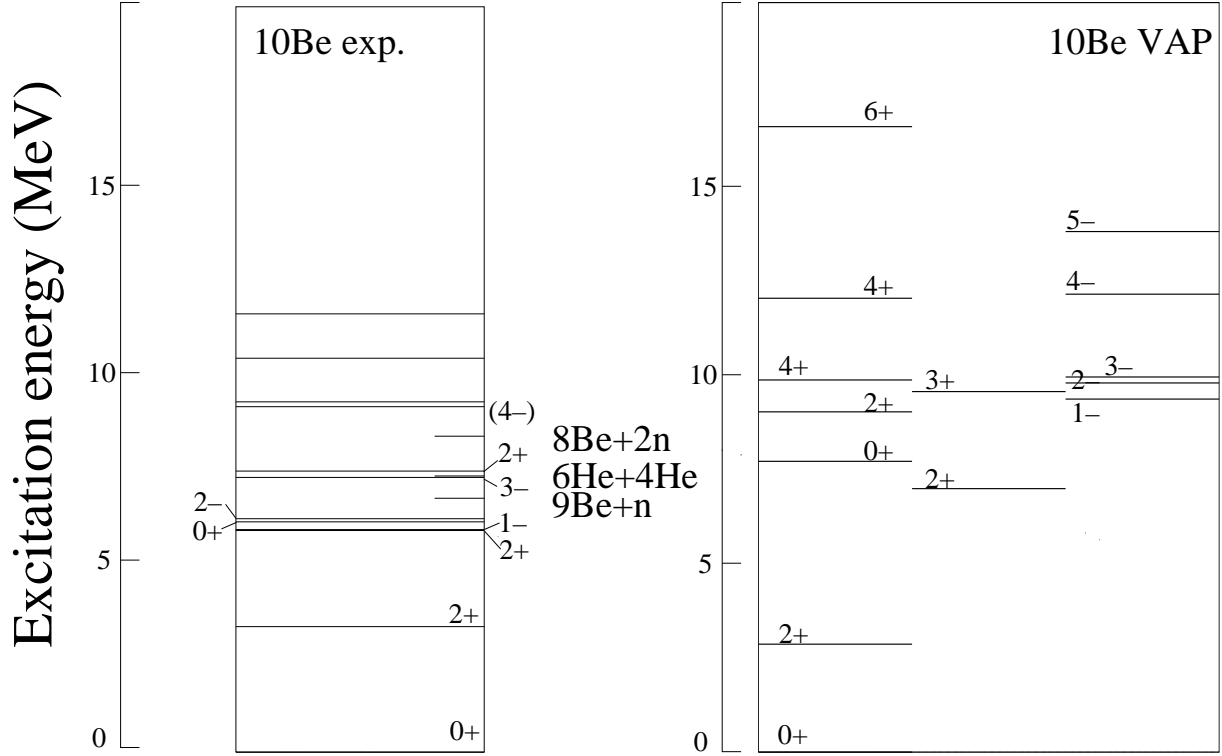}}
\end{figure}

The binding energy obtained with the case(g) interactions is 61.1 MeV, 
and the one with case(h) is 61.3 MeV.
The excitation energies of the results are displayed in 
Fig.\ref{fig:be10sped}. By
diagonalization of the Hamiltonian matrix 
the excited states $4^+_2$, $6^+_1$ are found in the rotational band 
$K^\pi=0^+_2$, and $5^-$ state
is seen in the $K^\pi=1^-$ band.
Comparing with the experimental data, the level structure is well reproduced 
by theory. Although it is difficult to estimate the width of resonance 
within the present framework, the theoretical results suggest
the existence of $3^+$, $4^+$, $6^+$ and $5^-$ states 
which are not experimentally identified yet.
The excited levels can be roughly classified as the rotational bands
$K^\pi=0^+_1$, $2^+$, $0^+_2$ and $1^-$ which consist of 
($0^+_1$, $2^+_1$, $4^+_1$), 
($2^+_2$, $3^+_1$), 
($0^+_2$, $2^+_3$, $4^+_2$, $6^+_1$) and
($1^-$, $2^-$, $3^-$, $4^-$, $5^-$), respectively.
The intrinsic structures of these rotational 
bands are discussed in detail in the next section.

%\subsection{transition strength}
The data of transition strength are of great help to investigate the 
structures of the excited states. The results 
with the interaction case(g) and the experimental data
of $E2$ and $E1$ transition strength are listed in Table \ref{tab:be10e2}.
The theoretical values well agree with the experimental data.  
The strength $B(E2)$ for $^{10}$C;$2^+_1\rightarrow 0^+_1$
is simply calculated by the wave function of $^{10}$C 
supposed to be mirror symmetric with $^{10}$Be.
The present result for $E2$ strength of 
$^{10}$C;$2^+_1\rightarrow 0^+_1$ is better than the work with simple AMD
calculations (see Fig. \ref{tbl:cqmom}) \cite{ENYOd}.
As for the values with a shell model, 
$(0+2)\hbar\omega$ shell model calculations from the reference \cite{NAKADA} 
with effective charges $e_\pi=1.05e, e_\nu=0.05e$ are also listed.
The shell model calculations qualitatively reproduce some experimental data
of the $E2$ properties of low-lying levels.

\begin{table}
\caption{\label{tab:be10e2} $E2$ and $E1$ transition strength. The 
theoretical results of VAP calculations
 with the interaction case (g) are compared with
the experimental data \protect\cite{AJZEN}. 
The shell model calculations are quoted from the work
with the $(0+2)\hbar\omega$ shell model in the reference 
\protect\cite{NAKADA}.
}

\begin{center}
\begin{tabular}{ccccc}
 transitions & Mult. & exp.  & present VAP  & shell model\\
\hline
$^{10}$Be;$2^+_1\rightarrow 0^+_1$ &
 $E2$ & 10.5$\pm$1.1 (e fm$^2$)  & 11 (e fm$^2$) &  16.26 (e fm$^2$)\\
$^{10}$Be;$0^+_2\rightarrow 2^+_1$ &
 $E2$ & 3.3$\pm$2.0 (e fm$^2$) & 0.6 (e fm$^2$) & 7.20 (e fm$^2$) \\
$^{10}$Be;$0^+_2\rightarrow 1^-_1$ &
 $E1$ & 1.3$\pm$0.6$\times 10^{-2}$ (e fm) & 
 0.6$ \times 10^{-2}$(e fm)&\\
\hline
$^{10}$C;$2^+_1\rightarrow 0^+_1$ &
 $E2$ & 12.3$\pm$2.0 (e fm$^2$)  & 9 (e fm$^2$) &  15.22 (e fm$^2$)\\
\end{tabular}
\end{center}
\end{table}

The strength of the $\beta$ decays of Gamow-Teller(GT) type 
transitions can be deduced from the
cross sections at the $0^\circ$ forward angle 
of the charge exchange reactions which have been measured recently
\cite{FUJIWARA}. 
These new data for the Gamow-Teller type $\beta$ transitions are 
very useful probes to discuss the
structures of the excited states of unstable nuclei.
Table \ref{tab:be10beta} shows the values of $B(GT)$.
The experimental values for the $\beta$ transitions from $^{10}$B(3+) to 
$^{10}$Be$^*$ are deduced from the data of the reaction 
$^{10}$B(t,$^3$He)$^{10}$Be.
In the present calculations, the wave functions for the neighbor nucleus 
$^{10}$B are calculated with VAP where 
$(J^\pm, K')$ is chosen to be ($3^+,-3$) for the ground $3^+_1$ state 
and ($1^+,-1$) for the $1^+_1$ state.
$^{10}$Be and $^{10}$B are calculated with the case (g) and
(h) interactions. 
The theoretical values reasonably match to the experimental data.
Since the data for 
$^{10}$B$(3^+)\rightarrow^{10}$Be(9.4MeV) well correspond
to the theoretical value of 
$^{10}$B$(3^+)\rightarrow^{10}$Be($3^+_1$),
the excited level of $^{10}$Be at 9.4MeV is considered to be
the $3^+_1$ state.
The strength of these GT transitions from $^{10}$B$(3^+)$
is governed by the configuration of the ground state of $^{10}$B
which is understood as the state $3^+$ with $|K|=3$ in $p$-shell in the simple
shell model limit.
It is natural that the transitions to 
$2^+_2$ and $3^+_1$ states in the $K^\pi=2^+$ bands of $^{10}$Be are strong
while the transitions to the states in the $K^\pi=0^+$ bands are weaker.
That is why the predicted $B(GT)$ for
$^{10}$B$(3^+)\rightarrow^{10}$Be($2^+_3$) is small
because $^{10}$Be($2^+_3$) is the state in the 
$K^\pi=0^+_2$ band constructed by the linear structure with the 
developed clustering.
The result of $B(GT)$ for 
$^{10}$Be$(0^+_1)\rightarrow^{10}$B($1^+$) is consistent with
the experimental value of the $\beta$ decay from the mirror nucleus 
$^{10}$C$(0^+_1)\rightarrow^{10}$B($1^+$). 

\begin{table}
\caption{ \label{tab:be10beta} $B(GT)$ values of $\beta$ decays which are
the square of the expectation values of Gamow-Teller operator.
The experimental data are the values$^{(a)}$ 
deduced from the cross sections of 
$^{10}$B(t,$^3$He)$^{10}$Be$^*$ at $0^\circ$ forward 
angle \protect\cite{FUJIWARA}
 and the 
one$^{(b)}$ from \protect\cite{CHOU}. 
The theoretical results are obtained with case (g) and case (h) interactions
for $^{10}$Be and $^{10}$B.}

\begin{center}
\begin{tabular}{ccccc}
&   & exp. &&\\ 
&initial & final & & \\
&$(J^\pi,E_x)$ (MeV) & $(J^\pi,E_x)$ (MeV) & B(GT) & \\
\hline
&$^{10}$B($3^+$,0) & $^{10}$Be($2^+_1$,3.37) & 0.08$\pm$ 0.03$^{a)}$ &\\
&$^{10}$B($3^+$,0) & $^{10}$Be($2^+_2$,5.96) & 0.95$\pm$ 0.13$^{a)}$ &\\
&$^{10}$B($3^+$,0) & $^{10}$Be($2^+$ or $3^+$, 9.4) & 0.31$\pm$ 0.08$^{a)}$ &\\
&$^{10}$C($0^+$,0) & $^{10}$B($1^+$, 0.72) & 3.44 $^{b)}$ &\\
\hline
\hline
& & theory case(g)& &\\
\hline
&$^{10}$B($3^+$) & $^{10}$Be($2^+_1$) & 0.02 &\\
& & $^{10}$Be($2^+_2$) & 1.1 &\\
& & $^{10}$Be($3^+_1$) & 0.40 &\\
& & $^{10}$Be($4^+_1$) & 0.08 &\\
& & $^{10}$Be($2^+_3$) & 0.03 &\\
&$^{10}$Be($0^+_1$) & $^{10}$B($1^+$) & 2.9 &\\
\hline
\hline
& & theory case(h)& &\\
\hline
&$^{10}$B($3^+$) & $^{10}$Be($2^+_1$) & 0.00 &\\
& & $^{10}$Be($2^+_2$) & 0.92 &\\
& & $^{10}$Be($3^+_1$) & 0.38 &\\
& & $^{10}$Be($4^+_1$) & 0.10 &\\
& & $^{10}$Be($2^+_3$) & 0.00 &\\
&$^{10}$Be($0^+_1$) & $^{10}$B($1^+$) & 2.5 &\\
\end{tabular}
\end{center}
\end{table}

\subsubsection{Intrinsic structure}
Here we discuss the structure of the excited states
by analyzing the wave functions.
Even though the state calculated by VAP mixes with each other 
after the diagonalization 
of the Hamiltonian matrix, the state 
$P^{J\pm}_{MK'}\Phi_{AMD}({\bf Z}^{J\pm}_n)$ projected from a Slater 
determinant obtained in VAP with $(J^\pm,K',n)$ is the main component
in the final result of the $J^\pm_n$ state.
In this section we consider  
the Slater determinant $\Phi_{AMD}({\bf Z}^{J\pm}_{n})$ 
as the intrinsic state 
for the $J^\pm_n$ state.

\begin{figure}%----------------------------------------
\noindent
\epsfxsize=0.5\textwidth
\centerline{\epsffile{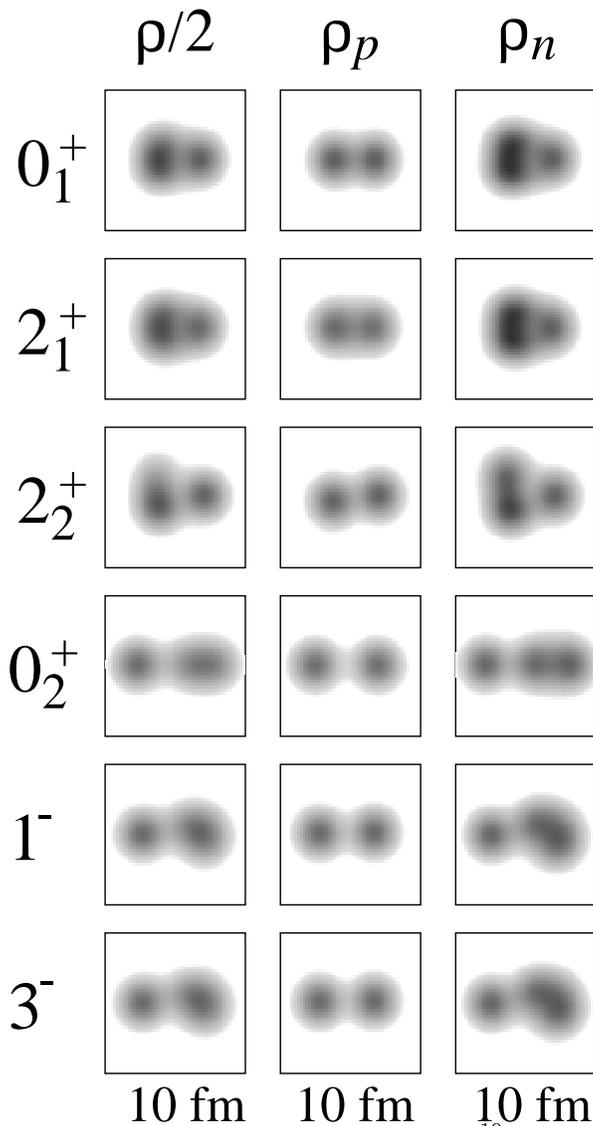}}
\caption{\label{fig:be10dens}
The intrinsic structure of the excited states of $^{10}$Be
obtained by VAP calculations.
The density distribution of matter, protons and neutrons
of the intrinsic states are shown at left, middle and right, respectively.
The density is integrated along the axis perpendicular to adequate planes.
The figures are for the results with the interaction (g).
}
\end{figure}

In the excited states, various kinds of structures are found.
Here we analyze the structures of the intrinsic states
$\Phi_{AMD}({\bf Z}^{J\pm}_{n})$.
%Analyzing the intrinsic structures of the excited states, 
It is found that the excited levels are classified into rotational bands
as $0^+_1$, $2^+_1$, $4^+_1$ states in $K^\pi=0^+_1$ band, 
$2^+_2$, $3^+_1$ in $K^\pi=2^+$ band, $0^+_2$, $2^+_3$, $4^+_2$, $6^+_1$
in the second $K^\pi=0^+_2$ band and $1^-$, $2^-$, $3^-$, $4^-$, $5^-$ 
in $K^\pi=1^-$ band. Particularly the 
molecule-like states with  
the well-developed 2$\alpha$ cores construct the rotational bands
$K^\pm=0^+_2$ and $1^-$ in which the level spacing is small because of the 
large moments of inertia.
The density distributions of matter, 
protons and neutrons in the intrinsic states
$\Phi_{AMD}({\bf Z}^{J\pm}_{n})$ are presented in Fig.\ \ref{fig:be10dens}.
We found the $2\alpha+2n$ structures in most of the intrinsic states.
The density of protons indicates that the clustering structure develops 
more largely in $1^-$ than $0^+_1$ and most remarkably in $0^+_2$. 
As seen in Fig.\ \ref{fig:be10dens}, the intrinsic structure of $0^+_2$ 
state has an axial symmetric 
linear shape with the largest deformation, while 
$1^-$ state has an axial asymmetric shape because 
of the valence neutrons.
The structures of $0^+_1$ and $1^-_1$ states are similar to 
the ones of the previous work with the simplest version of 
AMD \cite{ENYOdoc}. The increase of the degree of  
the deformation along $0^+_1$, $1^-$ and $0^+_2$ is consistent
with the previous works such as \cite{OGAWA,DOTE,OERTZEN}. 
In the $K^\pm=1^-$ band, the deformation toward the prolate shape 
shrinks as the total spin $J$ increases.

In the $K^\pi=0^+_1$ band,
the development of the $2\alpha$ cores 
weaken with the increase of the total spin
due to the spin-orbit force. The reduction of the clustering structure
is more rapid in the case of interaction (h) with the stronger spin-orbit
force. As a result, 
the $2^+_1$ and $4^+_1$ states in the case (h) interactions
contain the dissociation of $\alpha$.
Regarding the dissociation of the $\alpha$ cores,
the structures of those states are sensitive to the strength of the 
spin-orbit force.

\subsection{Behavior of valence neutrons}

Even though we did not assume the existence of any clusters in the
model, we have found that the most of the intrinsic states 
of $^{10}$Be contain
$2\alpha+2n$ structures. 
We study the behavior of the valence neutrons surrounding $2\alpha$
by analyzing the single-particle wave functions
to understand the role of the valence neutrons in the neutron-rich Be nuclei.
Considering that an intrinsic state is approximately 
written by a Slater determinant $\Phi_{AMD}$, 
the single-particle wave functions and the single-particle energies 
of an intrinsic state are determined 
by diagonalizing the single-particle Hamiltonian
which has been already explained in \ref{subsub:nskin}.

\begin{figure}%----------------------------------------
\noindent
\epsfxsize=1.0\textwidth
\centerline{\epsffile{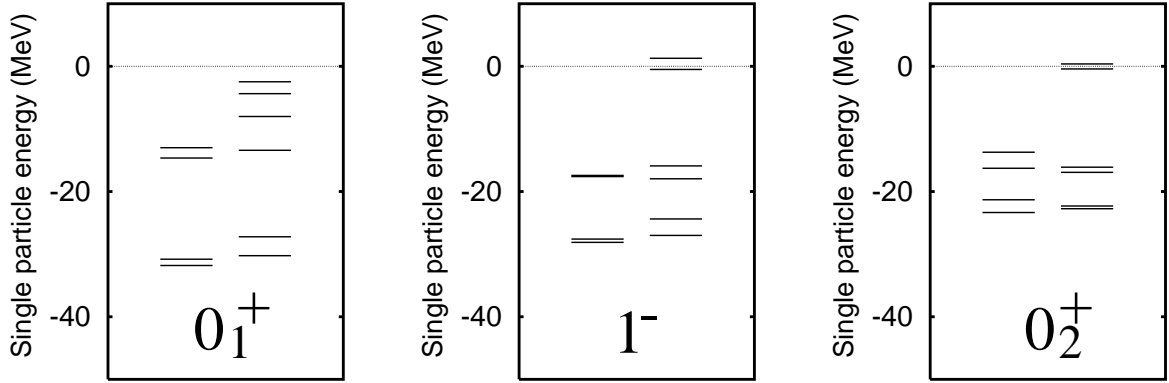}}
\caption{\label{fig:be10hfe}
Single-particle energies in the intrinsic systems of the 
$0^+_1$, $1^-$ and $0^+_2$ states. The energies of protons(neutrons)
are displayed in the left(right) side in each figure.
}
\end{figure}

\begin{figure}%----------------------------------------
\noindent
\epsfxsize=0.5\textwidth
\centerline{\epsffile{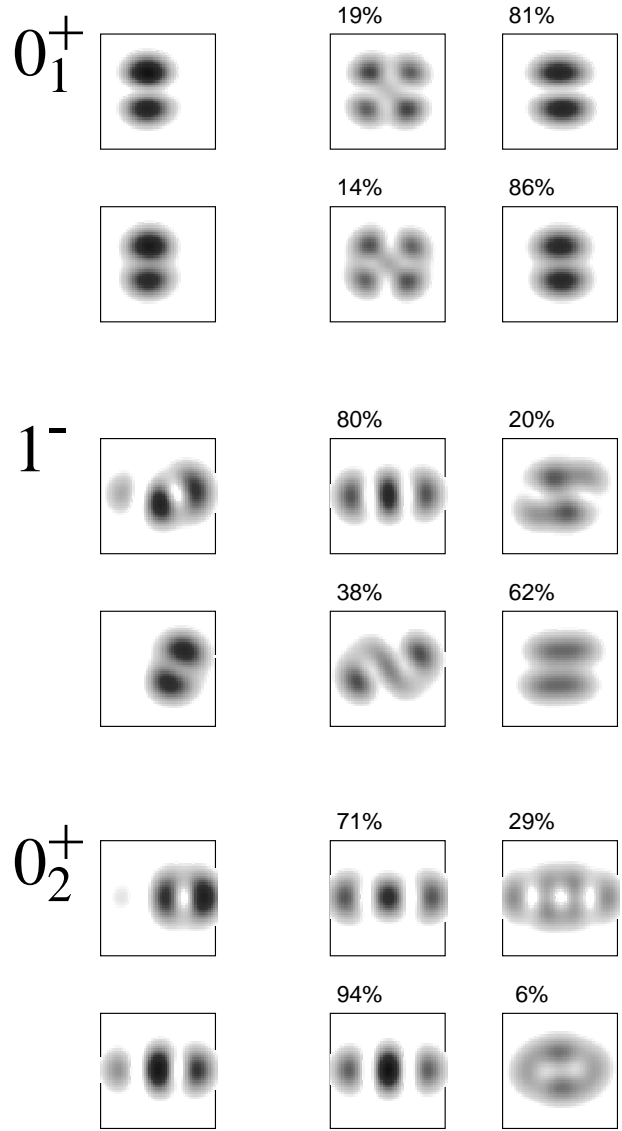}}
\caption{\label{fig:be10sing}
Density distribution of single-particle wave functions of the 
valence two neutrons
in the intrinsic states of $0^+_1$, $1^-$ and $0^+_2$
(left column).
The method to extract the single-particle wave functions are 
explained in the text.
The middle and right columns are for the density of the positive parity 
and negative parity components 
projected from the single-particle wave functions, 
respectively. The wave functions projected 
into the parity eigen states are 
normalized in the presentation.
}
\end{figure}

\begin{figure}%----------------------------------------
\noindent
\epsfxsize=0.5\textwidth
\centerline{\epsffile{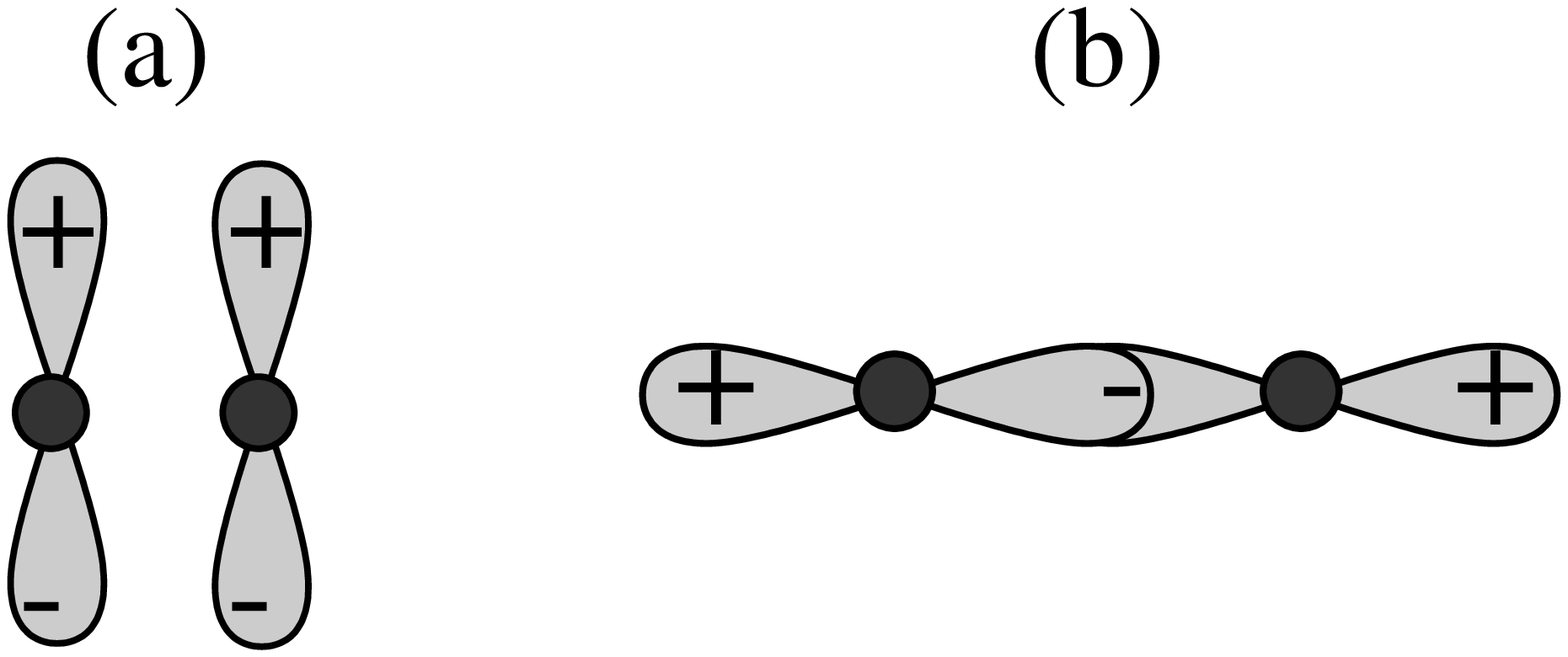}}
\caption{\label{fig:sigmapi}
Schematic figures of the molecular orbits $\pi$ bond (a)
and $\sigma$ bond (b) surrounding
2$\alpha$ clusters.
}
\end{figure}

The single-particle energies in the 
$0^+_1$, $1^-$ and $0^+_2$ states are shown 
in Fig.\ \ref{fig:be10hfe}. In each state four neutron orbits 
from the bottom correspond to those for the neutrons in the 
$2\alpha$ clusters.
The level spacing of these four lower orbits becomes 
smaller in $1^-$ than in $0^+_1$ and smallest in the 
$0^+_2$ state with the increase of the distance between clusters.
We consider the last two neutrons in the higher orbits as the 
valence neutrons surrounding 2$\alpha$ cores.
We display the density distributions of 
the single-particle wave functions for the two valence neutrons
in the left column of Fig.\ \ref{fig:be10sing}. 
Figures in the middle and right columns 
of Fig.\ \ref{fig:be10sing} 
are for the normalized density of the positive 
and the negative parity components
projected from the single-particle wave functions, 
respectively.
By analyzing the single-particle wave functions it is found that
two valence neutrons of the $0^+_1$ states contain the 
negative parity components more than $80\%$
(the right column in figure\ \ref{fig:be10sing}) which seem to be 
$\pi$ bonds(Fig.\ \ref{fig:sigmapi}(a))
in terms of molecular orbits.
On the other hands in the case of $0^+_2$ band, the last 
2 neutrons are predominantly in the positive parity orbits, 
which are analogous 
to the $\sigma$ bonds (Fig.\ \ref{fig:sigmapi} (b)).
In the $1^-$ band, each valence neutron contains both the positive parity 
component like $\sigma$ and the negative parity one similar to $\pi$.
Since the parity of the total system is negative 
in the $1^-$ band,
the states are considered to have
one neutron in $\sigma$ orbit and the other neutron in $\pi$ orbit
after the parity projection. Therefore
roughly speaking, the $0^+_1$, $1^-$ and $0^+_2$ states
are understood as $2\alpha$ and two valence neutrons in $\pi^2$,
$\sigma\pi$ and $\sigma^2$ orbits, respectively. 
The interesting point is that the valence neutrons play important roles to 
develop the clustering structure in the excited bands.
We can argue that the clustering develops in the $1^-$ band 
and mostly in $0^+_2$ band owing to the $\sigma$ orbits of 
the valence neutrons, because the $\sigma$ orbit prefers the prolately 
deformed system as to gain its kinetic energy.
This idea originates from the application of 
the two centers shell model to the $2\alpha$ dimer model of Be isotopes 
by W. von Oertzen \cite{OERTZEN}
and consistent with the argument in the 
work with the method of AMD+HF by Dot\'e et al. \cite{DOTE}. 

From the view points of the shell model, 
the levels $J^\pi=$$0^+_1$, $2^+_1$, $4^+_1$ 
in the $K^\pi=0^+_1$ band and $2^+_2$, $3^+_1$ in the $K^\pi=2^+_1$
 are the $0\hbar\omega$ states. 
These states are dominated by the components with 
the same [442] spatial symmetry 
in the supermultiplet limit.
On the other hands, the negative parity states 
in the $K^\pi=1^-$ are dominated by 
the $1\hbar\omega$ configurations as for the neutron $p$-shell. 
In the excited states in the $K^\pi=0^+_2$ band, 
the $2\hbar\omega$ configurations with 2 neutrons excited
are prime.
It is consistent with the rotational band 
seen in the $2\hbar\omega$ states of shell model calculations
\cite{BROWN}.
However the well-developed clustering structures in the $K^\pi=1^-$ 
and the $K^\pi=0^+_2$ bands
contain considerably the higher configurations 
which are not included explicitly 
in the wavefunction of the $1\hbar\omega$ model space and 
$(0+2)\hbar\omega$ model space of the shell model.
Also in the 
Skyrme Hartree-Fock calculations \cite{TAKAMI},
the $0^+_2$ state is found to be a state with 2 neutrons in $sd$-orbits. 


\subsection{Results of $^{12}$Be}

In $^{12}$Be, one of the interesting problem is 
the vanishing of the magic number $N=8$
which is closely related with the abnormal parity
of the ground state of the $^{11}$Be.
Another subject on $^{12}$Be is the molecule-like structure 
in the excited states which is reported in Ref. \cite{FREER}.

\begin{figure}%----------------------------------------
\noindent
\epsfxsize=0.5\textwidth
\centerline{\epsffile{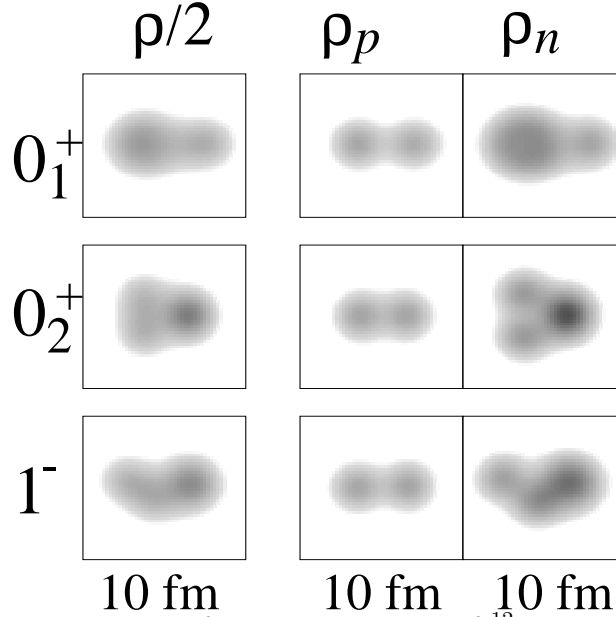}}
\caption{\label{fig:be12dens}
The intrinsic structure of the excited states of $^{12}$Be
obtained by VAP calculations with the interaction (h).
The density distribution of matter, protons and neutrons
of the intrinsic states
are shown at left, middle and right, respectively.
}
\end{figure}

In present calculations, we adopt the interaction (h) with which we can
reproduce the spin parity of the ground state 1/2$^+$ of $^{11}$Be.
For comparison, we also use the interaction (g) which contains
the weaker spin-orbit force than the interaction (h) and fails to
reproduce the parity inversion in $^{11}$Be.
In the theoretical results of the energy levels we found many 
rotational bands such as $K^\pi=0^+_1$, $K^\pi=1^-_1$
and $K^\pi=0^+_2$.
We present the density distributions of the intrinsic structures
in Fig.\ref{fig:be12dens} for the results with the interaction (g).
It is surprising that 
the intrinsic state of the  
ground band $K^\pi=0^+_1$  is the largely deformed molecule-like state
with the dominant 2p-2h state in terms of neutron $p$-shell
which is analogous to the second $0^+$ state in $^{10}$Be,
while the second band $K^\pi=0^+_2$ is dominated by 
the normal $0\hbar\omega$ state with the rather spherical shape due to the
neutron closed shell. 
It means the abnormal phenomena 
that the $sd$-orbits energetically intrude in the $p$-orbits
and the magic number of $N=8$ vanishes in the $^{12}$Be.
On the other hands, adopting the interaction (g) we obtain the 
rather spherical ground state with the closed neutron $p$-shell,
and the deformed structure with 2p-2h appears
in the second $0^+$ state lying a few MeV above the ground state. 
Such a difference in results 
between the adopted interactions (g) and (h) originates
from the strength of spin-orbit force.
The stronger spin-orbit force in (h) prefers the closed 
$p_{3/2}$-shell and breaks the $p$-shell in Be isotopes.
With both interaction (g) and (h) we find 
the $K^\pi=1^-$ band with the developed clustering structure 
in the low-lying levels.
The $K^\pi=0^+$ band with the 2p-2h configurations
reaches to the $J^\pi=8^+$ state, and 
the band terminal of $K^\pi=1^-$ is found to be a $J^\pi=6^-$ state.
The spins $8^+$ and $6^-$ at the band terminal are the same as the 
the highest spins in the corresponding 
configurations $(0p)^2_\pi(0p)^4_\nu (0d1s)^2_\nu$ and
$(0p)^2_\pi(0p)^5_\nu (0d1s)^1_\nu$, respectively.

The $2\alpha+4n$ clustering structures are seen in many levels in $^{12}$Be. 
By analyzing the single-particle wave functions of the intrinsic structures
in similar way to $^{10}$Be, we can roughly 
consider the $0^+_1$,$1^-$, and $0^+_2$ states
as the state with 4 neutrons in $\pi^2\sigma^2$, $\pi^3\sigma^1$
and $\pi^4$ orbits surrounding $2\alpha$ in terms of the 
molecular-orbits $\pi$ and $\sigma$ for the valence neutrons. 
The clustering develops as the number of the valence 
neutrons in the $\sigma$ orbit increases.

The experimental value for Gamow-Teller transition strength 
$B(GT)$ from the $^{12}$Be ground state to the $^{12}$B ground state 
is 0.59  \cite{CHOU} deduced from experimental data of $\beta$ decay.
The weaker GT decay from $^{12}$Be than expected
with the shell-model calculations in the $0p$-shell model space
has been discussed in the relations with the possible 
abnormal structure of the ground state \cite{TSUZUKIa}.
The experimental $B(GT)$ value is the useful probe to 
investigate the structure of the $^{12}$Be ground state.
One possible reason for the weak decay is considered to be that the
$^{12}$Be ground state has more intruder admixture of the higher
shell orbits. However there was no microscopic calculation
which can describe the weak decay.
To evaluate the $B(GT)$ we obtain the wave function for the final state
by performing VAP calculations of $^{12}$B with $(J^\pi,K')=(1^+,-1)$.
The calculated $B(GT)$ with the interaction (h) is 0.8 which well agrees 
to the experimental data 0.59.
Before the diagonalization for the obtained states,
the GT strength from 
the $0^+_1$ state 
with large $2\hbar\omega$ components is 0.2, 
and that from the $0^+_2$ state with $0\hbar\omega$ is 3.0.
The reason for the former value is that the state
with $2\bar\omega$ configurations is forbidden to decay to the 
$^{11}$B ground state which is almost the normal $0\hbar\omega$ state. 
The latter value is as much as the simple shell-model calculation
and is much larger than the experimental data.
After the diagonalization for the spin-parity eigen states projected from the 
obtained intrinsic states, the $0^+_1$ state
and the $0^+_2$ state are mixed to
redistribute the decay strength for the $0\hbar\omega$ state
into other states. As a result, $B(GT)$ for
$^{12}$Be($0^+_1$)$\rightarrow$$^{12}$B($1^+$) is 0.8
and the one for  $^{12}$Be($0^+_2$)$\rightarrow$$^{12}$B($1^+$)
is 2.1.
Is it found that 
the main reason for the weak Gamow-Teller transition is that the $^{12}$Be 
ground state is dominated by the abnormal state with 2$\hbar\omega$
configurations.
On the other hands, the calculations with the interaction (g) give
much larger $B(GT)$ strength for
$^{12}$B($0^+_1$)$\rightarrow$$^{12}$B($1^+$) than the experimental data
because in this case the main component of 
the ground state of $^{12}$Be is the $0\hbar\omega$ state.

The abnormal $2p-2h$ structure of the ground band of $^{12}$Be may be
observed in the strength of $E2$ transitions. By using the interaction
(h) which give the abnormal state for the ground band of $^{12}$Be, 
the $E2$ transition strength in the second band 
$B(E2;2^+_2\rightarrow 0^+_2)$ is as small as the simple AMD calculations,
however the $E2$ strength in the ground band 
$B(E2;2^+_1\rightarrow 0^+_1)$ is predicted to be 
almost twice as large as the simple AMD result.

\section{Mechanism of the developments of clustering structure in
 Be isotopes
}\label{sec:cluster}

In the study with the simplest version of AMD and with the VAP calculations,
it is found that 2$\alpha$-cluster cores are seen in most of the low 
energy states of Be isotopes heavier than $^7$Be.
In this section, we try to understand the mechanism of the
 clustering developments in Be isotopes systematically.
We discuss this problem from two view points. 
First we take care of the valence neutrons in the molecular orbits.
In the second subsection we give the discussion from the other 
view point of the two-center clusters.

\subsection{Molecular orbits of valence neutrons}

  In the study of Be isotopes with the simplest version of AMD, 
we have already shown the dependence of the development 
of the clustering structure on the neutron number (Fig. \ref{fig:berpp}).
We have discussed the clustering of the lowest states of the 
normal parity states with $0\hbar\omega$ configurations 
and the non-normal parity states with $1\hbar\omega$.
In the study of the excited states with VAP calculations in the framework
of AMD, we have found the rotational bands. The interesting point is that 
the rotaional bands can be classified by the number
of the neutrons in the $sd$ orbits.

As already mentioned in the subsection \ref{subsec:be10},
in the case of $^{10}$Be, three kinds of the intrinsic states
construct the rotational bands.
The intrinsic states are characterized by the 
orbits of the two valence neutrons surrounding $2\alpha$ cores.
The lowest positive parity state in the $K=0^+_1$ band 
corresponds to the state with 2 neutrons in $p$-orbits. 
On the other hands, one neutron occupies the $sd$-orbit 
in the excited states in th $K=1^-_1$ band.
Furthermore the states with two neutrons in the $sd$-orbits 
appear in the second $K=0^+_2$ band.
The clustering develops larger and larger from $K=0^+_1$
to $K=1^-_1$ and to $K=0^+_2$.

According to the VAP calculations, 
we found these three kinds 
of the intrinsic states with $2\alpha$ cluster cores
in the other neutron-rich nuclei $^{11}$Be
and $^{12}$Be as well as $^{10}$Be .
Namely the most of the excited states in the low-energy region
are classified into the states with the all the valence neutrons 
in the $p$-orbits, the states with one neutron in $sd$-orbit, and the states 
with two neutrons in $sd$-orbits which surround the $2\alpha$ cores.
Strictly speaking the single-particle motion of the valence neutrons surrounding
$2\alpha$ is not necessarily the ideal independent one
but also the correlation between neutrons should be important.
However, we think it useful to discuss the structure of Be isotopes from the 
viewpoints of molecule-like orbits surrounding $2\alpha$
because the component that neutrons occupy the
molecule-orbits is significant in the neutron-rich Be isotopes 

\begin{figure}%----------------------------------------
\noindent
\epsfxsize=0.7\textwidth
\centerline{\epsffile{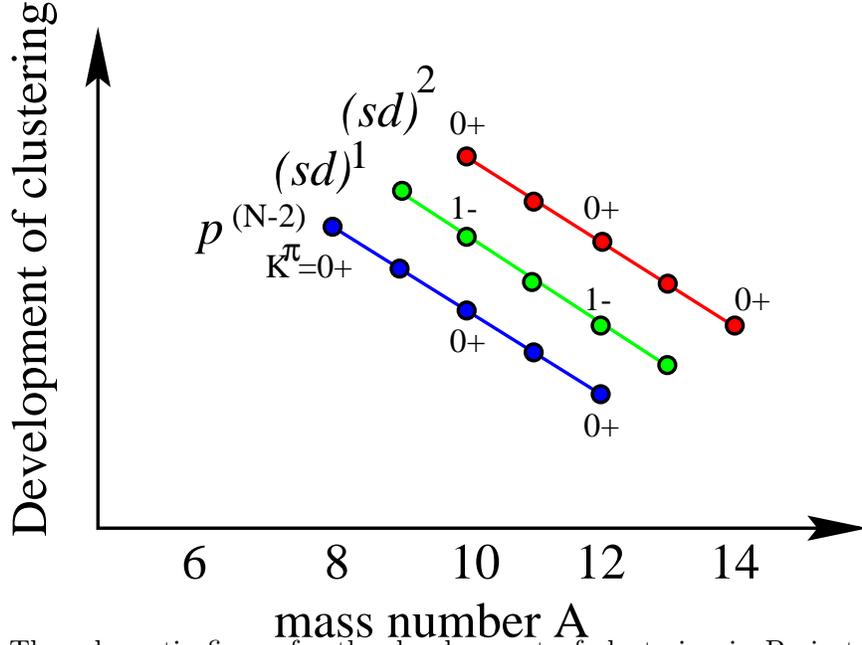}}
\caption{\label{fig:berpp1}
The schematic figure for the development of clustering
in Be isotopes. The lines correspond to the cluster development 
of the lowest states in the given configurations
(all the valence neutrons in the $p$-orbits, one neutron in the $sd$-orbits,
and two neutrons in the $sd$-orbits)
as the function of the mass number.
The intrinsic states of the lowest states well correspond to the 
ones in the rotational bands.
}
\end{figure}

Here we improve the figure \ref{fig:berpp} for the development 
of the clustering in Be isotopes from the view points of the 
orbits of the valence neutrons surrounding $2\alpha$ cores.
The Fig.\ref{fig:berpp1} presents the schematic figure for the 
cluster development of the lowest states in the given configurations. 
The lower line is for the state with all the 
valence neutrons surrounding $2\alpha$ in the $p$-orbits.
The middle and upper lines correspond to the state with 
the one neutron and two neutrons in the $sd$-orbits, respectively. 
With the help of the figure \ref{fig:berpp1} we can understand the 
development of the clustering in the Be isotopes systematically.

In the states with the same number 
of the valence neutrons in $sd$-orbits, the clustering development 
decreases gradually with the increase of the neutron number. 
The mechanism of the clustering development is easily understood 
by the idea of molecule-like orbits surrounding $2\alpha$ cores.
In Be isotopes, 
the single particle wave functions of the the valence neutrons 
surrounding $2\alpha$ cores 
are analogous to the molecule-like orbits, $\pi$  
and $\sigma$ orbits, which correspond to the $p$-orbit and $sd$-orbit in
terms of the shell-model respectively, 
as shown in the subsection \ref{subsec:be10}.
In that sense the additional neutrons in each lines occupy the $\pi$-orbit 
as the neutron number increases. 
Therefore it seems that the valence neutrons in the $\pi$-orbits
effect as to weaken the spatial clustering.

In the each Be nuclei, as the number of the neutrons in $sd$ orbits
increases the clustering develops larger and larger. In the case of 
Be isotopes, the $sd$ orbits correspond to the $\sigma$ orbits.
It is naturally understood that the valence neutrons in $\sigma$
orbits prefer the developed clustering states so as to gain 
the their kinetic energy.
That is why the clustering develops as the increase of the number of
neutrons in the $\sigma$ orbits. 

\subsection{Viewpoint of the di-cluster structure}

As mentioned above, one of the viewpoints for
the mechanism of the clustering development is the explanation with the
molecule-like orbits $\sigma$ and $\pi$.
Here we try to understand the clustering in the excited states of 
$^{10}$Be from the other viewpoint
of the two-center clustering model.

The reason for the clustering development 
in the ordinary nuclei has been
considered as the system gains the kinetic energy 
with the development of clustering
even if the potential energy looses. 
In order to understand the mechanism of clustering 
in light unstable nuclei, we think it helpful 
to investigate the competition of the 
kinetic and the potential energies in the molecule-like states
of $^{10}$Be.

\begin{figure}%----------------------------------------
\noindent
\epsfxsize=0.7\textwidth
\centerline{\epsffile{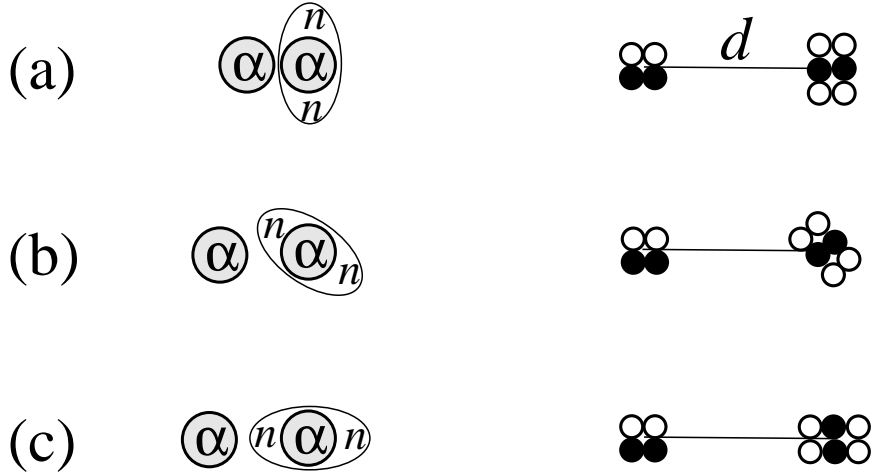}}
\caption{\label{fig:be10dim}
Schematic figures for the intrinsic structure of $0^+_1$, $1^-$ and
$0^+_2$ from the view point of the two-center cluster model 
are shown in the left columns of (a), (b) and (c), respectively.
In the right columns we display three kinds of the configurations of 
the centers of Gaussians 
in the simplified AMD wavefunctions 
which indicate the two-center $^6$He+$\alpha$ cluster model
for the excited states $0^+_1$(a), $1^-$(b) and
$0^+_2$(c) of $^{10}$Be.
The black(white) circles correspond to the centers of Gaussians
of the single-particle wave functions of protons(neutrons).
}
\end{figure}

\begin{figure}%----------------------------------------
\noindent
\epsfxsize=.45\textwidth
\centerline{\epsffile{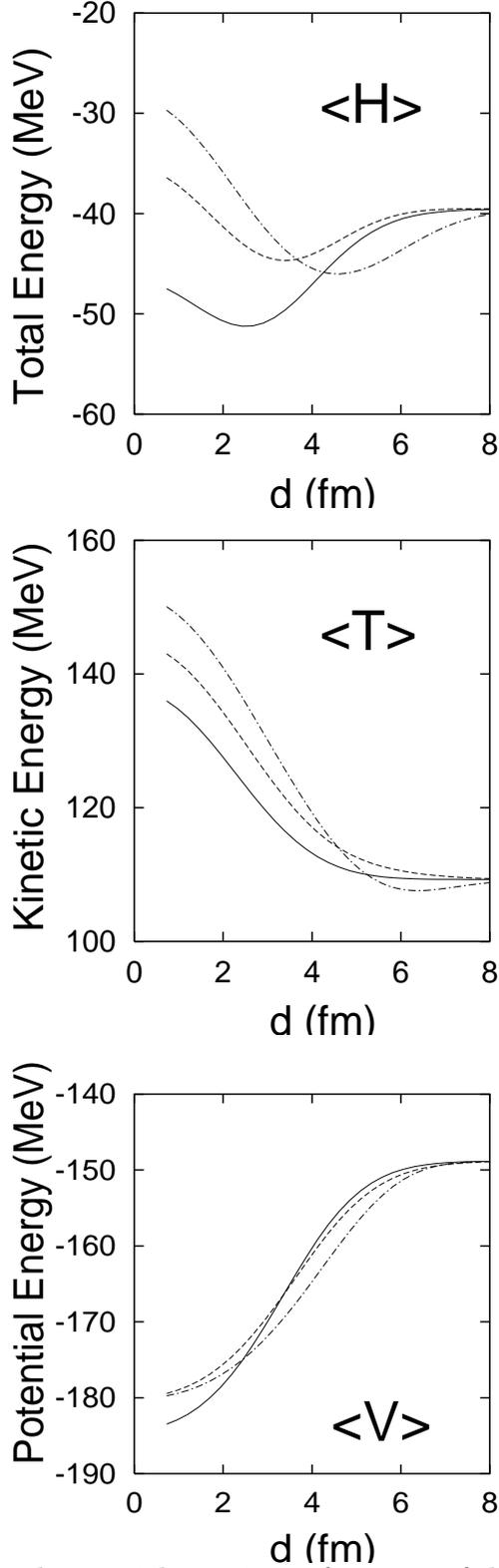}}
\caption{\label{fig:be10htv}
Total, kinetic and potential energies as functions of 
the distance between clusters
in the simplified $^6$He+$\alpha$ cluster models for the excited states of 
$^{10}$Be.
The adopted interaction is the MV1 force with $m=0.62$, and 
the spin-orbit and the Coulomb forces are omitted.
The distance $d$ between clusters and the configuration for the 
$0^+_1$, $1^-$ and $0^+_2$ states are defined in the text.
The solid lines, the dashed lines and the dot-dashed lines 
correspond to the energies
in the system for $0^+_1$, $1^-$ and $0^+_2$ states of $^{10}$Be, 
respectively. It is found that the minimum point shifts
outward from $0^+_1$ to $0^+_2$ 
because of the repulsive kinetic force in the small $r$ region.
}
\end{figure}

In the density distribution of VAP results, the  
structures of the $0^+_1$, $1^-$ and $0^+_2$ states seem to be
the two-center clustering structures which consist of $^6$He+$\alpha$
(see figure \ref{fig:be10dim}). 
To estimate the dependence of the kinetic and potential energies on the
degree of the spatial clustering development we represent the three kinds of 
configurations for the $^6$He+$\alpha$ system
 corresponding to the $0^+_1$, $1^-$ and $0^+_2$ states of $^{10}$Be 
by the simplified AMD wave functions $\Phi_{AMD}({\bf Z})$ 
as follows. The intrinsic spins of the single-particle wave 
functions are fixed to be up or down
for simplicity.
For the $^6$He+$\alpha$ system with the inter-cluster distance $d$ (fm), 
the centers of single-particle Gaussian wave functions are located around  
two points $\vec{a}_1=(-3d/5\sqrt{\nu},0,0)$ and 
$\vec{a}_2=(2d/5\sqrt{\nu},0,0)$.
The three kinds of configurations of the centers for 
$0^+_1$, $1^-$ and $0^+_2$ states are 
shown in the right figures of Fig.\ \ref{fig:be10dim}. 
We put the centers for $p\uparrow$, $p\downarrow$, $n\uparrow$, 
$n\downarrow$ at the point $\vec{a}_1$ and $p\uparrow$, $p\downarrow$ at 
$\vec{a}_2$.
The centers for the last 4 neutrons are located at the points very close
to $\vec{a}_2$ as $\vec{a}_2\pm\vec{\delta}$ where enough small 
$\vec{\delta}$ is chosen so that the angle $\theta$ between $\vec{\delta}$
and $\vec{a}_2$ is $\theta$=
$\pi/2$, $\pi/4$ and $0$ corresponding to the $0^+_1$, $1^-$ 
and $0^+_2$ states, respectively. 
For the parity eigen states projected from 
these three kinds of the simplified AMD wave functions 
$\Phi_{AMD}({\bf Z})$
we calculate the expectation values of total, kinetic and potential 
energies as the function of the inter-cluster distance $d$;
\begin{eqnarray} 
& &
\langle H \rangle\equiv 
{\langle(1\pm P)\Phi_{AMD}({\bf Z})|H|(1\pm P)\Phi_{AMD}({\bf Z})\rangle
\over \langle(1\pm P)\Phi_{AMD}({\bf Z})|(1\pm P)\Phi_{AMD}({\bf Z})\rangle}\\
& &
\langle T \rangle\equiv 
{\langle(1\pm P)\Phi_{AMD}({\bf Z})|T|(1\pm P)\Phi_{AMD}({\bf Z})\rangle
\over \langle(1\pm P)\Phi_{AMD}({\bf Z})|(1\pm P)\Phi_{AMD}({\bf Z})\rangle}\\
& &
\langle V \rangle\equiv 
{\langle(1\pm P)\Phi_{AMD}({\bf Z})|V|(1\pm P)\Phi_{AMD}({\bf Z})\rangle
\over \langle(1\pm P)\Phi_{AMD}({\bf Z})|(1\pm P)\Phi_{AMD}({\bf Z})\rangle},
\end{eqnarray}
where we omit the spin-orbit and the Coulomb forces for simplicity. 
In Fig.\ \ref{fig:be10htv} we present 
the total energy, the kinetic energy and the potential energy
for the three kinds of clustering states 
$0^+_1$, $1^-$ and $0^+_2$ as the function of the distance $d$ between
$^6$He and $\alpha$ clusters. As shown in the figure for 
the total energy $\langle H \rangle$,
 the optimum distances indicate that
the clustering structure develops in the system for the $1^-$ state
and is most remarkable in the state for $0^+_2$, which is consistent with the
present results of VAP calculations.
The shift of the minimum point of the total energy is
understood by the energetically advantage of the kinetic part as follows.   
It is found that the kinetic energies in the small $d$ region
are sensitive to the configurations,
while in the case of the potential energies significant differences 
are not seen in the three configurations.
When the cluster approaches each other, 
the kinetic energy in the configuration for the $1^-$ state
becomes larger than $0^+_1$ by ${1\over 2}\hbar\omega$ 
(about 7 MeV for $\nu=0.17$)
because all the neutrons in $0^+_1$ 
are in $0s$ and $0p$ shells. On the other hands,
in the negative parity state for $1^-$ one valence neutron  
must rise to the higher $sd$ shell in the small $d$ region.   
Since in the small $d$ limit the wave function is almost same as 
the harmonic oscillator shell model wave function due to the 
antisymmetrization, the kinetic energy of 
the $sd$ shell is larger by $1/2\hbar\omega$ than the one of $p$ shell. 
Thus the system for $1^-$ loses the kinetic 
energy in the small distance $d$, 
as a result, the minimum point of total energy shifts to the larger 
$d$ region than the one in the $0^+_1$ state.
In the case of the linear configuration for the excited $0^+_2$ state, 
the two valence neutrons occupy the $sd$ orbit in the small $d$ limit and
the kinetic energy is larger than $0^+_1$ by $\hbar\omega$. That is why
the optimum point $d$ in $0^+_2$ is the largest of the three.
In other words, when the cluster approaches each other, the system feels
the repulsive force in kinetic part because of the Pauli principle. 
That is the reason why the clustering structure remarkably develops in the 
$0^+_2$ state. 
In the analysis with this simplified two cluster model, we can conclude
that the clustering develops so as to gain the kinetic energy.
It is compatible with the viewpoint of the molecular $\sigma$ orbit.

\section{Summary}\label{sec:summary}

We studied the structures of the ground and the excited states of light 
unstable nuclei with the theoretical method of AMD.

Li, Be, B and C isotopes were investigated with the simplest version
of AMD. 
We showed the calculated results of 
binding energies, energy levels, radii,
magnetic moments, electric moments, and transition strengths 
comparing with the experimental data.
The AMD calculations systematically reproduce a lot of experimental data 
of many nuclei.
By analyzing the intrinsic wave functions, we found 
the drastic changes between shell-model-like structures and 
clustering structures.
The theoretical results suggest some new features 
 in unstable nuclei such as the clustering
structures, the molecular orbits, the opposite deformation between protons
and neutrons, the neutron skin structures, et al.
We discussed these phenomena with the help of the 
experimental informations.

We adopted VAP calculations to study the excited states of 
neutron-rich Be isotopes.
Many excited levels were reproduced and were predicted by theory.
We showed the theoretical results of the $E1$, $E2$, and $\beta$ 
transition strength, which agree well with the experimental data.
By analyzing the structures of excited states in $^{10}$Be and $^{12}$Be, 
rotational bands 
$K^\pi=0^+_1$, $1^-$, $0^+_2$ are found. 
Particularly $K^\pi=1^-$ and $0^+_2$ bands have the largely 
deformed states with the molecule-like structures which consist of 
the developed $2\alpha$ clusters
and surrounding neutrons. 
We extracted the single-particle wave functions to discuss the behavior 
of the valence neutrons, and found that the valence neutrons in molecule-like
orbits surrounding $2\alpha$ play important roles in the clustering
structure.

In order to understand the mechanism of the clustering development,
we discussed the clustering structures of Be isotopes from 
two view points, the view point of the molecular orbits of 
the valence neutrons and the one of the two-center cluster model. 
We concluded that the clustering development  in neutron-rich nuclei 
can be understood by the roles of the excess neutrons.

Although AMD calculations describe many new phenomena seen in
unstable nuclei, however, future problems remain to be solved. 
For example, 
the extremely large radii due to the neutron halo structures
are not reproduced because the AMD wave functions written by Gaussians
are not enough to represent the long tail of the single particle 
wave functions.
It is not obvious which the effective interaction parameters
should be adopted in AMD framework. 
We must more carefully check the properties of the effective 
interactions such as the spin-orbit force and the central force.
At least when the study goes to the heavier nuclear systems, 
we will soon meet the problems of the saturation property.
That is to say that 
it is difficult to reproduce all the binding energies of 
the stable nuclei in the wide mass-number region
with a set of parameters of the adopted interactions 
in the present work.
Owing to the progress of the computational power, it becomes
possible to make AMD calculations for the heavier unstable nuclei,
where we may discover the other exotic structures such as the 
polymers of $\alpha$ clusters and the super deformations.

Finally we stress again that, as far as we know, 
the present work is the first theoretical study
which can systematically describe the various clustering structures 
and shell-model-like structures in the ground and the excited states 
of many light unstable nuclei in one framework 
without assuming any inert cores nor the existence of clusters. 

The computational calculations of this work were partially supported by 
the Research Center for Nuclear Physics in Osaka University.
We owe the calculations to 
the system in Yukawa Institute for Theoretical Physics in Kyoto University,
and the super computer systems in High Energy Accelerator Research 
and in the Institute of Physical and Chemical Research.

\end{document}